\documentclass[useAMS,usenatbib]{mn2e}
\usepackage{graphicx}
\usepackage{lscape}
\title[Automated Merger Detection Using Residuals.]{A New Automatic Method to Identify Galaxy Mergers I. Description and Application to the \textsc{STAGES} Survey.
\thanks{Based on observations made with the NASA/ESA Hubble
Space Telescope, obtained at the Space Telescope Science Institute,
which is operated by the Association of Universities for Research in
Astronomy, Inc. under  NASA contract  NAS 5-26555. These observations
are associated with programme GO10395.}}

\author[Carlos Hoyos \textit{et al}.]{Carlos Hoyos$^{1,2}$, Alfonso Arag\'on-Salamanca$^{1}$, Meghan E. Gray$^{1}$,\newauthor
David T. Maltby$^{1}$, Eric F. Bell$^{3}$, Fabio D. Barazza$^{4}$, Asmus B\"ohm$^{5}$, \newauthor
Boris H\"au\ss ler$^{1}$, Knud Jahnke$^{6}$, Sharda Jogee$^{7}$, Kyle P. Lane$^{1,8}$,\newauthor
Daniel H. McIntosh$^{9}$, Christian Wolf$^{8}$.
\\$^{1}$School of Physics and Astronomy, The University of Nottingham, University Park, Nottingham, NG7 2RD, UK.
\\$^{2}$Departamento de F\'isica Te\'orica, Facultad de Ciencias, Universidad Aut\'onoma de Madrid, Cantoblanco, 28049 Madrid, Spain.
\\$^{3}$Department of Astronomy, University of Michigan, 500 Church St., Ann Arbor, MI 48109, USA.
\\$^{4}$Department of Physics, University of Basel, Klingelbergstrasse 82, 4056 Basel, Switzerland.
\\$^{5}$Institute of Astro and Particle Physics, University of Innsbruck, Technikerstr. 25/8, A-6020 Innsbruck, Austria.
\\$^{6}$Max-Plank-Institut f\"ur Astronomie, K\"onigstuhl 17, D-69117, Heidelberg, Germany.
\\$^{7}$Department of Astronomy, University of Texas at Austin, 1 University Station C1400 RLM 16.224. Austin, TX 78712-0259, USA.
\\$^{8}$Department of Physics, University of Oxford, Denys Wilkinson Building,  Keble Road. Oxford, OX1 3RH, UK.
\\$^{9}$Department of Physics, University of Missouri-Kansas City, 5110 Rockhill Road, Kansas City, MO 64110, USA.
}

\begin{document}

\date{\today}
\pagerange{\pageref{firstpage}--\pageref{lastpage}} \pubyear{2011}
\maketitle
\label{firstpage}
\newcommand{\MBY}
{
\citep{DavidMaltby2009}
}
\newcommand{\TABSAMPLE}
{
\ref{tab:trainsample1}
}

\small

\begin{abstract}
We present a new automatic method to identify galaxy mergers  
using the morphological information contained in the residual images 
of galaxies after the subtraction of a smooth S\'ersic model.
The removal of the bulk signal from the host galaxy light is done with 
the aim of detecting the much fainter and elusive minor mergers.
The specific morphological parameters that are used in the
merger diagnostic suggested here are the Residual Flux Fraction (\textit{RFF}), and
the asymmetry of the residuals ($A(\mathrm{Res})$).
The new diagnostic has been calibrated and optimized so that the resulting merger sample is very complete. 
However, the contamination by non-mergers is also high.
If the same optimization method is adopted for combinations of other structural parameters such as the CAS system, the merger 
indicator we introduce yields merger samples of equal or higher statistical quality than the samples obtained through 
the use of other structural parameters.
%The statistical quality of a sample is here defined in an unambiguous and objective way.
We investigate the ability of the method presented here to select minor mergers by 
identifying a sample of visually classified mergers that would not have been picked up by the 
use of the CAS system, when using its usual limits.
However, given the low prevalence of mergers among the general population of galaxies and
the optimization used here, we find that the merger diagnostic introduced in this work is best 
used as a \emph{negative} merger test, i.e., it is very effective at selecting non-merging galaxies.
In common with all the currently available automatic methods, the sample of merger
candidates selected is heavily contaminated by non-mergers, and further steps are needed to produce a clean
merger sample.
This merger diagnostic has been developed using the HST/ACS F606W images of the 
A901/02 multiple cluster system($z=0.165$) obtained by the STAGES (Space Telescope A901/02 Galaxy Evolution
Survey) team. In particular, we have focused on a mass and magnitude 
limited sample ($\log M/M_{\odot}>9.0$, $R_{\mathrm{Vega, Total}}\leq 23.5\mathrm{mag}$)
which includes 905 cluster galaxies and 655 field galaxies
of all morphological types.
\end{abstract}

\begin{keywords}
methods: data analysis, galaxies: clusters: individual: A901/902; galaxies: evolution; galaxies: interactions; galaxies: structure
\end{keywords}

\normalsize

\section{Introduction}
\label{seccion:intro}
%Que son los mergers?
Mergers are the most extreme type of galaxy interaction, as the final
product of a merger event can be totally different from the original objects involved.
Considerable efforts have been devoted towards the understanding
of the physical processes that regulate galaxy mergers starting from the very early
work of \cite{1951ApJ...113..413S} or the seminal simulational works presented in
\cite{1972ApJ...178..623T,1977egsp.conf..401T}.
These work made it clear that, even though the stars rarely collide with each other during
a merger process, such episodes can have dramatic consequences for the gaseous component 
of the galaxies involved.
Later works such as those of \cite{1991ApJ...370L..65B,1992ARA&A..30..705B,1996ApJ...471..115B,2002MNRAS.333..481B,2005A&A...437...69B,2007MNRAS.375..805W,2005ApJ...622L...9S,2008MNRAS.389L...8B,2009MNRAS.397..802H,2009ApJ...702..307S,2010A&A...518A..61C} have helped address
specific issues of merger processes such as the internal structure of the remnants, the relevance
of the orbital parameters or the impact of the gas fraction on the possible regeneration of 
galactic discs after a merger episode.

However, the study of mergers is not only relevant because of the physics involved. The evolution of
the massive early type galaxies that populate the red sequence cannot be explained using 
passive evolution models only, and mergers have been found to play a key role
in their evolution. In particular, the evolution of the luminosity function and colours 
of galaxies since $z\simeq1.0$ observed in the COMBO-17 (Classifying Objects by Medium-Band Observations in 17 Filters)
\citep{2003A&A...401...73W,2004ApJ...608..752B} and the phase 2 of the DEEP (Deep Extragalactic Evolutionary Probe) survey
\citep{2007ApJ...665..265F} suggest that the merger episodes have a huge impact
on the evolution of early-type galaxies, increasing the stellar mass by a factor of two
over the last 8Gyr. More recently, the importance of mergers has been highlighted
in \cite{2010ApJ...719..844R}, who conclude that the evolution of massive, red galaxies
depends strongly on their merging history. Further studies have helped ascertain the 
impact of mergers in specific aspects of the evolution of red galaxies, including masses \citep{2010ApJ...709.1018V}, sizes
\citep[see][among others]{2006MNRAS.373L..36T,2007MNRAS.382..109T,2007NCimB.122.1209G,2008ApJ...687L..61B,2008ApJ...677L...5V}
and velocity dispersions \citep{2009ApJ...696L..43C}.

The identification of mergers in deep astronomical images is thus a very important issue
for galaxy evolution studies, and the huge number of galaxies
observed in modern surveys creates the need of reliable automated merger detection mechanisms.
A reliable merger identification technique is the key element in the calculation
of the merger fraction. The merger fraction is defined as the fraction of galaxies with a recognizable  ongoing 
merger episode that is found in any given (often mass-limited) sample. It is the first step towards the 
\emph{comoving} merger rate, which is the number of merger \emph{events} 
per $\mathrm{Mpc}^{-3}\mathrm{Gyr}^{-1\phantom{1}}$.

Several automatic identification techniques have been developed
to single out mergers from non interacting galaxies. These methods use morphological 
criteria 
\citep[CAS and $G-M_{20}$ systems,][]{2003ApJS..147....1C,2008MNRAS.386..909C,2004ApJ...612..679L,2005MNRAS.357..903C,2004AJ....128..163L,2008MNRAS.391.1137L,2009ApJ...697.1971J},
kinematical and spatial close pairs
\citep{2000ApJ...536..153P,2002ApJ...565..208P,2004ApJ...617L...9L,2007ApJ...666..212D,2008ApJ...681..232L,2010MNRAS.407.1514E,2010ApJ...719..844R},
or even the correlation function \citep{2006ApJ...652..270B,2006ApJ...644...54M}.
The morphological techniques are based on the fact that the objects involved in a merger episode will be gravitationally disturbed.
The CAS (Concentration, Asymmetry, clumpineSs) system measures these specific aspects of
the surface brightness distribution of galaxies in order to identify mergers. Objects
with high asymmetries are usually taken to be mergers by this method.
On the other hand, the $G-M_{20}$ system measures whether the galaxies appear to be shredded or not, since
both the Gini and the $M_{20}$ numbers measure whether and how is the light concentrated in any given object.
In this system, the more shredded galaxies are selected as mergers.
The pairing techniques look for pairs of galaxies whose relative positions and velocities should be conducive
to strong interactions in a relatively short timescale after observation.
Each of these methodologies are sensitive to different time scales, mass ratios, orbital parameters, and 
gaseous content of the galaxies involved. For instance, \cite{2006ApJ...638..686C,2008MNRAS.391.1137L} conclude
that the CAS parameters are sensitive to roughly a timescale of $0.4 \to 1.0 \times 10^{9}$ years, while
the time sensitivity of the pairing techniques depend on the projected separation
between the galaxies.

This work contributes to the morphological automated detection of mergers.
Here, the morphological parameters \emph{of the residual images}
after the subtraction of a smooth S\'ersic model 
\citep[see][for a definition of this profile]{1963BAAA....6...41S} are explored.
This should, at least in principle, better reveal the impact of the gravitational
interaction on the morphology of galaxies.
This was done with the aspiration of detecting minor mergers.
An isolated galaxy will, with time, adopt an approximately symmetric profile whereas an interacting galaxy will appear to be
more asymmetric. The removal of an intrinsically symmetric profile such as the S\'ersic model, which
could be regarded as the quiescent, underlying galaxy, will
more clearly expose the asymmetric signature of the light from an interacting galaxy.
Thus, the structure of the residuals is investigated with the aim of
finding the combination of structural parameters that produces merger samples
of better statistical quality. This optimization step is done in an unambiguous
way, by using an objective criterion to grade the performance
of the diagnostics tried. The specific criterion used here encourages completeness 
at the expense of a fairly high contamination by non-mergers, and
the resulting merger sample needs to be cleaned afterwards.

%As an application, the method is then used to investigate the impact of environment and morphology in the merger fraction. This part of the work uses the HST/ACS F606W images of the A901/902 multiple cluster of galaxies obtained by the Space Telescope A901/902 Galaxy Evolution Survey (\textsc{STAGES}).

In this paper, \S \ref{sec:obsdata} presents the observational data used together 
with the galaxy samples selected to derive and then test the proposed method. Section \S \ref{sec:numeros} 
describes the data processing techniques employed, and the structural parameters used. Section \S \ref{sec:ident}
presents the objective method introduced here to determine what combination of structural
parameters produces the merger sample of highest statistical quality.
The precise definition of the ``statistical quality'' of a sample that is selected from a parent population 
is given in \S \ref{subsec:Fscore}.
Section \S \ref{sec:further} presents a visual analysis of the 
objects selected as potential mergers by the method presented here, focusing on the contamination
of the resulting sets of merger galaxies by non-mergers (\S \ref{subsec:contaminacion}) and on the ability to detect 
minor mergers (\S \ref{subsec:minors}).
Finally, \S \ref{sec:conclusiones} presents the conclusions of this work.

%Throughout this paper, the concordance cosmology with $\Omega_{m}=0.3$, $\Omega_{\Lambda}=0.7$ and $\mathrm{H}_{0}=70.0 \mathrm{km\ s}^{-1}\mathrm{Mpc}^{-1}$ is used. In this cosmology, 1 arcsecond corresponds to 2.83 kpc at $z=0.165$.

\section{Data}
\label{sec:obsdata}

%Lo primero, una cepillada a STAGES.
The data used to illustrate this method is provided by the HST/ACS F606W observations that were obtained as a part of the
Space Telescope A901/902 Galaxy Evolution Survey \textsc{(STAGES)}\footnote{http://www.nottingham.ac.uk/astronomy/stages}
\citep{2009MNRAS.393.1275G}.
\textsc{STAGES} is a multiwavelength project that was designed to explore the impact of environment on galaxy evolution.
Its main target is a multiple cluster of galaxies located at $z \simeq 0.165$ that harbors different environments 
with different densities. At the distance of this multiple cluster, 1 arcsecond corresponds to 2.83 kpc. 
This survey also includes X-ray XMM-Newton, UV GALEX, IR Spitzer, spectroscopic 2dF, Radio Giant 
Metrewave Telescope, and optical COMBO-17 observations.

%Ahora, hablar un poco de las observaciones HST.
The HST/ACS \textsc{STAGES} observations form an 80-tile mosaic that covers almost $30 \times 30\ \mathrm{arcmin}^{2}$
in the F606W filter, with an average exposure time of around 2 ks. The observations were reduced using an output
pixel scale of 0.03 arcsec, and a \textit{pixfrac} of 0.8, in order to keep the PSF ellipticity as stable as possible
for weak lensing studies. The PSF Full-Width-Half-Maximum (FWHM) is 3.12 pixels. The point source completeness limit of 
these images is $F_{606W,AB}=28.5\mathrm{mag}$.
These observations are deep enough so as to probe most of the luminosity function of galaxies in clusters, as it is possible
to recover reliable structural information for galaxies up to an absolute magnitude of $M_{F606W,AB}=-15.0\mathrm{mag}$. This limit
is just two magnitudes brighter than the most luminous globular clusters, and is typical of dwarf
elliptical systems.

\subsection{Sample Selection \& Morphologies}
\label{subsec:samples}

This work makes use of a mass and magnitude selected sample 
($9.0 \leq \log M/M_{\odot}$, $R_{\mathrm{Vega}} \leq 23.5\mathrm{mag}$)
which is very similar to the sample used in \cite{DavidMaltby2009}.
The galaxies are all in the $0.05 \leq z_{\mathrm{Phot}} \leq 0.30$ redshift range, with
a relative dearth of sources in the lower end of this interval.
The mass limit ensures that the sample is complete in stellar mass for both the blue cloud and the 
red sequence, as was shown in \cite{2006A&A...453..869B}. The magnitude limit guarantees 
reliable visual morphologies, since all the sources show extended images in the HST/ACS data. The
sample includes 1560 galaxies distributed among both the Field and the Cluster environments.
The sample can also be divided into four different morphological 
classifications labeled as ``E'', ``S0'', ``Sp'', and ``Oth''. The ``E'' bin is made of course of elliptical 
systems, the ``S0'' bin gathers the lenticular galaxies, the ``Sp'' bin comprises the spiral 
galaxies, and the ``Oth'' bin includes a mixture of irregulars, compact, and highly disturbed sources
that do not fit into any of the other galaxy classes. The ``Oth'' bin includes 235 irregular galaxies, 41 disturbed
galaxies, and 13 compact sources. This bin is thus dominated by irregular systems.
This sample was selected for the current work due to the following reasons.

\begin{enumerate}

\item[Redshift Uniformity:]{Most of the galaxies are found in the fairly narrow 
$0.16 \leq z \leq 0.30$ range, and only $7\%$ of the objects are in the
$0.05 \leq z \leq 0.16$ interval. The lookback time is thus approximately the 
same for most objects and no age related systematic 
effects are expected. The STAGES HST/ACS observations will also probe similar
rest frame wavelength intervals redwards of the 4000\AA\ break, observing the 
same stellar populations.
The spatial resolution is approximately the same for most sources, too. There is no need 
to apply \emph{morphological} $k-\mathrm{corrections}$ or distance corrections to the structural parameters.
It is important to keep in mind, though, that the total $k-\mathrm{correction}$ spread could be 
relatively large, ranging from 0.0 mags for the nearest blue objects to 1.0 mags for the farthest red galaxies.
The $(1+z)^{4}$ cosmological dimming could also introduce a bias of around 0.5 mag between the
high redshift objects and those located at the cluster's distance.
However, the sample used in this work is a mass limited sample, and
the faint magnitude cut applied is only made in order to ensure that 
all the relevant structural parameters can be measured reliably.
This makes it possible to estimate the fraction of sources that 
could be affected by the $k-\mathrm{correction}$ and cosmological dimming biases
previously mentioned. This is done by adding 1.5 mag to the observed magnitudes and calculating
the fraction of sources that would have been excluded from the sample in this situation.
This simple estimate indicates that 90\% of the galaxies in the current 
sample should be free from these systematics.}

\item[Variety of Environments and Morphologies:]{This sample
includes a good number of galaxies of several morphological types residing in very different environments.
This makes it ideally suited to explore whether the performance of our merger indicators is sensitive 
to these variables or not.
%For instance, despite the Coma cluster being closer than the A901/02 multiple cluster, most of its members are either elliptical or lenticular systems due to its higher density and thus the statistics for spiral galaxies would be poorer.
}

\end{enumerate}

%These key features make the HST/ACS \textsc{STAGES} observations an optimal dataset to investigate the relationship between the morphology and environment of galaxies and the incidence of merger episodes. Many of the issues that could affect this work are discounted by the very nature of the data and/or sample selection.

All the objects of this sample also have good morphological information, which
will be presented in detail in a forthcoming paper (Gray \textit{et al}. in prep.).
The morphological catalogue was created using visual classification from seven 
\textsc{STAGES} team members who were first trained on a consistent subsample of 
previously classified galaxies.  All 5090 galaxies with $R<23.5\mathrm{mag}$ and $z_{\rm phot} < 0.4$ 
were classified blind without knowledge of their cluster or field 
membership.  Galaxies were randomly assigned to classifiers who used a revised 
Hubble T-type scale and weighting scheme described in \cite{2007MNRAS.378..716L}. Each 
galaxy in the sample was classified by three people, while a subset of 786 
bright galaxies previously studied in \cite{2005A&A...443..435W}
received classifications from all seven.

In addition to the revised Hubble T-type, classifiers were also able to note 
further information about the galaxy structure within certain well-defined 
parameters. The dynamical state involved an interpretation of the probable 
cause of any observed disruption, where the possibilities were a tidal 
interaction with a neighbour (I); tidal interaction suggesting a merger (M); 
tidal feature without obvious cause (T); and chaotic systems (C).  The degree 
of disruption was quantified by a disturbance parameter which was allowed 
values of 0, indicating little or no disturbance;  [1,2] indicating moderate/strong asymmetry 
(e.g. an HII region); [3,4] showing a moderate or strong distortion (e.g. a tidal 
tail).

The precise sample definition and its breakdown are now summarized.

\begin{enumerate}

\item[The Cluster Sample:]{This sample is defined by a redshift 
interval $z_{\mathrm{Phot}}=[0.17-\Delta (R) ,0.17+\Delta (R) ]$, where 
the photometric redshifts have been calculated from the COMBO-17 data, and the half-width \\ 
$\Delta (R) = \sqrt{0.015^{2} + 0.0096525^{2} \times (1 + 10^{0.6 \times (R_{\mathrm{Tot}}-20.5)})}$\\
is allowed to vary with apparent $R$ band magnitude. This redshift interval 
varies with apparent brightness due to the higher precision of the COMBO-17 photometric 
redshifts for the brighter sources, \cite[see][for a more detailed explanation.]{2009MNRAS.393.1275G}.
The photo-$z$ half-width distribution was normalized so the completeness was $\geq 90\%$ at any magnitude.
For the faint end, there is also some sample contamination from field galaxies.
This was calculated using the counts of the smooth models seen in Fig. 14 of \cite{2009MNRAS.393.1275G}.
When estimating the contamination, the field distribution was assumed to be consistent with the average 
galaxy counts $N(z, R)$ outside the cluster. The average contamination is 30\%.
This subsample thus contains cluster galaxies with good photometry for which the HST/ACS observations 
show an extended source. This cluster sample contains 905 galaxies.}

\item[The Field Sample:]{This includes galaxies in the redshift intervals
$z_{\mathrm{Phot}} = [0.05, 0.14]$ and $z = [0.22, 0.30]$.
In this subsample, only HST/ACS extended sources with good photometry are included.
This field sample contains 655 galaxies.}

%\item[Cluster Core Subsample:]{The third subsample is made of cluster galaxies found in the densest parts of the cluster. Comparison of this subsample with the field sample is likely to provide insight into the effects of the environment in the morphology of galaxies or their merger history, since environmental effects will be more apparent in this comparison. Galaxies where $\log \Sigma_{\mathrm{300\,kpc}}^{M}(>10^{9}{\mathrm{M_{\odot}}})>12.5$ were included in this subsample. The $\log \Sigma_{\mathrm{300\,kpc}}^{M}(>10^{9}{\mathrm{M_{\odot}}})$ number is the local mass density of galaxies more massive than $10^{9}M_{\odot}$ measured in apertures of 300 physical kpc. The Cluster Core Subsample contains 203 galaxies.}

\end{enumerate}
% Creating Reliable Samples %

This leaves a final sample of 655 Field galaxies (100 E, 60 S0, 318 Sp, 177 Oth) and 905 
Cluster galaxies (192 E, 216 S0, 383 Sp, 114 Oth).
%, 203 of which are Cluster Core galaxies (64 E, 67 S0, 59 Sp, 13 Oth).
As before, the ``Oth'' galaxies represent a mixture of irregulars, compact, and highly disturbed systems.
This classification is dominated by irregular galaxies.
The \textsc{STAGES} morphological catalogue has good morphological information for all objects 
in this sample.
The source detection runs presented in \S \ref{subsec:sextractor} and the structural 
parameter analysis presented in  \S \ref{subsec:galfit} yielded structural information for 
a total of 1537 sources. Thus, there is structural information available for 98.5\% of the 
full sample.

\section{Data Processing}
\label{sec:numeros}

This section presents the \textsc{SEXTRACTOR}\footnote{Version 2.5.0} \citep{1996A&AS..117..393B} runs and 
subsequent morphological analysis carried out here.
The main ingredients this work needs are an error or weight image and a mask image.
The error image is needed by both \textsc{SEXTRACTOR} and \textsc{GALFIT} \citep{Peng02}
in order to modulate the importance of each pixel in the photometry and 
morphological properties. The mask image, which is created
by \textsc{SEXTRACTOR}, is later needed by \textsc{GALFIT} so that it
fits galaxies in a way that is designed to identify mergers.
The basic idea is to produce a simple S\'ersic model that is as focused
on the main body of the target galaxies as possible. The putative
mergers should not therefore have any effect on these fits and will be either
simultaneously fit or masked out.

This S\'ersic model should be an appropriate description of the surface brightness distribution
of elliptical galaxies. However, it has to be noted that in the case of lenticular and spiral 
galaxies, bulge+disc decompositions (B+D) would produce better descriptions.
However, there are a number of reasons that make simple S\'ersic fits preferable. 
\begin{itemize}
\item{A simple Sersic profile will remove a fraction of the smooth, symmetric signal from
any given galaxy, regardless of its morphology. This statement remains 
true for disc galaxies.}
\item{The goal of the current work is to develop an automatic and fork free method.
The use of B+D decompositions together with simple S\'ersic fits would require
an additional step in order to decide what functional form describes best
each galaxy. There are objective ways to do this, such as the Bayesian method
presented in \cite{2011MNRAS.414.3052D} or the Akaike information
criterion, but this is well beyond the scope of the current work.}
\item{B+D decompositions need to be re-examined for their internal consistency, as their
final results might depend on the initial point in parameter space chosen for the galfit
minimization.}
\end{itemize}

Thus, it is not possible to use the \textsc{SEXTRACTOR} and structural 
catalogue presented in \cite{2009MNRAS.393.1275G} because the source 
detection and fits shown there were made with the goal of creating a robust catalogue 
of structural parameters derived from single S\'ersic fits.
Instead, the \textsc{SEXTRACTOR} configuration used in the current work (see \S \ref{subsec:sextractor})
is fairly sensitive to the detection of faint, small features, as it is set to detect the faintest sources 
that can be said to be detectable in the images used.
More importantly, the deblending parameters are selected so that \textsc{SEXTRACTOR} will tell apart sources
with up to a three magnitude brightness difference, incorporating much of the \texttt{1:3} to \texttt{1:10} mass ratio
range defining minor mergers, assuming that their $M/L$ ratios are similar.
These \textsc{SEXTRACTOR} parameters are thus not only intended to separate the objects
involved in a major merger, with luminosity ratios between \texttt{1:1} and \texttt{1:3}, but
are also geared towards the deblending of the objects involved in a minor merger episode, with luminosity 
ratios between \texttt{1:3} and \texttt{1:10}. Also, the very 
sensitive detection threshold employed makes it more likely
that the less luminous segment of any given deblended object will be engulfed by the
Kron aperture of the larger segment. This is key in the current analysis, since
it is hypothesized here that such less luminous segments could be
the less luminous galaxies in merger episodes. However, these smaller segments
could just be \textsc{HII} regions or simply objects along the same line of sight, which
will naturally lead to a contamination of the merger sample. This will be discussed in
\S \ref{subsec:contaminacion}. The larger segments in deblended
sources would then be the more luminous galaxies.
The Kron aperture \citep{1980ApJS...43..305K}
is defined to contain a specific fraction of the light of a galaxy.
Such fraction depends on the intrinsic profile of galaxies, but
it ranges from 90\% for the steepest profiles to 95\% for exponential 
disks. Both percentages depend on the observed surface brightness, though, in the
sense that \textsc{SEXTRACTOR} misses more flux from the dimmest objects. See \cite{2010arXiv1005.3300H}
for a description of this systematic error.

The larger and more luminous object is then fit by a smooth S\'ersic model which is created using 
\textsc{GALFIT} according to the rules explained in \S \ref{subsec:galfit}.
This model is then subtracted from the original image in order to estimate
how the image would look without the more luminous galaxy of a merger event.
This \emph{residual} image still contains most of the signal from the less luminous sources
that were found within the observed Kron aperture of the larger object.
This ensures that these smaller sources will have a great impact on the structural
properties of the \emph{residual} images. This is indeed one interesting
property of these \emph{residual} images that will be exploited.

The structural parameters of both the original image and of the residual image are then
calculated within the aforementioned Kron aperture. The specific morphological parameters 
used are shown in \S \ref{subsec:extendedcas}. These are mostly
based in previous works. The structural parameters of the
residual image within the Kron aperture are then expected to be very sensitive to the 
smaller and less luminous member of the merger.

\subsection{\textsc{SEXTRACTOR} Runs}
\label{subsec:sextractor}

\textsc{SEXTRACTOR} was run twice on each of the 80 tiles that compose the \textsc{STAGES} HST/ACS F606W mosaic.
The first pass was performed for the sole purpose of obtaining an empirical map of the background variance and of the
filtered poissonian signal. Such images can be produced by \textsc{SEXTRACTOR} as \emph{output} images. These images 
were then combined to produce an appropriate weight image to be used
in the second pass. The first \textsc{SEXTRACTOR} run is presented in \S \ref{subsubsec:sextractor1}, and it can 
be safely skipped by users providing their own error or weight images.
The second run is presented in \S \ref{subsubsec:sextractor2}, which presents the \textsc{SEXTRACTOR}
configurations used to produce the final source lists and mask images.

\subsubsection{Preliminary \textsc{SEXTRACTOR} run}
\label{subsubsec:sextractor1}

The most relevant \textsc{SEXTRACTOR} parameters used in the first run are summarized in 
table~\ref{tab:sextractor1}.
Table~\ref{tab:sextractor1} also presents the main results obtained from this first 
\textsc{SEXTRACTOR} run, averaged over the 80 tiles. 
The average background sigma is an estimate of the noise that is observed in the image areas in
which there are no galaxies. This is set by the readout noise of the Analog-to-Digital converters of the ACS camera, the
poissonian noise of the sky background and the use of the \textsc{MULTIDRIZZLE} technique.
It was estimated as the average of the median values of the background \textit{RMS} images created by \textsc{SEXTRACTOR} in this
first pass.
The average effective gain $\langle G_{\mathrm{eff}} \rangle $ roughly measures the growth in the photometric errors caused by the intrinsic
poissonian nature of photon counting measurements. It was measured using the background
subtracted filtered frame and a very rough empirical estimate of the per-pixel 
\textit{RMS} image obtained by means of a loose adaptation of the method
presented in \cite{2006A&A...449..951G}, which shows precise formulae to calculate 
the \textit{RMS} of an image in the case of noise correlation. The effective gain image
is then.

\begin{equation}
\langle G_{\mathrm{eff}} \rangle=\frac{I}{RMS^{2}}
\end{equation}

\noindent
where I is the background subtracted filtered image, and \textit{RMS} is the empirically
derived uncertainty.
\indent

The $\langle G_{\mathrm{eff}} \rangle$ number is yielded by the value of this
image in the brighter areas of the images dominated by the poissonian noise.

The measured value of $\langle G_{\mathrm{eff}} \rangle $, reported in table~\ref{tab:sextractor1} agrees very well with its 
expected value of around 1445, which can be theoretically estimated as:

\begin{equation}
G_{\mathrm{eff}}=g\times T \times \left(\frac{0.03}{0.05}\right)^{2} \simeq 1445.0
\end{equation}

\noindent
where $G_{\mathrm{eff}}$ is the effective gain, $g$ is the original detector's gain (2, for the \textsc{STAGES} observations), T is 
the total exposure time, and 
the fraction is the ratio between the effective areas of the pixels before and after the operation of 
the \textsc{MULTIDRIZZLE} technique. The latter value of 1445.0 was adopted for use in the second run of \textsc{SEXTRACTOR}.
The measured value of $G_{\mathrm{eff}}$ was then merely used as a sanity check to ensure that the errors in the input image indeed 
behave as expected.
\indent

The first \textsc{SEXTRACTOR} run thus produces an estimate of the background $\sigma_{\mathrm{Bkg}}$ and a background subtracted
filtered frame containing the poissonian signal $S$. These were combined according to the usual CCD error equation in order to obtain a
weight image for use in the second \textsc{SEXTRACTOR} run:

\begin{equation}
\mathrm{Weight}=\frac{1}{\sigma^{2}_{Bkg}+S/G_{\mathrm{eff}}}
\end{equation}

\noindent
where Weight is the final weight image, $\sigma_{Bkg}$ is the background \textit{RMS} image, $S$ is the background subtracted filtered image
created by \textsc{SEXTRACTOR}, and $G_{\mathrm{eff}}=1445.0$. The final weight image was later processed
using the \textsc{WEIGHTWATCHERS}\footnote{See \texttt{http://www.astromatic.net/software/weightwatcher.}} code 
to ensure that problematic pixels with either zero exposure time (typical of the
image edges) or with saturated signal were assigned zero weight. Less than 2\% of the pixels had to be discarded in this final procedure.
\textsc{WEIGHTWATCHERS} also produces a flag image that was used in the second pass of \textsc{SEXTRACTOR}.
\indent
This final weight frame is then used in the second run of \textsc{SEXTRACTOR} to supress the detection of
objects in low weight pixels and give appropriate pixel weights for the photometry.

\begin{table*}
\begin{minipage}{110mm}
\begin{tabular}{|ll|ll|} \hline
\texttt{DETECT\_MINAREA}                           & 7     &  \texttt{WEIGHT\_TYPE}                     & BACKGROUND \\
\texttt{DETECT\_THRSHLD}                           & 0.75  &  \texttt{BACK\_TYPE}                       & AUTO       \\     
\texttt{ANALYSIS\_THRSHLD}                         & 0.75  &  \texttt{BACK\_SIZE}                       & 256        \\      
\texttt{FILTER\_FWHM}                              & 3.0   &  \texttt{BACK\_FILTERSIZE}                 & 5          \\      
\texttt{DEBLEND\_NTHRESH}                          & 32    &  \texttt{BACKPHOTO\_TYPE}                  & LOCAL      \\     
\texttt{DEBLEND\_MINCONT}                          & 0.005 &  \texttt{BACKPHOTO\_THICK}                 & 64         \\    \hline
Avg. Bck. Sigma ($\sigma_{\mathrm{Bkg}}$).           & $3.73\times 10^{-3}$ &  Avg. Eff. Gain ($\langle G_{\mathrm{eff}} \rangle$).      & $1.6 \times 10^{3}$   \\    \hline
\end{tabular}
\caption{\textsc{SEXTRACTOR} parameters used for the first pass of this code on the
HST/ACS F606W images. This first run was used to obtain an empirical estimate of the
final error image. The main results obtained from this first pass are also presented in this table.}
\label{tab:sextractor1}
\end{minipage}
\end{table*}

\subsubsection{Second \textsc{SEXTRACTOR} run}
\label{subsubsec:sextractor2}

This second pass was used to obtain the final segmentation image and a \textsc{SEXTRACTOR} catalogue, which are key to
the \textsc{GALFIT} analysis of the images. The segmentation images separate object pixels from background pixels, and provide 
basic photometric information to be used as initial conditions for \textsc{GALFIT}.
It is highlighted here that the \emph{input photometric catalogue} of target sources is not produced
by this \textsc{SEXTRACTOR} run. The list of objects to fit and study is defined in
\S \ref{subsec:samples}, and \textsc{SEXTRACTOR} is run here with the sole purpose of obtaining
basic photometric information about this pre-defined sample.
The new configuration used is presented in Table~\ref{tab:sextractor2}.

As it can be seen from Table~\ref{tab:sextractor2}, the new configuration is fairly aggressive. It is more sensitive than the
``hot'' configuration used in \cite{2009MNRAS.393.1275G}. The minimum nominal integrated $S/N$ of the detections is 8.3. This is 1.7 times
higher than the usual $S/N$ limit of 5.0 which is usually accepted for a point source detection. The new \textsc{SEXTRACTOR}
parameters were chosen this way because, to zero-th order, the effect of noise correlation is an artificial increase of
the $\sigma$ image with respect to the inverse square root of the weight image. Thus, a stronger signal is needed 
in order to spawn a genuine detection. Given the number
of images that were multidrizzled together and the target resolution of the HST/ACS images used, experience shows that 
the nominal $S/N$ has to be multiplied by around 0.7, as is shown in \cite{2010arXiv1005.3300H} and \cite{2000AJ....120.2747C}.
This is the reason behind the additional factor of 1.7.
The second \textsc{SEXTRACTOR} configuration makes use of a gaussian filter with a FWHM of 3 pixels, and a minimum detection
area of 12 pixels. This effectively removes
spurious noise peaks, which are generally much smaller than the instrumental FWHM.
As for the deblending parameters, the ones shown in Table~\ref{tab:sextractor2} would, in principle, make \textsc{SEXTRACTOR}
separate objects with a factor of 100 difference in flux. However, the value of the
\texttt{DEBLEND\_NTHRESH} parameters also affects the deblending process.
With the choice of 32 deblending thresholds, the child detections tend to bear a larger fraction 
of the total flux than just 1\% of the parent source.
This was done this way with the idea of separating minor mergers in a very late stage, with a mass 
ratio close to \texttt{1:10}.
%Calor, y cielo.
The \textsc{SEXTRACTOR} parameters used in the second pass also ensure that the source catalogue will include detections
all the way up the real, almost point source detection limit of the images. At the same time, the minimum nominal $S/N$ ratio required to foster
a detection and the \texttt{DETECT\_MINAREA} value adopted ensure that the number of spurious sources caused by 
\emph{the multidrizzle algorithm and noise peaks} will be kept at a minimum. This configuration, however, is open
to the inclusion of spurious extended sources caused by \emph{statistical fluctuations.} Also, this configuration makes \textsc{SEXTRACTOR} 
include pixels with poor signal within the isophotal area of the detections which guarantees that the sky portion 
of the segmentation image is free from most of the flux originating from the 
uncovered sources. This also has the important effect of making the Kron apertures as large as they could possibly be given
the data used. This maximizes the probability of a small galaxy in a minor merger event falling within the
Kron aperture of the more luminous object. This is particularly important since the structural parameters measurements
shown in \S \ref{subsec:extendedcas} are performed over this Kron aperture.
%Algunas estadisticas.
On average, \textsc{SEXTRACTOR} found 34600 objects in each of the frames. This number is to be compared with the \emph{total}
number of sources found by \cite{2009MNRAS.393.1275G}, which is 75805 \emph{in all 80 tiles.} This is explained by the fact
that the \textsc{SEXTRACTOR} configuration used in \cite{2009MNRAS.393.1275G} was optimized to find and fit $R_{\mathrm{ap}} \leq 24.0\mathrm{mag}$ 
counterparts from a previous catalogue obtained from the $R-\mathrm{band}$ COMBO-17 data, while the goals of the \textsc{SEXTRACTOR}
catalog used in this work encourage the detection of faint sources near the brighter ones. This does not mean that all the sources
found by this second \textsc{SEXTRACTOR} run are legitimate, bona fide detections. In fact, specific simulations
indicate that the majority of objects with \texttt{MAG\_ISO} dimmer than 27.0 mag are spurious objects caused by statistical
fluctuations, while the majority of the objects with \texttt{MAG\_ISO} brighter than 27.0 mag are real sources. This is not a 
problem for the target sources studied in this work, whose \texttt{MAG\_ISO} are all brighter than 24.0 mag. The fitting 
scheme used here, which is presented in \S \ref{subsec:galfit}, just discards these faintest detections and therefore they 
will not have any impact on the actual fits produced for the much brighter targets of interest. At the same time, this 
\textsc{SEXTRACTOR} configuration will detect \emph{and more importantly, isolate} sources in the
$24.0\mathrm{mag}<F606W(AB)<27.0\mathrm{mag}$ magnitude interval, which are much more likely to be real objects.
The comparison between the number of detections obtained here and in the \cite{2009MNRAS.393.1275G} work
merely reflects that the configuration used in the current work is much more sensitive to the smallest features, which increases
the chance of including spurious detections in the \textsc{SEXTRACTOR} catalog. 

\begin{table*}
\begin{minipage}{110mm}
\begin{tabular}{|ll|ll|} \hline     
\texttt{DETECT\_TYPE}          & CCD         &  \texttt{FLAG\_TYPE}        & OR           \\    
\texttt{DETECT\_MINAREA}       & 12          &  \texttt{KRON\_FACT}        & 2.5          \\
\texttt{THRESH\_TYPE}          & RELATIVE    &  \texttt{MIN\_RADIUS}       &  3.5         \\
\texttt{DETECT\_THRSHLD}       & 0.80        &  \texttt{DEBLEND\_NTHRESH}  &  32          \\ 
\texttt{ANALYSIS\_THRSHLD}     & 0.80        &  \texttt{DEBLEND\_MINCONT}  &  0.01        \\  
\texttt{FILTER\_FWHM}          & 3.0(Gauss)  &  \texttt{CLEAN}             &  Y           \\ 
\texttt{BACK\_TYPE}            & AUTO        &  \texttt{CLEAN\_PARAM}      &  1.0         \\  
\texttt{BACK\_SIZE}            & 256         &  \texttt{STARNNW\_NAME}     & default.nnw  \\      
\texttt{BACK\_FILTERSIZE}      & 3           &  \texttt{MASK\_TYPE}        & CORRECT      \\         
\texttt{BACKPHOTO\_TYPE}       & LOCAL       &  \texttt{INTERP\_TYPE}      &  ALL         \\         
\texttt{BACKPHOTO\_THICK}      &  64         &  \texttt{INTERP\_MAXXLAG}   &  16          \\         
\texttt{WEIGHT\_TYPE}          & MAP\_WEIGHT &  \texttt{INTERP\_MAYYLAG}   &  16          \\         
\texttt{WEIGHT\_GAIN}          & N           &  \texttt{SATUR\_LEVEL}      & 40000.0      \\             
\texttt{BACK\_TYPE}            & AUTO        &  \texttt{MAG\_ZEROPOINT}    & 26.49113     \\         
\texttt{BACK\_FILTERSIZE}      & 3           &  \texttt{PIXEL\_SCALE}      & 0.03         \\         
\texttt{BACK\_SIZE}            & 256         &  \texttt{GAIN}              & 1445.0       \\              
\texttt{BACKPHOTO\_TYPE}       & LOCAL       &  \texttt{SEEING\_FWHM}      &     0.106    \\  
\texttt{BACKPHOTO\_THICK}      & 64          &                             &              \\ 
\hline
\end{tabular}
\caption{\textsc{SEXTRACTOR} parameters used for the second pass of this code on the
HST/ACS F606W images. This second pass makes use of the empirical error image created using the
\emph{output} images of the first \textsc{SEXTRACTOR} run. The flag image 
created by \textsc{WEIGHTWATCHERS} was also used to exclude pixels 
with saturated signal or zero effective exposure time.}
\label{tab:sextractor2}
\end{minipage}
\end{table*}

\subsection{Galaxy Fitting: The GalP-Hyt Wrap Scripts}
\label{subsec:galfit}

The main contribution of the current work is that it begins to explore whether the morphological information
contained in the \emph{residual} images of galaxies can be used to assess if a galaxy is involved in a
merger episode. Such residual images are created by subtracting smooth models
of the target galaxies from the original images while leaving most of the signal
from other, possibly interacting objects in these \emph{residual} images. There 
are a number of codes capable of producing models of
galaxies in astronomical images by performing two-dimensional model fits to their surface brightness distributions.
The most commonly used ones are \textsc{GIM2D} \citep{1998ASPC..145..108S} and \textsc{GALFIT}. Even though 
these codes are different in their specific details, their
basic principles are very similar. They both try to minimize a possibly weighted $\chi ^{2}$ value that depends on the structural parameters
of the object being fit. This residual sum of squares is formed from the difference between the observational data and a trial function
that is created according to a user-supplied set of rules. Chief among these rules are an error image, a mask image, and a PSF image.
The error image regulates the relative weight that the different pixels should be given. Thanks to this image, it is possible to prevent
saturated pixels from having any weight in the figure-of-merit that the codes are set out
to minimize. This image can also be used to reduce the impact of areas with lower or no 
exposure time in mosaic images created using the \textsc{MULTIDRIZZLE} technique.
The mask image can be used to modify how the different fitting codes treat the different areas of the input image.
For instance, it can help the fitting codes to tell what pixels belong to the target object being analysed, and
what pixels belong to other objects and should therefore be discarded. The PSF image is also a key ingredient in
the structural analysis of the surface brightness distribution of galaxies.
It is also possible to constrain the parameter space region that the codes are allowed to explore.

The present work uses the \textsc{GALFIT} code. The setup with which \textsc{GALFIT} runs
is created by a \texttt{python} code called \textsc{GALP-HYT}\footnote{Pronounced Galp-Hit.} written by CH.
The \textsc{GALFIT} setup used is designed to create and subtract a smooth model of the primary or target sources, while 
leaving the signal of nearby objects largely intact in the residual images. In this context, the target
sources are the galaxies that are being studied, and close sources are the objects
found within the Kron aperture of the primary galaxy. Given the
\textsc{SEXTRACTOR} parameters used, these close sources could be the less luminous galaxies in a minor merger event, or 
one of the protagonists in a major merger episode. Thus, \textsc{GALFIT} is configured to
produce a residual image in which only the model for the primary source has been removed. This 
is done in order to guarantee that the Kron aperture in the residual images will contain most of 
the information from the merged sources, while being free from the effect
of the target galaxy. The \textsc{GALP-HYT} code is now briefly described.

For each science image or tile it is given, it takes the following information as input:

\begin{itemize}
\item{The science image itself. These are the same images used in the structural parameter catalog presented in
\cite{2009MNRAS.393.1275G}.}
\item{The weight image that \textsc{SEXTRACTOR} used during its second pass. This is converted into a 
$\sigma-$image by taking its inverse square root. In this process, pixels with zero weight are given 
a very large value of $\sigma$.}
\item{The \textsc{SEXTRACTOR} list of sources and segmentation image created during the second pass
of the detection software.} 
\item{A list of primary targets. This list of targets is given in a separate text file, one object per line. In this file, targets 
are identified by their \textsc{SEXTRACTOR} number IDs.}
\item{A PSF imagelet. This was again taken from the work done in \cite{2009MNRAS.393.1275G}. This is a PSF imagelet
that was built from many different non saturated stars found across a number of tiles and hence its $S/N$ ratio is
very high.}
\end{itemize}

For each object in the list of targets the \textsc{SEXTRACTOR} catalogue is examined and objects
are classified with respect to the primary object according to the following set of rules:

\begin{itemize}
\item{The target object itself. It is fit using a single S\'ersic model with a free floating
diskiness-boxiness ``C0'' parameter. This extra freedom allows the model for the target object
to take into account a larger fraction of the symmetric, undisturbed signal of the primary source.
At the same time, this extra parameter does not complicate the interpretation of the fits 
since it is unlikely to introduce important degeneracies. Also, the evolution of ``C0'' as a function
of the number of minor merger events is a prediction of the \cite{2005A&A...437...69B,2008MNRAS.389L...8B} 
models.}
\item{Objects up to two magnitudes fainter than the target source whose centres lie within the Kron aperture 
of the target source are tagged as ``A''. In this context, the relevant magnitudes are the \texttt{MAG\_ISO}
magnitudes. The two magnitude difference was used since this is approximately the expected magnitude difference 
between the galaxies involved in a minor merger with a mass ratio of \texttt{10:1}, assuming that
the $M/L$ ratios are similar. These sources are fit with a
single S\'ersic profile with elliptical isophotes. The centres of the ``A'' components are fixed
to the values found by \textsc{SEXTRACTOR}. This constraint prevents degenerate fits, as the ``A'' components
will not blend and shift towards the primary object.}
\item{Objects more than two magnitudes fainter than the target source whose centres lie within the Kron aperture 
of the target source are tagged as ``B''. Again, the magnitudes used in this criterion are the \texttt{MAG\_ISO}
magnitudes. These sources are fit as simple exponential models with elliptical isophotes. This ansatz is 
less flexible than the one used for the ``A'' sources, but this is justified by the fact that
there is less information available for these dimmer sources. The use of exponential
profiles also helps to increase the execution speed. Again, the centres of the ``B'' components are directly
taken from the values found by \textsc{SEXTRACTOR} in order to prevent degenerate fits.}
\item{Objects up to three magnitudes fainter than the target source outside the Kron aperture 
of the target source are tagged as ``E'' (for external). As before, \texttt{MAG\_ISO} measurements were used.
These objects are fit using simple exponential disks. Objects more than three magnitudes fainter than the 
source of interest are ignored. The centres, ellipticities and position angles of the ``E'' components
are again taken from the \textsc{SEXTRACTOR} catalogue.}
\end{itemize}

The segmentation image produced by \textsc{SEXTRACTOR} is also modified so that the pixels belonging
to the target source and to the ``A'' objects are given a value of 0. This ensures that these 
pixels are taken into account by \textsc{GALFIT} when calculating its figure-of-merit. Pixels belonging 
to ``B'' and ``E'' objects are not nulled, and thus the pixels
contained within their \texttt{ISOAREAS} have no weight in the fit. Only the extended tails
of the ``B'' and ``E'' objects that go beyond their \texttt{ISOAREAS} are fit by the exponential profiles.
It is important to note that there will be many objects three or more magnitudes fainter than the target 
source that will not fit any of the three categories above. These faint objects are not 
fit in any way and have no effect on the fits. This is important because the integrated magnitudes
of the sources of interest are all brighter than $F606W(AB)=24.0 \mathrm{mag}$, and therefore the
large number of spurious sources with magnitudes fainter than $F606W(AB)=27.0 \mathrm{mag}$ will have
no effect on the fits.
%However, most of the flux from these faint detections is indirectly removed from the fits when processing the segmentation image and therefore their overall impact in the fit is minimized at a minimum CPU cost.The light from these very faint sources is however still present in the residual image.
The sky is left as a free floating parameter. This is justified by a number of reasons. First of all, the
\textsc{SEXTRACTOR} configuration used guarantees that the isophotal apertures of the detected objects
reach to fairly faint surface brightnesses, and thus the sky pixels in the segmentation image
contain very little residual signal from the detected objects. Also, the masking process
implies that the majority of the flux from the ``B'' and ``E'' objects is not taken into account
by the fits and only their tails are fit using exponential profiles. This naturally means that whatever 
their contribution might be to the average sky level affecting the target object, it will be approximately
corrected for by the exponential fits. Finally, the ``A'' objects are fully fit.
For these reasons, it is appropriate to leave the sky as a free floating parameter.

The size of the fitting box is defined in two steps. In the first step, the maximum between 150 pixels
and the circularized Kron diameter is used. In the second and final step, this area is expanded
so that it encompasses the centres of all ``E'' objects whose Kron apertures intersected the
first box.

The total number of additional sources that have to be fit in conjunction with a primary target
is around 20. It is highlighted here that each \textsc{SEXTRACTOR} detection is fit by a
simple, solid profile. In each case, the target object is fit by a single S\'ersic model, and 
the majority of the remaining components corresponds to ``E'' objects and do not merge with the 
main body of the target galaxy.
This number of extra components is not unusual, see e.g. \cite{2007ApJS..172..615H}.
The main target galaxy is fit using a single Sersic model. In most cases, the majority of the 
remaining components are external (E) objects whose main purpose is to help galfit compute a 
good sky value. The A and B components are the putative minor mergers, and their role
is to allow the Sersic profile to provide a good fit to the main body of the target galaxy.

Although this is very expensive in terms of CPU time, the \textsc{GALP-HYT} code was written to be run
in High Performance Computers with thousands of CPUs, as it features a semi-intelligent built-in system
of organizing its internal data flow. It was run in the HPC computer of the 
University of Nottingham\footnote{See \texttt{http://www.nottingham.ac.uk/hpc/}}.
Initial values and constraints are taken from the source catalog created by \textsc{SEXTRACTOR}.

\subsection{Additional Structural Parameters. The CAS Indices and the Gini-$\textrm{M}_{20}$ System}
\label{subsec:extendedcas}

One of the common ways to tackle the automated detection of mergers is based on the signatures that merger events leave on 
the morphology of galaxies. As shown in \cite{2003ApJS..147....1C,2005ApJ...631..101P}, the structures 
of galaxies contain important information about their past star formation modes, and they can also shed light 
on their interaction history. These ideas are used to identify mergers according to their
morphological properties.
In the current work, the structural properties of the residual images
after the subtraction of the S\'ersic model described in \ref{subsec:galfit}
are explored and used as merger diagnostics.

There are two systems which are currently in use for the morphological identification
of mergers. The first one is the use of the Concentration, Asymmetry, ClumpineSs (CAS) numbers, which 
was introduced in \cite{2000AJ....119.2645B,2003ApJS..147....1C}. This system has been used in many works aimed at the 
study of the merger fraction in many different contexts, see for 
instance \cite{2009MNRAS.394.1956C}, \cite{2009A&A...501..505L}, and \cite{2009ApJ...697.1971J}.

This system makes use of three different indices:

\begin{description}
\item[$\bullet \; \mathbf{C}$:]{The concentration index ``C'' measures to what extent the light in the galaxy is concentrated towards
its centre. It is defined as:
\begin{equation}
C=5.0 \times \log \left(\frac{r_{80}}{r_{20}}\right)
\end{equation}
\noindent
where $r_{80}$ is the circular radius containing 80\% of the total light from the galaxy, and
$r_{20}$ is the radius of the circular aperture that encloses 20\% of the total light of the target galaxy.
\indent
The concentration index ``C'' takes values between 2.2 and 5.0. If it is calculated for simple S\'ersic models
it depends mainly on the S\'ersic index and then on the ellipticity.
}

\item[$\bullet \; \mathbf{A}$:]{The asymmetry ``A'' measures to what extent any given image changes under a 180 degree rotation around the
point that minimizes the asymmetry of that image. It is defined as:
\begin{equation}
A=\left(\frac{\sum_{i,j}|I_{i,j}-I^{180}_{i,j}|}{\sum_{i,j}|I_{i,j}|}\right)-\left(\frac{\sum_{i,j}|B_{i,j}-B^{180}_{i,j}|}{\sum_{i,j}|I_{i,j}|}\right)
\end{equation}
\noindent
where $I_{i,j}$ represents the original image, and $I^{180}_{i,j}$ is a 180 degree rotated version of the original image.
In the same manner, $B_{i,j}$ is a patch of background, and $B^{180}_{i,j}$ is a 180 degree rotated version of this patch of background.
This contribution from the background is minimized independently in the same manner.
This asymmetry measurement is defined even for images whose average value is 0.0, as it is normalized
to the sum of the \emph{absolute} values of the fluxes from each pixel.
\indent
The rotation centre is optimized so that the value of the first term in the subtraction is a minimum.
Here, the rotation centre is allowed to lie at most 9 pixels away from the
\textsc{SEXTRACTOR} defined centre. However, when calculating the assymetry of the residual images, the rotation
center is only allowed to move 4 pixels. The second term 
in the subtraction also undergoes this optimization process and it removes the contribution to the asymmetry from
the background.
The asymmetry index of real images of galaxies can take values between 0.0 and approximately 0.8. Most objects 
have asymmetries lower than 0.2, though.
The asymmetry index of the residual images of galaxies after the subtraction of a single S\'ersic model
ranges between 0.4 and 1.6.
}

\item[$\bullet \; \mathbf{S}$:]{The clumpiness ``S'' quantifies the fraction of light in a galaxy that is contained in clumpy
distributions. Large values of $S$ imply that the light if the galaxy is accumulated in few, distinct structures.
Low values of $S$ indicate that the light distribution is smooth. It is defined as:
\begin{equation}
S=10 \times \left(\left(\frac{\sum_{i,j}(I_{i,j}-I^{\sigma}_{i,j})}{\sum_{i,j}I_{i,j}}\right)-\left(\frac{\sum_{i,j}(B_{i,j}-B^{\sigma}_{i,j})}{\sum_{i,j}I_{i,j}}\right)\right)
\end{equation}
\noindent
where $I_{i,j}$ again represents the original image, and $I^{\sigma}_{i,j}$ represents a blurred version of it, which is produced
by convolving the original image with a two-dimensional circular gaussian kernel with a typical dispersion of $\sigma$.
It is usually correlated with the size of the target galaxy.
\indent
The residual image after this subtraction only includes signal that is included in high frequency features of the galaxy.
Also, the convolution procedure is applied to a blank patch of sky in the image. This ensures that the contribution
from the background noise is discounted from the final value of $S$.
The ``S'' parameter can take values between -0.5 and 1.5, although it depends on the size of the convolution kernel used. 
}
\end{description}

The second system is based on the use of the Gini and $M_{20}$ parameters, which 
were originally introduced by \cite{2003ApJ...588..218A} and \cite{2004AJ....128..163L}.
The Gini coefficient $G$ measures the 
light concentration, like the $C$ parameter, but it is insensitive to any particular centre. It is calculated 
according to the following formula:
\begin{equation}
G=\left(\frac{1}{\bar{|f|} \times n \times (n-1)} \sum_{i=1}^{n}(2\times i-n-1)\times |f_{i}| \right)
\end{equation}
\noindent
where n is the number of pixels, $f_{i}$ is the flux observed in the $i^{\mathrm{th}}$ resolution element, and the sum
is made in ascending order of fluxes, so that $f_{i-1}\leq f_{i}\leq f_{i+1}$.
\indent
This $G$ index tells whether the light is evenly distributed among the different resolution elements of an image.
The $G$ index has a value of 0.0 for flat light distributions, and it has a value of 1.0 for light distributions
in which all the light is contained in a single pixel. In practical terms, the Gini coefficient of real galaxy images
lies in the [0.35,0.85] interval.

The $M_{20}$ parameter is based on the second-order moment of the light distribution $M_{\mathrm{Tot}}$. It is defined as:
\begin{equation}
\scriptstyle{ M_{20}=\log \left( \frac{\sum_{i=1 \in A }^{K:\sum_{l=1 \in A}^{K}f_{l}=0.2\times L_{\mathrm{Tot}}}f_{i} \times ((x-x_{c})^{2}+(y-y_{c})^{2})}{M_{\mathrm{Tot}}}\right) }
\end{equation}
\noindent
where A is the aperture within which this number is obtained, $f_{i}$ is the flux of the $i^{\mathrm{th}}$ resolution 
element, $L_{\mathrm{Tot}}$ is the total
apparent luminosity contained in the aperture used, $x_{c}$ and $y_{c}$ are the coordinates of the barycentre
of the light distribution for which the index is being calculated, and $M_{\mathrm{Tot}}=\sum_{i \in A}f_{i} \times ((x-x_{c})^{2}+(y-y_{c})^{2})$.
In this definition, and in contrast with the definition of $G$, $f_{i-1} \geq f_{i}\geq f_{i+1}$.
\indent

$M_{20}$ measures how far from the galaxy centre it is possible to find the brightest features
of the surface brightness distribution of the light. The $M_{20}$ number can go from -3.0 for very concentrated
objects to -0.4 for objects with shredded light distributions.

In addition to the aforementioned indices, the residual flux fraction \citep[\textit{RFF}, see][]{ArticuloIII} is used. It is here defined as:

\begin{equation}
RFF=\frac{\sum_{i,j \in A}|I_{i,j}-I^{\mathrm{GALFIT}}_{i,j}|-0.8 \times \sum_{i,j \in A}\sigma_{\mathrm{Bkg }i,j}}{\sum_{i,j \in A}I^{\mathrm{GALFIT}}_{i,j}}
\end{equation}
\noindent
where A is the particular aperture used to calculate this index, and $I^{\mathrm{GALFIT}}$ is the model created by \textsc{GALFIT}.
\indent

The \textit{RFF} as defined here measures the fraction of the signal contained in the residual image 
that can not be explained by fluctuations of the background. The 0.8 factor included in the
definition of the \textit{RFF} ensures that the expectation value of the \textit{RFF} of 
a purely gaussian noise error image of constant variance (as opposed to a spatially varying variance)
is 0.0. This fact arises from the following integral:

\begin{equation}
0.8=\sqrt{\frac{1}{2 \pi}} \int _{-\infty}^{\infty}|x| \times e^{-x^{2}/2}\, dx.
\end{equation}

\noindent
which calculates the expectation value of the absolute value of a gaussian random variable.
\indent

In the current work, the structural indices presented above were calculated for three different images.
These are the imagelet containing the target object cropped by \textsc{GALFIT}, the simple S\'ersic model created by 
\textsc{GALFIT} and the \emph{residual} image obtained by subtracting the second image from the first one.
It is stressed that this latter image is only stripped from the signal of the main body of the target galaxy.
The indices were also calculated for an artificial image simulating the background noise that affects the primary 
galaxy, which will be described later on. This image is needed to calculate the background terms in the $A$ and $S$ indices, but
it is also used to estimate the errors in the derived morphological parameters.

In all cases, the aperture used was the Kron aperture calculated by \textsc{SEXTRACTOR}. This aperture was chosen for the
following reasons:
\begin{itemize}
\item{This aperture is designed to trace an elliptical region in the image in which the contribution in flux
from the source of interest to which it is associated is either dominant or noticeable. Outside the Kron aperture
there is still flux from the primary source, but it is heavily affected by noise. It is therefore 
pointless to go much further since morphological perturbations of the target galaxy at these levels
will be impossible to measure reliably.}
\item{The Kron aperture misses a definite fraction of the light from most common profiles.
This fraction depends on the intrinsic shape or profile of the target source and on its
effective surface brightness as well. Thus, the Kron aperture is a $S/N$ matched aperture that grabs
a more or less constant fraction of the total flux for objects of similar $S/N$, with a slight dependence on its
profile.}
\item{This aperture has a radius that is typically 60\% larger than the Petrosian radius defined in
\cite{1976ApJ...209L...1P}. This is a little bit smaller than the typical
aperture of choice in which the structural parameters defined above are calculated 
in other studies, which is twice the Petrosian radius. The choice
of the Kron aperture will therefore provide higher $S/N$ measurements
of the morphological parameters, losing only a minimal amount of information about
the outer structure of the studied objects.}
\end{itemize}

The whole Kron aperture is used to calculate the CAS, $G$, and $M_{20}$ numbers for 
the real image, the model and the background noise frame. 
However, when calculating the structural parameters for the residual image, a small area 3 pixels
in diameter is removed from the centre of the Kron ellipse so that the indices are not biased by the
fit uncertainties, PSF mismatches, and resampling problems in these complicated regions.
This latter aperture in which the centre is excluded is also used to calculate the \textit{RFF} 
and \textit{S} structural parameters.
Also, the rotation axis for the calculation of the asymmetry parameter of the \emph{residual}
images is not allowed to drift from the optimal centre found for the original image
by more than 4 pixels.
In all cases, the background was subtracted from the images using the sky value yielded by
\textsc{GALFIT}. This was done with the purpose of minimizing the impact of the 
background terms in the $A$ and $S$ numbers.

For the calculation of the $S$ parameter, a gaussian kernel with $\sigma= 0.2 \times R_{K}$ was used.
In this expression $R_{K}$ is the radius of the Kron aperture calculated by \textsc{SEXTRACTOR}.
%However, when calculating the $S$ parameter for the residual image, the gaussian kernel had $\sigma=0.3 \times 0.2 \times R_{K}$. This is done because the residual image will be already stripped of most of its large scale structure.

The background noise image that is needed in order to calculate the background terms of 
the $A$ and $S$ numbers was created from the \textit{RMS} image that \textsc{GALFIT} used.
It was created on an object by object basis, so that each object has an individualized background noise image.
The first step is to change the data number values of the pixels in the \textit{RMS} image that, according to
the segmentation image created by \textsc{SEXTRACTOR}, have been flagged as 
object\footnote{This includes all objects, not only the galaxy of interest.} pixels.
Their new value is then set to the median value of the \emph{remaining} sky pixels in the \textit{RMS} image.
This modified image is then multiplied by a white noise image with $\sigma = 1.0$. This final step creates
an image that is a good representation of the underlying noise that affects the measurements. It would be ideal if
this image included correlated noise, but since the target galaxies are all larger than the error correlation 
length this image was deemed sufficient.

The noise image was also used to calculate the errors in the structural parameters.
This was done by using this frame to recreate ten realizations of the original image.
The structural parameters were then recalulated for this set of realizations 
and an error estimate is obtained by the very simple prescription of removing the smallest one and the largest one.
According to Chebyshev's inequality this produces an interval whose upper limit is $2.3 \to 3.0 \times \sigma$. The actual
value depends on the assumed underlying distribution. Here, the error distribution is assumed to be gaussian for all
structural parameters, and robust estimates of the $1-\sigma$ uncertainty are obtained by
doubling the Chebyshev error estimate.

The first term of the $A$ index was also calculated for the \textsc{GALFIT} model. This was 
used to estimate the systematic uncertainties in the asymmetries of the real galaxies, since 
they should be zero in the model image.
Deviations from zero thus reflect inaccuracies in the minimizing algorithm.
These deviations amount to less than 0.05 in almost all cases.

Table~\ref{tab:errores} presents the typical errors in all the derived parameters.
Although each object should have its own error for all the derived quantities, Table \ref{tab:errores}
gathers the median error for all the sources. 
It is seen that the typical errors are very small. This is caused by the fact that the 
Kron apertures deployed by \textsc{SEXTRACTOR} include of the order of $10^{4}$pixels, and 
hardly ever less than $10^{3}$pixels. The expected
errors in the structural parameters are therefore very small. The set of table entries above the horizontal line
in Table~\ref{tab:errores} are the statistical uncertainties. Table entries
below the horizontal line gather the systematic uncertainties for the asymmetry indices, derived
by calculating the asymmetry indices for the \textsc{GALFIT} models.

\begin{table}
\begin{tabular}{|ll|ll|} \hline
$\sigma C(\mathrm{Obj})$        & $1.0\times 10^{-4}$      & $\sigma C(\mathrm{MDL})$     & $1.0\times 10^{-4}$   \\
$\sigma A(\mathrm{Obj})$        & $4.0\times 10^{-2}$      & $\sigma A(\mathrm{Res})$     & $4.0\times 10^{-2}$     \\    
$\sigma S(\mathrm{Obj})$        & $6.0\times 10^{-2}$      & $\sigma S(\mathrm{Res})$     & $6.0\times 10^{-2}$     \\    
$\sigma G(\mathrm{Obj})$        & $1.0\times 10^{-3}$      & $\sigma G(\mathrm{Res})$     & $1.0\times 10^{-3}$   \\      
$\sigma M_{20}(\mathrm{Obj})$   & $1.0 \times 10^{-3}$      & $\sigma M_{20}(\mathrm{Res})$ & $1.0\times 10^{-3}$   \\      
$\sigma RFF$      & $2.0\times 10^{-2}$      & \ldots               & \ldots               \\   \hline
$\Delta A(\mathrm{Obj})$        & $6.0\times 10^{-2}$      & \ldots  & \ldots   \\       \hline 
\end{tabular}
\caption{Typical absolute errors in the structural parameters used in this work.
The left columns present the typical errors for the structural parameters
of the original galaxy image, while the columns to the right
show the uncertainities for the structural parameters of the residuals.
Errors above the horizontal line are the random errors, while the entry below the horizontal line 
is the systematic uncertainties for the $A(\mathrm{Obj})$ index.}
\label{tab:errores}
\end{table}

\section{Merger Identification Using Structural Parameters of the Residuals}
\label{sec:ident}

%Un poquito de introduccion.
As it was mentioned in \S\ref{seccion:intro}, the merger fraction is defined as 
the fraction of galaxies with an ongoing merger episode that is found in any given 
sample, which is often selected as a mass-limited one. This is a fundamental issue to galaxy evolution studies
and therefore the automated identification of mergers is of great importance.
%De que va esto, asi por toda la cara para empezar.
This section presents the main contribution from this paper. It shows that the use of the structural
parameters \emph{of the residual images} allows to identify mergers like the use of the structural 
parameters of the galaxies themselves. It also shows that the merger samples obtained using the properties 
of the residual images are of better or comparable
statistical quality than the samples that are culled using the morphological information of 
the original images.
The precise definition of the statistical quality of a sample, which is given in  \S \ref{subsec:Fscore}, is
used together with the set of galaxies defined in \S \ref{subsec:samples}, that were morphologically classified 
by the \textsc{STAGES} team. This key ingredient allows us to obtain a sample of visually detected mergers, which 
will be used as a training set for the method presented in this section.
%Ahora, hablemos de CAS-G y M20.
Many studies have used the morphological or structural properties of galaxies to estimate the merger fraction.
Usually, these works make use of the CAS system or of the $G$-$M_{20}$ system.
Each of these systems has its own pros and cons.
For instance, \cite{2008MNRAS.386..909C} concludes that the $G$-$M_{20}$ system discovers more mergers than 
the CAS methodology, although it also picks up more interlopers. It is however the case that all the structural 
approaches that have been presented so far have only taken advantage of the structural parameters of the real, direct
images of galaxies. In addition, CAS typically recovers a fairly high fraction 
(50\% to 70\% in \cite{2009ApJ...697.1971J,2009ApJ...705.1433H}) of visually classified mergers, but it is also
significantly contaminated by dusty, highly inclined non-interacting galaxies. These latter galaxies have low level asymmetries
caused by star formation episodes, as was noted in \cite{2009ApJ...697.1971J}.

\subsection{Statistical Quality of Samples}
\label{subsec:Fscore}

In science, one often confronts the problem of finding an algorithm or method to select a sample of items from a larger 
parent population with the condition that the selected items have to satisfy some requirements of scientific interest.
However, one rarely has a mechanism to retrieve \emph{all} the items in the parent population that satisfy the needed
requirements, and it is also very unlikely that the method is able to retrieve the required items \emph{only}.
One is then forced to speak about the \texttt{sensitivity} and \texttt{specificity} of the selection process.

The sensitivity, also known as the recall ratio, is defined as:

\begin{equation}
r=\frac{\rm \# True\ Positives}{{\rm \# True\ Positives}+{\rm \# False\ Negatives}}
\end{equation}

\noindent
This is more commonly known as the completeness in astronomical literature.
\indent

The specificity is defined as:

\begin{equation}
p=\frac{\rm \# True\ Negatives}{{\rm \# True\ Negatives}+{\rm \# False\ Positives}}
\end{equation}

\noindent
In the above definitions, a ``True Positive'' is a recovered item that did indeed present the required properties.
A ``False Negative'' is an item that was not retrieved by the culling algorithm but did present the needed properties. These 
latter errors usually reflect an excessive skepticism.
A ``True Negative'' is an item that was rightfully rejected by the selecting process since it did not have
the required properties.
A ``False Positive'' is an item that was incorrectly picked up by the sampling algorithm, but that does not have the properties of interest.
\indent

The sensitivity and the specificity can be combined into a single number, known as the 
$F-\mathrm{score}$, $F_{\beta}$ \citep{Fbeta}. This is a measure of sample purity, and it is just 
a weighted harmonic average of $r$ and $p$.

\begin{equation}
F_{\beta}=\frac{(1+\beta^{2}) \times p \times r}{(\beta^{2} \times p+r)}
\end{equation}

\noindent
where $\beta$ is a control parameter that regulates the relative importance of $r$ with respect to $p$. This is a user-supplied value that 
depends on the particular goals of the test\footnote{The reader can probably work out what is the value of $\beta$ used
by the managers of airport security screenings, where false positives represent a small additional test but false negatives
have disastrous consequences.}. In this work, a value of $\beta = 1.25$ is used, which can be thought of as weighing completeness 
more than the \emph{lack of} contamination. The use of this value will be justified in \S \ref{subsec:otrobeta}.
This choice leads to a galaxy sample that contains most mergers from that training set, although the corresponding contamination
is rather high. This will be further discussed in \S \ref{subsec:contaminacion}.
\indent

The $F-\mathrm{score}$ is used in the current study in order to grade the performance of a number merger diagnostics
at separating a merger sample from its parent population. Galaxies undergoing a merger episode play the role of the ``items
presenting the required properties'' discussed above, and the parent population used here is of course
the galaxy sample defined in \S \ref{subsec:samples}.

In this context, a ``merger diagnostic'' is defined as a two-dimensional diagram in which the 
parent population of galaxies described in \S \ref{subsec:samples} is presented.
In these plots, the horizontal axis is one structural 
parameter and the vertical axis is another morphological parameter, both
selected from the set of indices described in \S \ref{subsec:extendedcas}.
In these diagrams, merger galaxies should preferentially occupy specific regions. For instance, mergers 
should have large asymmetries and higher than average values of $G$. This is exploited by searching 
for the \textit{best} border that separates 
mergers from other galaxies in each of these diagrams. The ``border'' of a diagnostic is defined 
as a second order polynomial in the horizontal coordinate that maps to the vertical 
coordinate\footnote{The Gini-$M_{20}$ method, in which galaxies with too high values of G for
their $M_{20}$ value are classified as mergers, is a basic example of this approach.}.
Although it would be possible to use this same technique with more than two parameters, this would make
it more difficult to interpret the resulting 3-D parameter spaces. For this reason, the merger diagnostics
considered in the current work are simply 2-D.

Galaxies are then classified into four different types, depending on the side of the border 
in which they fall and on whether they are involved in a merger or not.

\begin{itemize}
\item{Mergers that fall in the merger side of the border are the ``True Positives''.}
\item{Mergers that do not fall in the merger side of the border are called ``False Negatives''.}
\item{Non mergers that however fall in the merger side of the border are regarded as ``False Positives''.}
\item{Non mergers that do not fall in the merger side of the border are of course ``True Negatives''.}
\end{itemize}

\noindent
In all the above definitions, the merger side of the border is to be understood as the zone in the diagram in which the
majority of mergers exist.
\indent

The \textit{best} border is then the border that maximizes the $F-\mathrm{score}$. In this
step, mergers serve as buoys or perhaps better as a training set in order to find the \textit{best} border.
This maximization is done by means of the Amoeba algorithm explained in \cite{Numerical}, using the polynomial coefficients as the 
problem parameters.

\begin{figure}
\includegraphics[scale=0.35,angle=0]{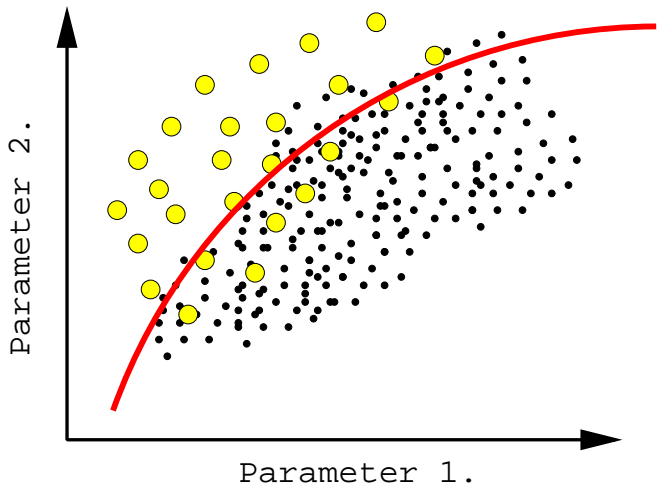}
\caption{Schematic graphical explanation of the $F-\mathrm{score}$ method used. The horizontal axis is one of the structural 
parameters calculated either for the original images or for the \emph{residual} images. The vertical axis is another
morphological parameter, regardless of its nature. The larger dots represent the galaxies that were marked as mergers in the
\textsc{STAGES} morphological catalog, and the smaller dots are galaxies that were not classified as mergers.
The thick line is then the border, which is fine tuned by the Amoeba algorithm so that most of the large dots
lie above it, while at the same time most of the small points are located underneath it.
In this representation, the merger side of the border is clearly the zone of the diagram above the
border.}
\label{fig:explicacion}
\end{figure}

The method is schematically presented in Figure \ref{fig:explicacion}. Both axes represent
a dummy structural parameter, which could be any one of the structural parameters defined
in \S \ref{subsec:extendedcas}. The large dots represent merging systems and the small dots
are non mergers. The thick line is a \textit{best} border found by the $F-\mathrm{score}$
maximization algorithm. This is a second order polynomial in the \texttt{Parameter \#1} into
the \texttt{Parameter \#2} dimension.

\S \ref{subsec:compa_diagnostics} presents this optimization for a number
of merger diagnostics. Some of the merger diagnostics presented are taken
from the literature, while others are introduced there for the first time.
The optimization process used tries to maximize the $F-\mathrm{score}$
number, which is derived from the $r$ and $p$ statistics. A meaningful assessment
of the resulting contamination requires an accurate estimate of the merger
fraction in the galaxy sample used here. The contamination ratios
for the best performing diagnostics is presented in \S \ref{subsec:contaminacion}.

\subsection{The Training Set for the $F-\mathrm{score}$ Technique}
\label{subsec:Train}

The next question is then what objects from the parent population of
galaxies are to be taken as \emph{true mergers} and therefore
used as the training set for the
 $F-\mathrm{score}$ maximization technique. 
presented in \S \ref{subsec:compa_diagnostics}.
To this end, the morphological information presented in the
\textsc{STAGES} morphological catalogue is used.

For the purposes of this study, galaxies classified as mergers by at
least two of the visual observers of the \textsc{STAGES} team are defined to be as mergers. This subset 
includes 39 objects and constitutes the training set.
On the other hand, sources which were classified as mergers by one or none 
of the \textsc{STAGES} observers were then considered as non-mergers. This subset
is made of 1498 sources. This subset includes 83 sources that were regarded as mergers
by only one of the \textsc{STAGES} visual classifiers. These objects were not included
in the training set in order to obtain a more robust merger training sample.

Figure~\ref{fig:showroom_trainingsample} is an image atlas
showing images of all the galaxies from this training set. This figure contains
both the angular and physical scale of each inset, the
\textsc{STAGES} \texttt{ID} of each source, the 
environment (Cluster ``C'', and Field ``F''), and the 
morphological type (Elliptical ``E'', Lenticular ``S0'', Spiral ``Sp'', and ``Oth'').
Each inset also includes the number of observers that agreed on whether that particular galaxy is
a merger or not. 
%Cluster Core ``CC''

%\input{atlas_mm2.tex}
\begin{figure*}
\caption{Image atlas of the merger training set. The \textsc{COMBO-17} \texttt{ID} is included in each panel. 
This figure shows the objects which were classified as mergers by at least two of the \textsc{STAGES} team visual observers.
Each panel shows three different insets. The first image is the direct image, the second image is the model
created by \textsc{GALFIT}, and the third image is the residual image.
\textbf{This latter panel is shown with an inverted look-up-table whose dynamic range is 15\% that of the other two images. This is don in order to enhance the visibility of the fainter features in the residual images.}
}
\label{fig:showroom_trainingsample}
\begin{tabular}{cc} \hline
\includegraphics[scale=0.64]{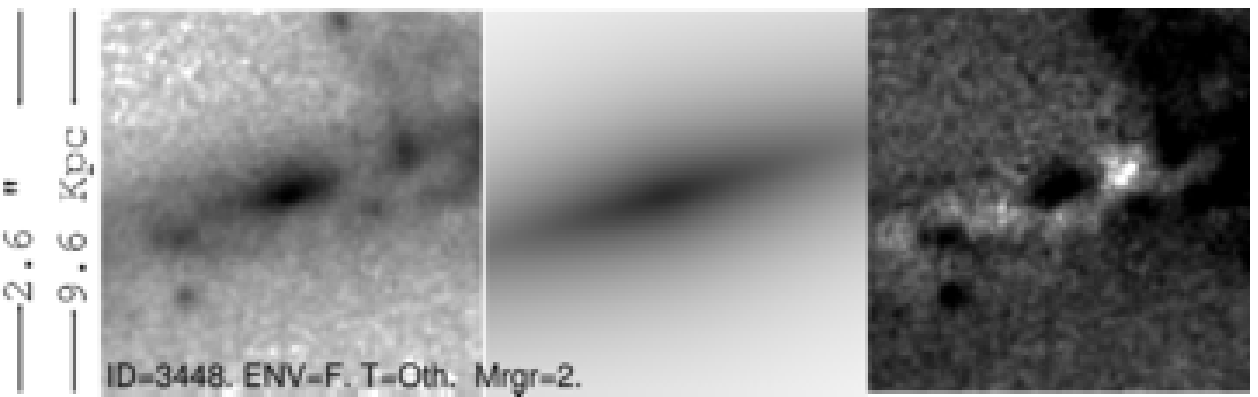}&\includegraphics[scale=0.64]{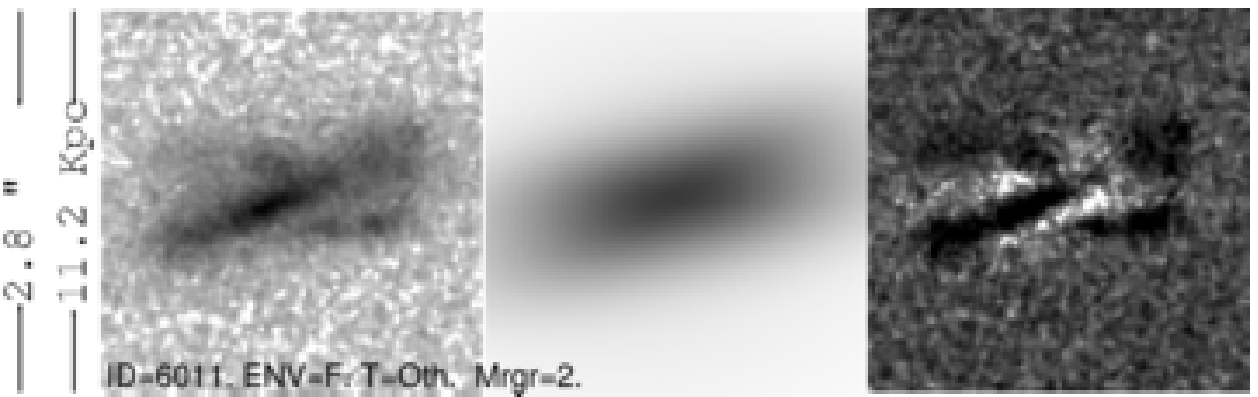}\\
\includegraphics[scale=0.64]{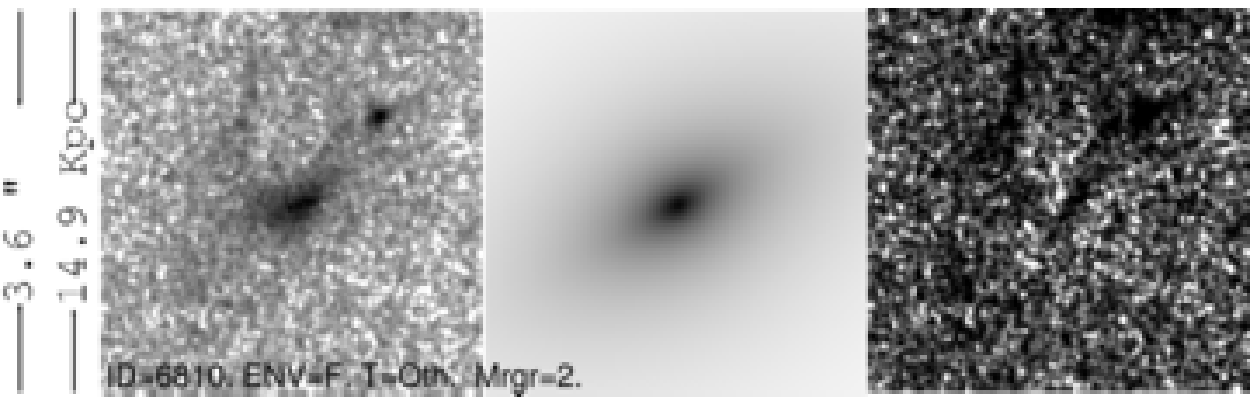}&\includegraphics[scale=0.64]{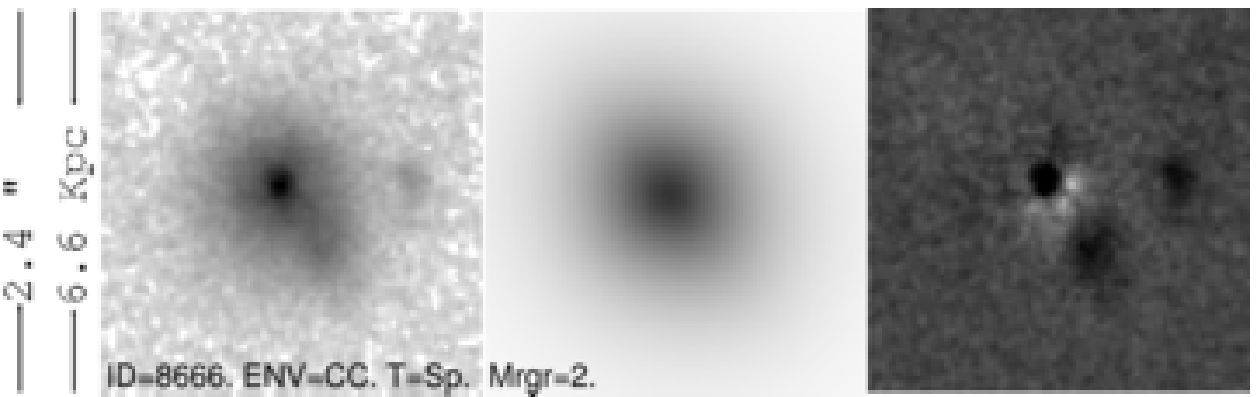}\\
\includegraphics[scale=0.64]{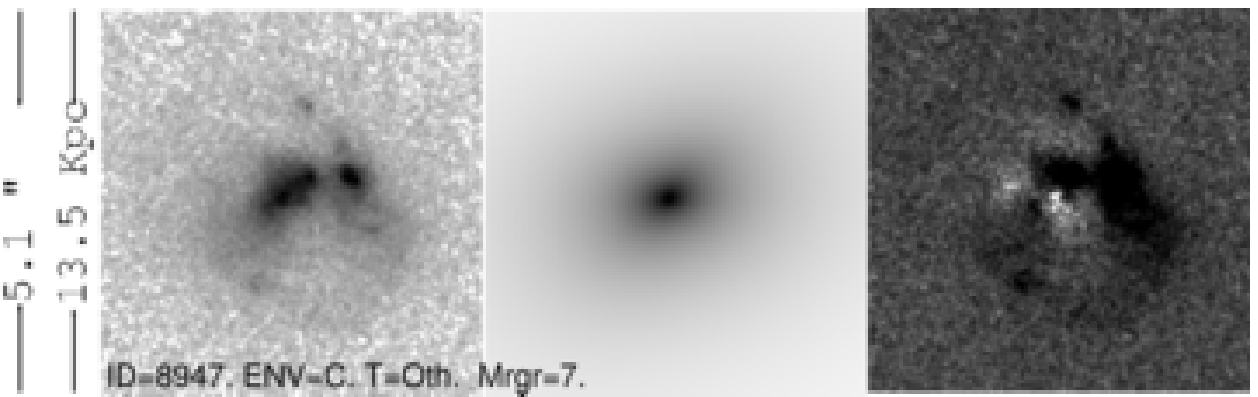}&\includegraphics[scale=0.64]{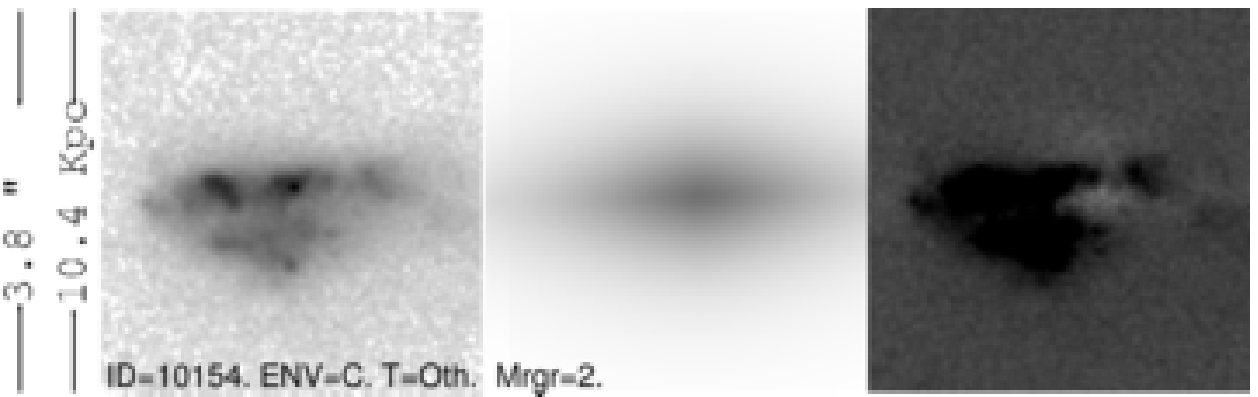}\\
\includegraphics[scale=0.64]{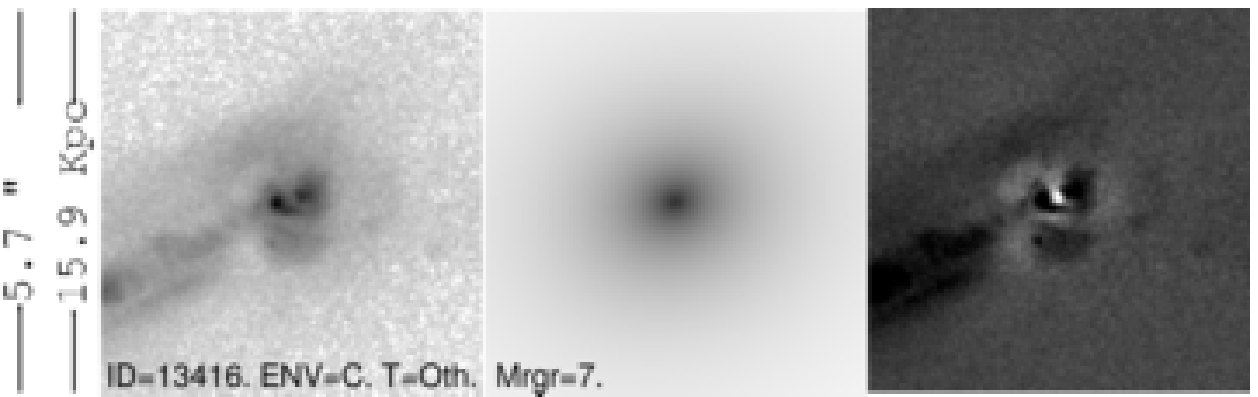}&\includegraphics[scale=0.64]{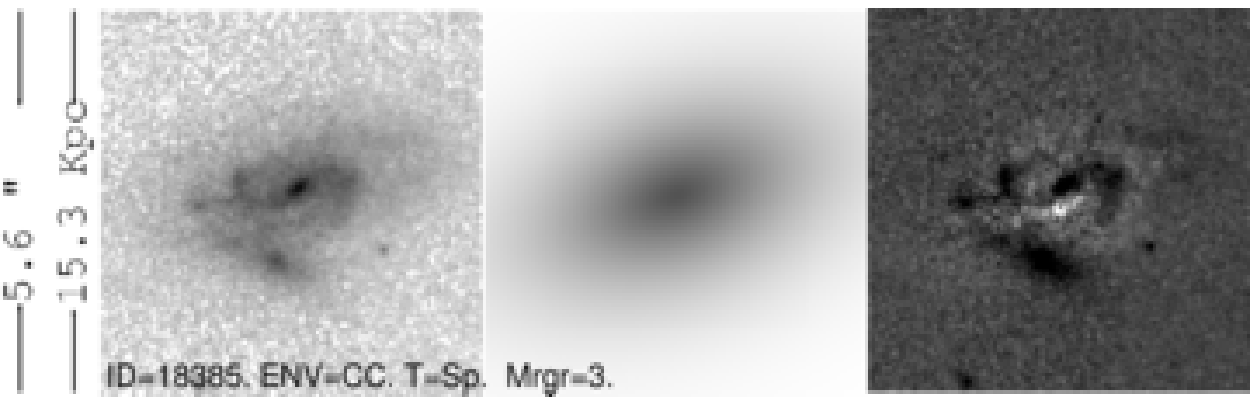}\\
\includegraphics[scale=0.64]{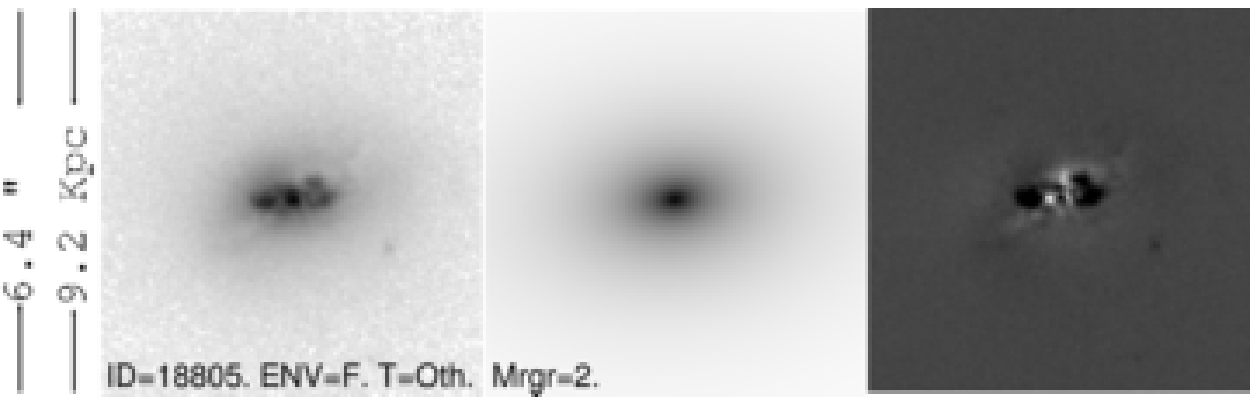}&\includegraphics[scale=0.64]{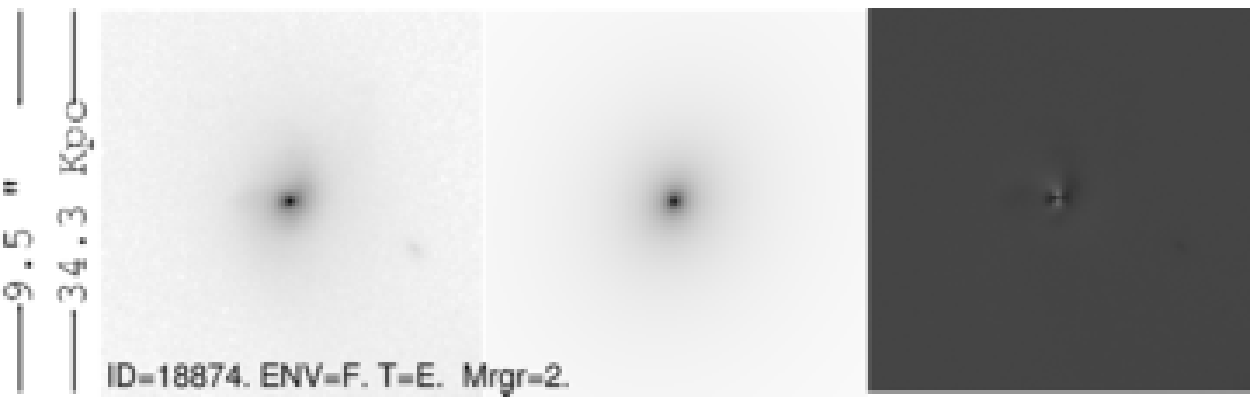}\\
\includegraphics[scale=0.64]{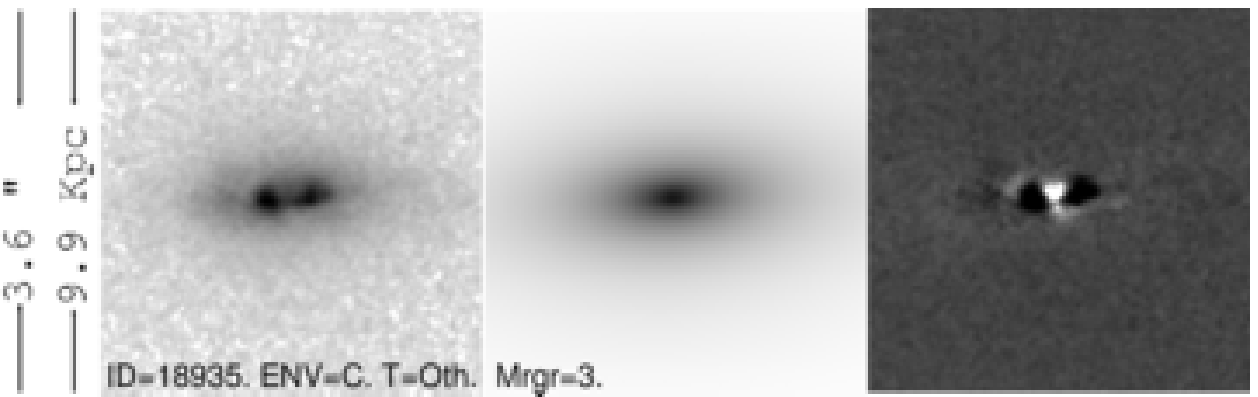}&\includegraphics[scale=0.64]{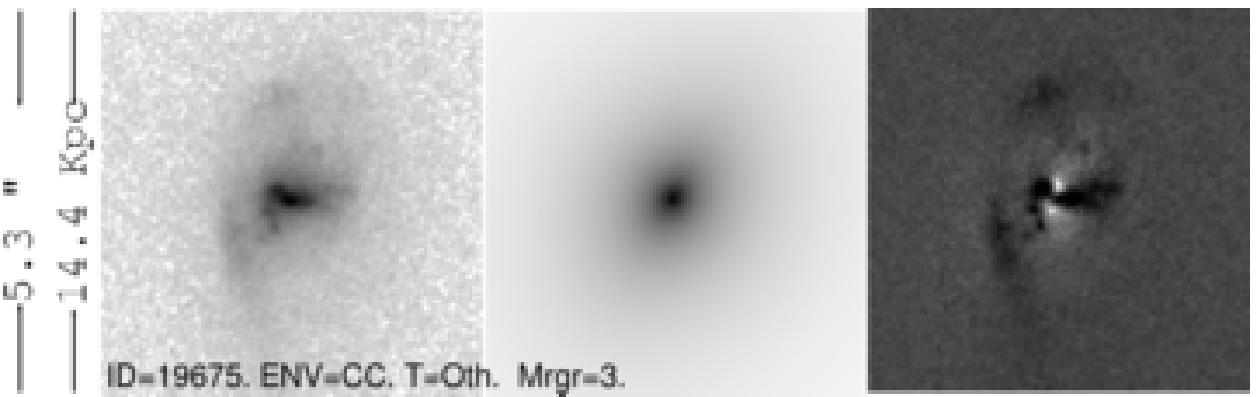}\\
\includegraphics[scale=0.64]{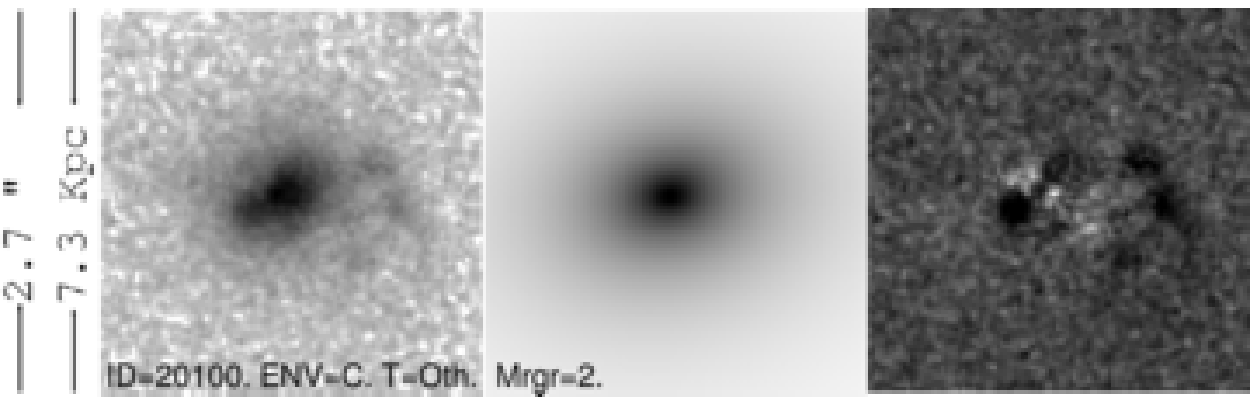}&\includegraphics[scale=0.64]{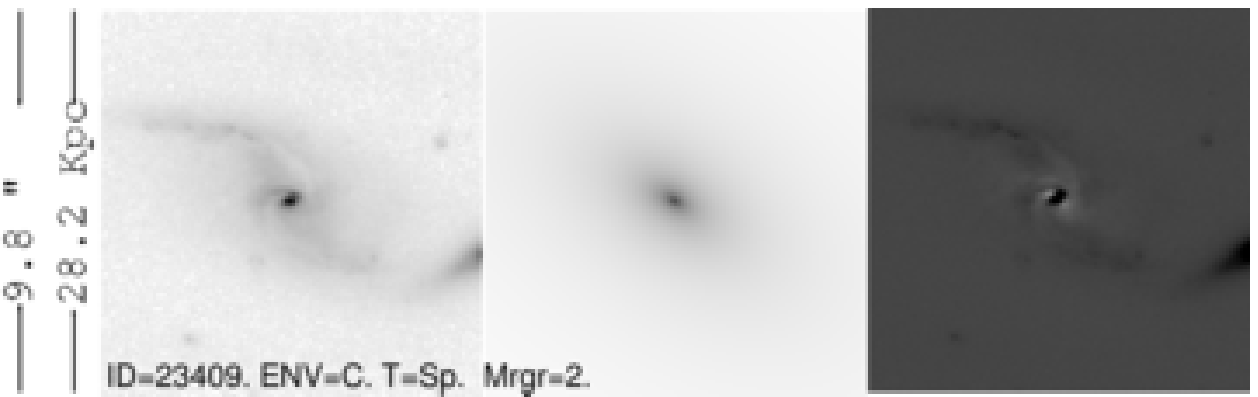}\\
\includegraphics[scale=0.64]{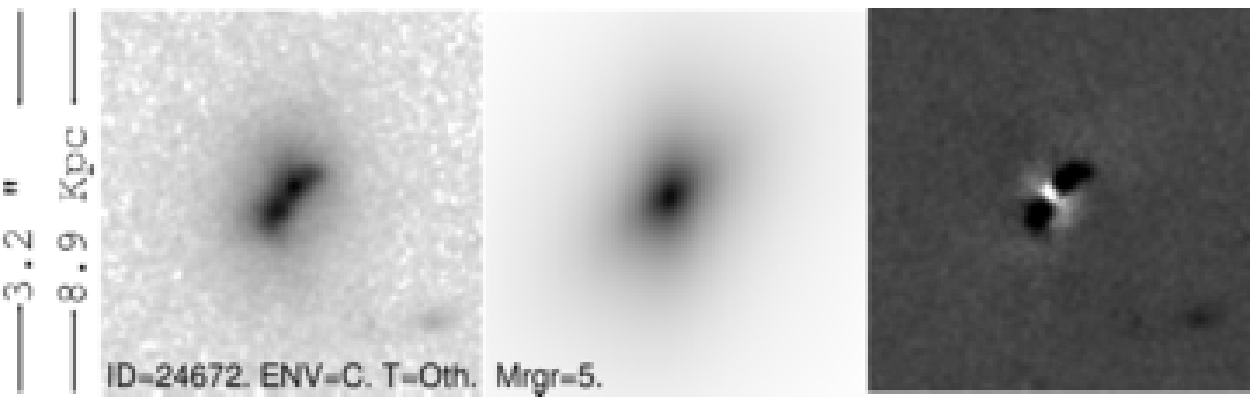}&\includegraphics[scale=0.64]{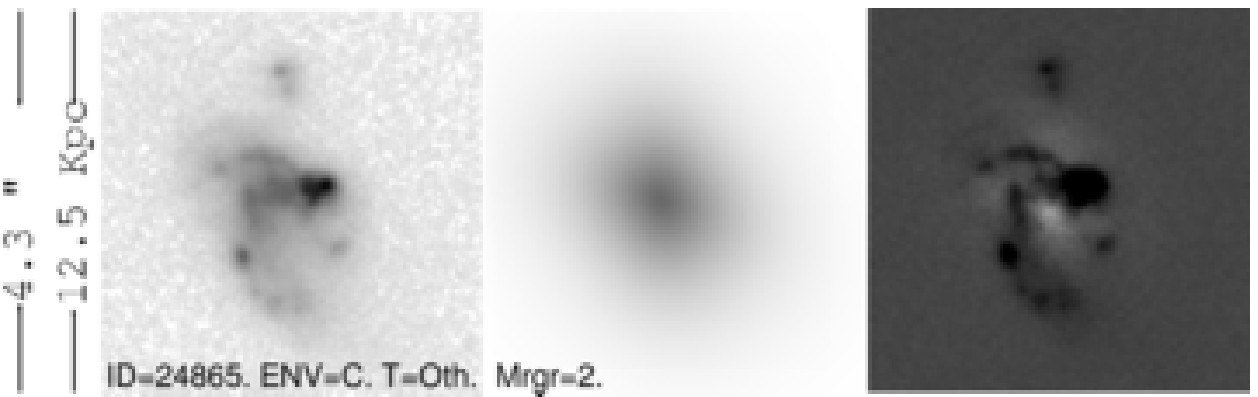}\\
\end{tabular}
%\vspace{0.2cm}
\end{figure*}

\begin{figure*}
\contcaption{Atlas of the training set galaxies.} 
\label{fig:showroom_trainingsample2}
\begin{tabular}{cc}
\includegraphics[scale=0.64]{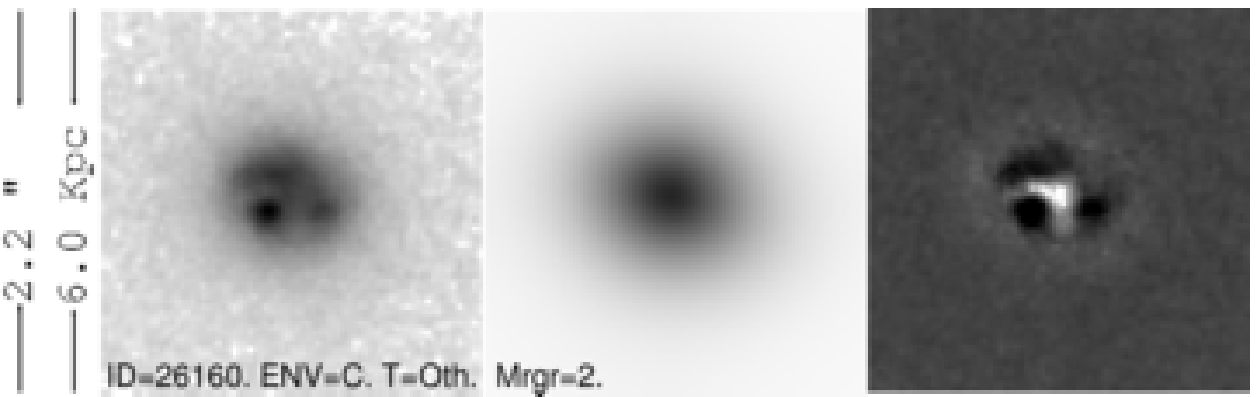}&\includegraphics[scale=0.64]{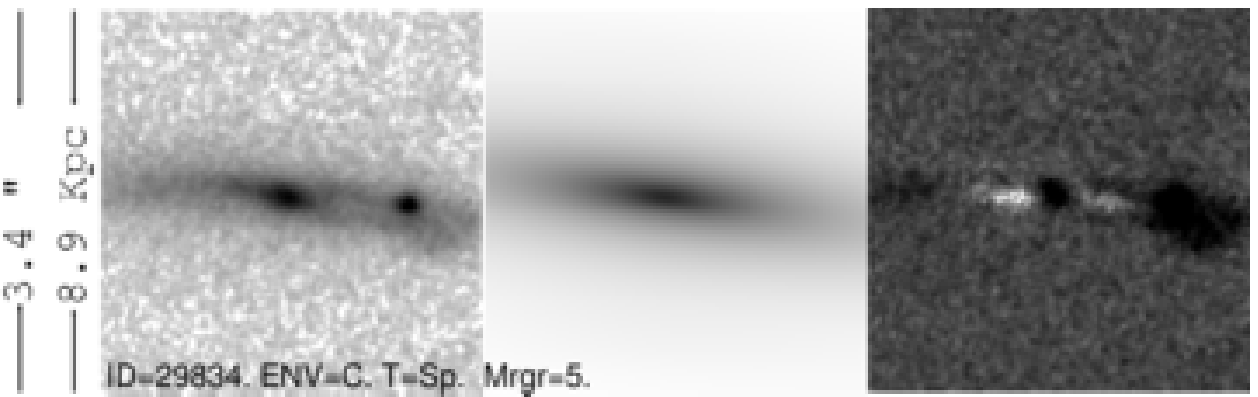}\\
\includegraphics[scale=0.64]{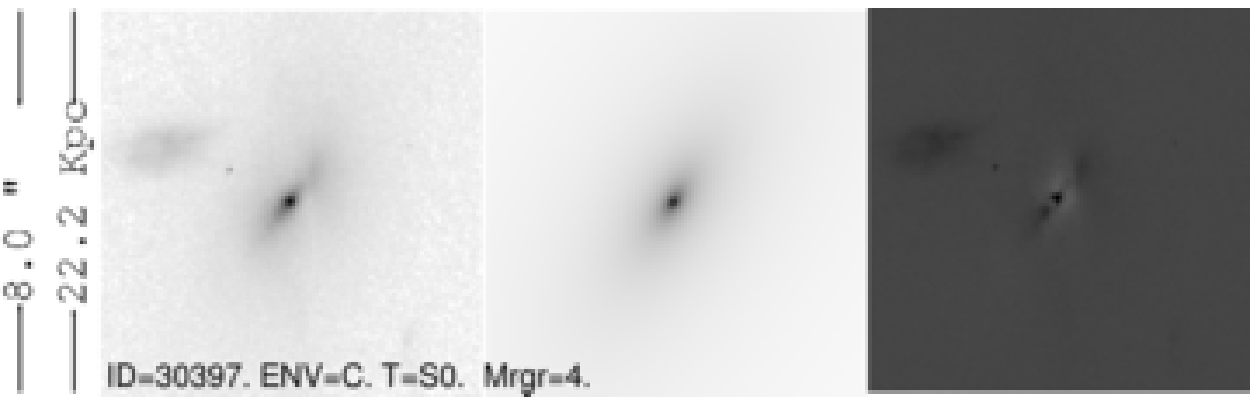}&\includegraphics[scale=0.64]{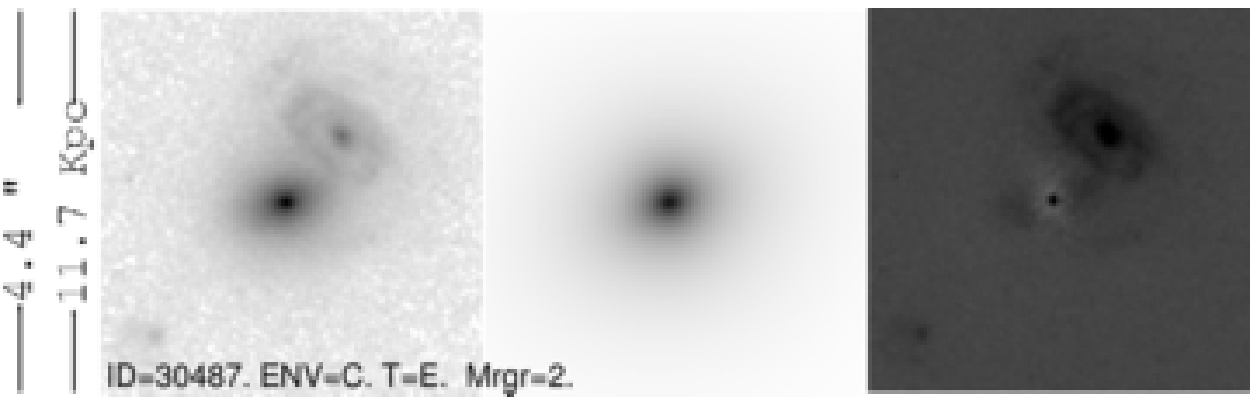}\\
\includegraphics[scale=0.64]{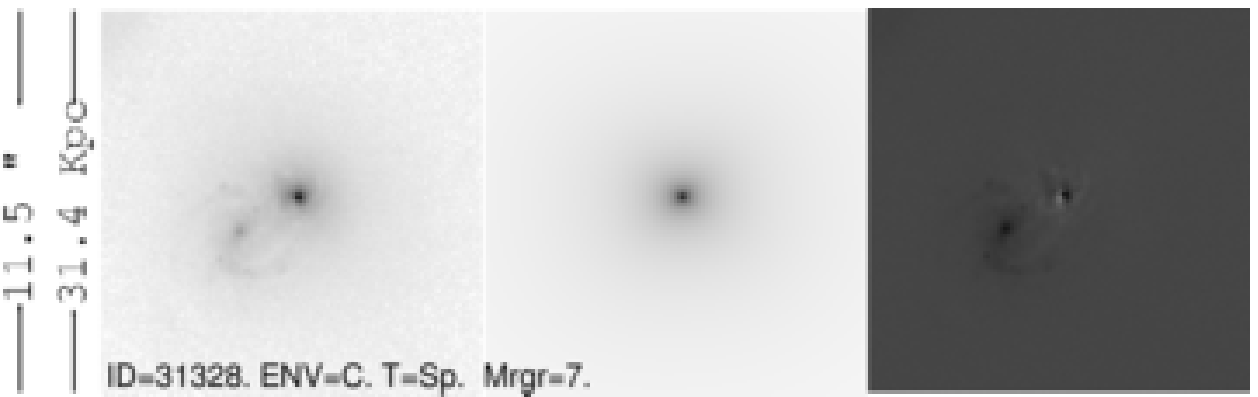}&\includegraphics[scale=0.64]{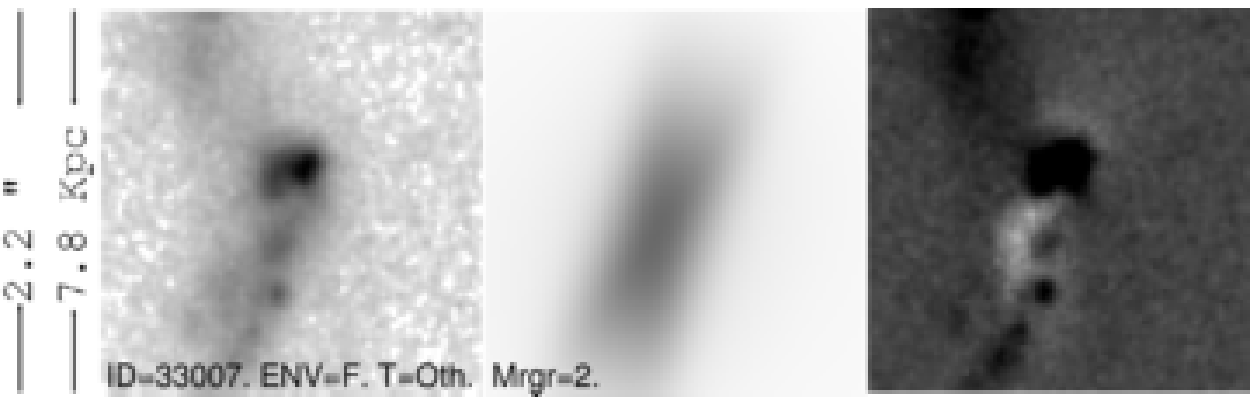}\\
\includegraphics[scale=0.64]{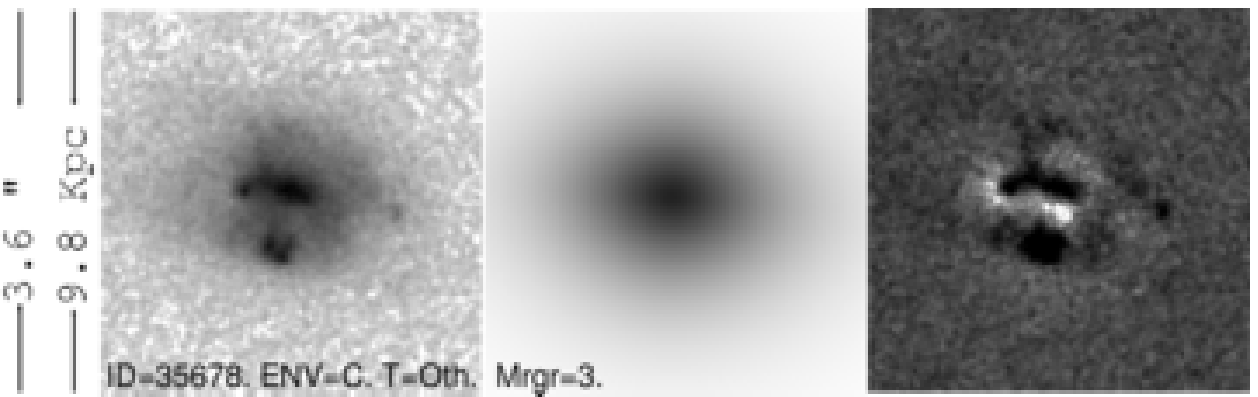}&\includegraphics[scale=0.64]{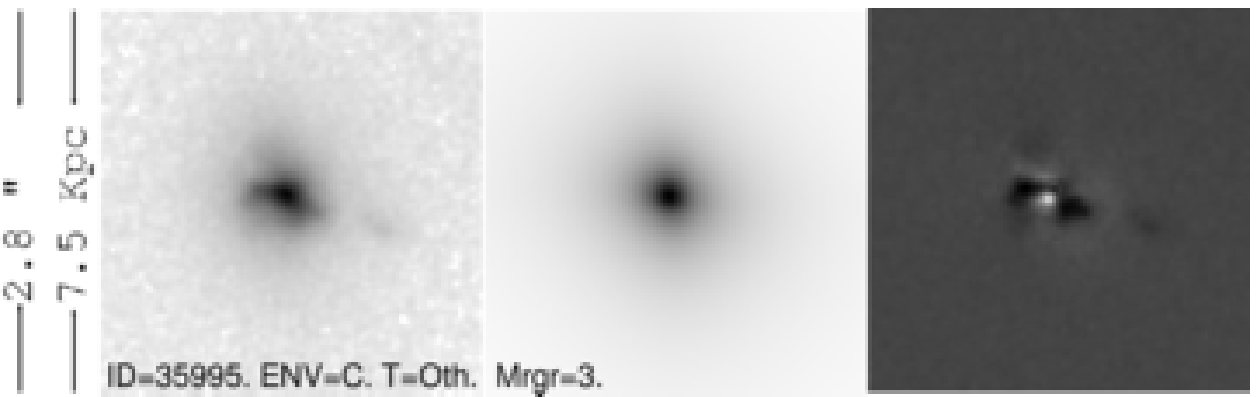}\\
\includegraphics[scale=0.64]{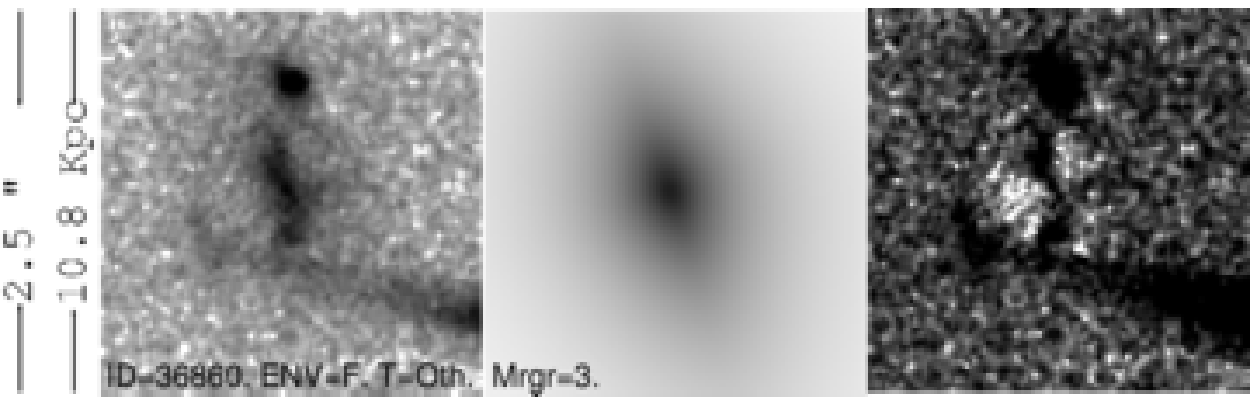}&\includegraphics[scale=0.64]{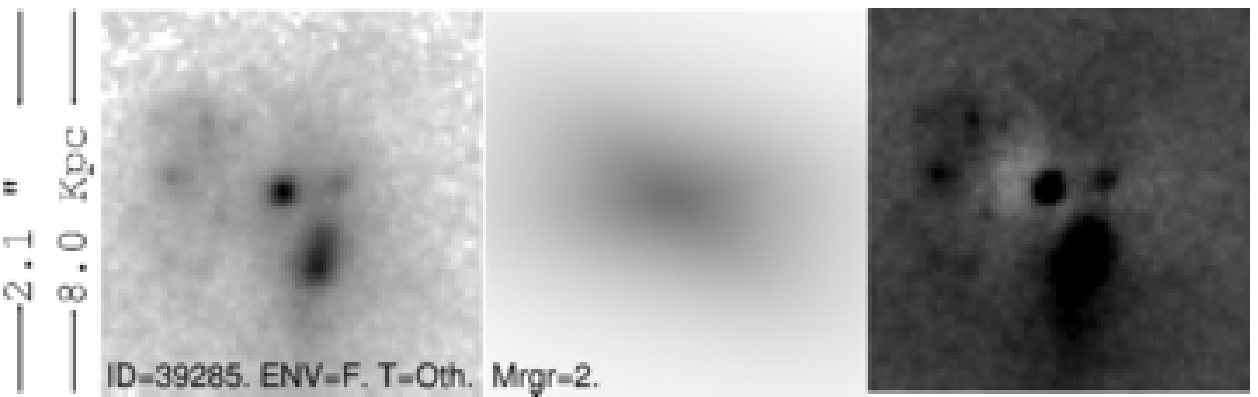}\\
\includegraphics[scale=0.64]{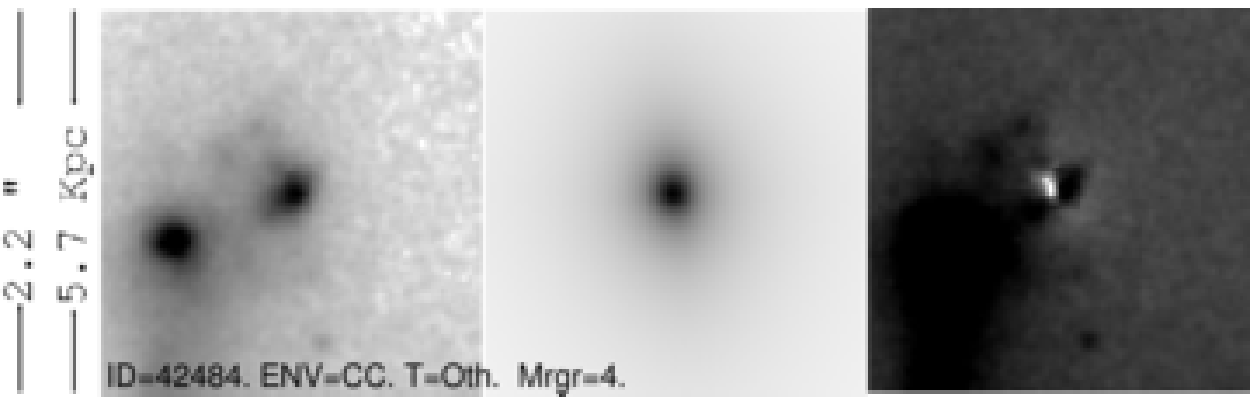}&\includegraphics[scale=0.64]{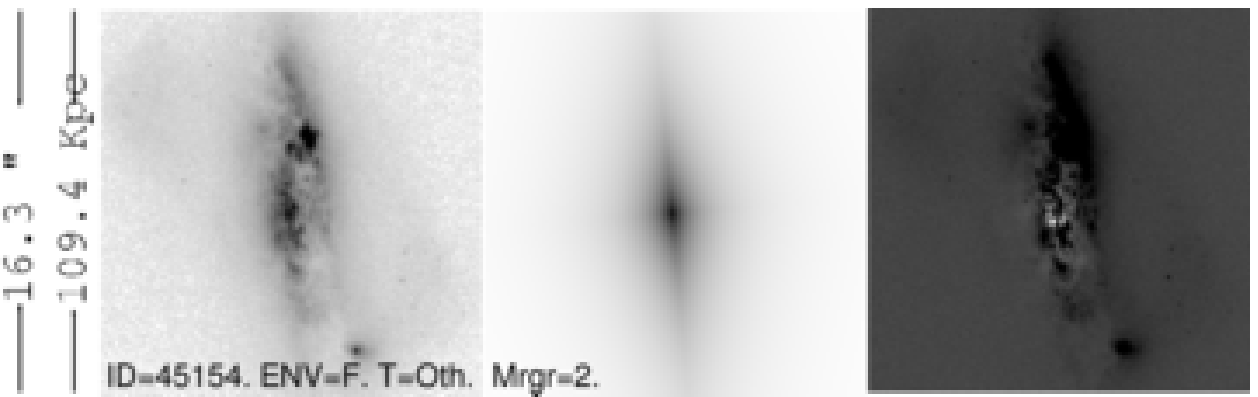}\\
\includegraphics[scale=0.64]{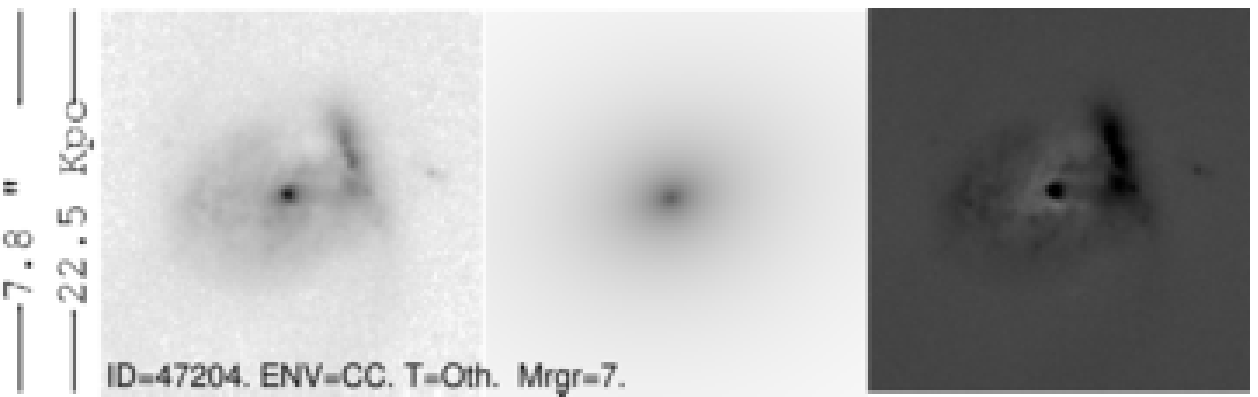}&\includegraphics[scale=0.64]{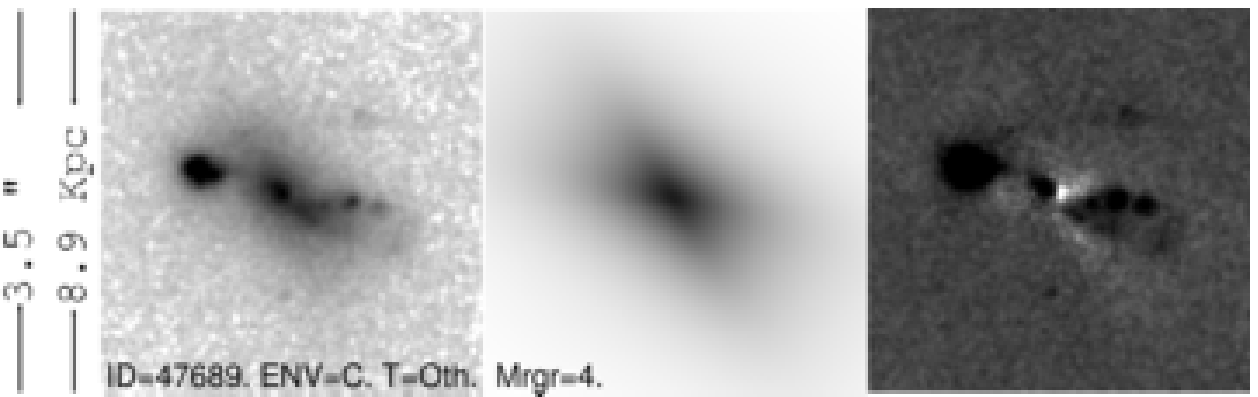}\\
\includegraphics[scale=0.64]{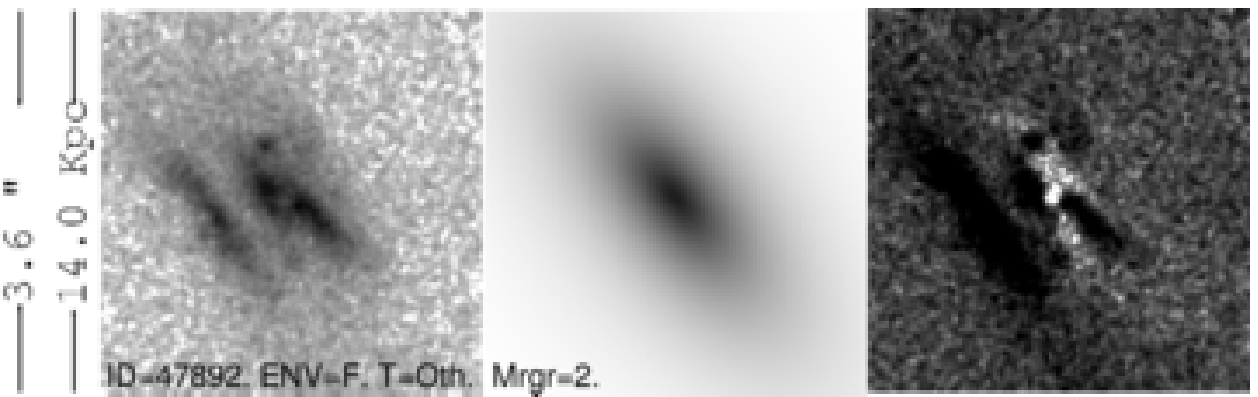}&\includegraphics[scale=0.64]{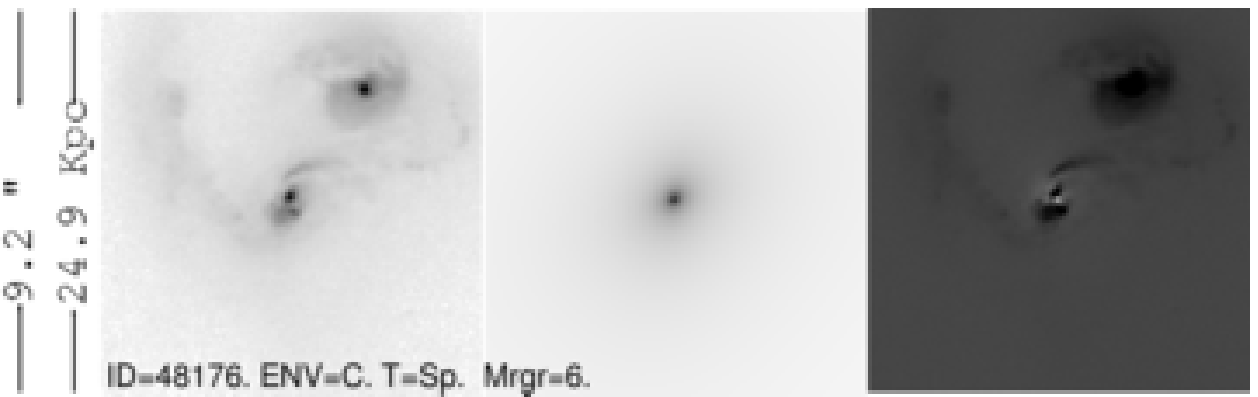}\\
\end{tabular}
\end{figure*}

\begin{figure*}
\contcaption{Atlas of the training set galaxies.} 
\label{fig:showroom_trainingsample3}
\begin{tabular}{cc}
\includegraphics[scale=0.64]{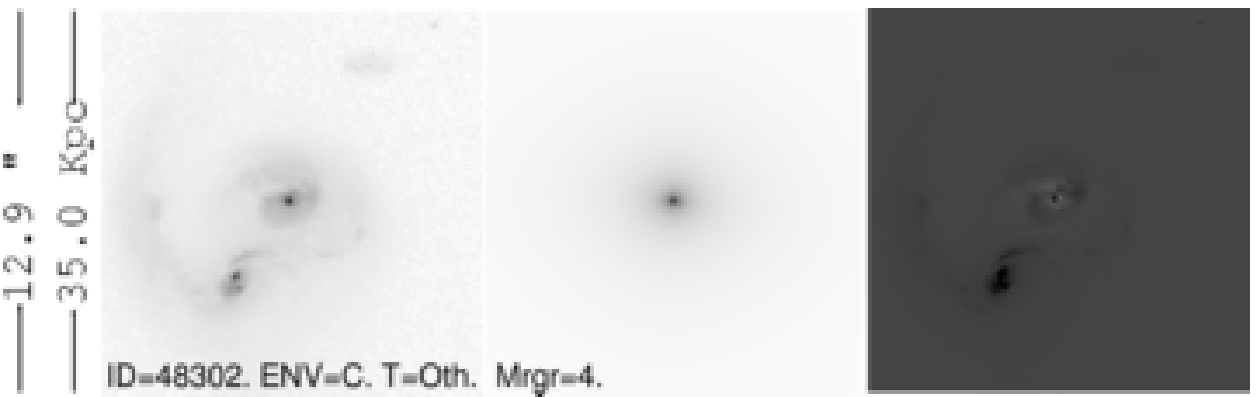}&\includegraphics[scale=0.64]{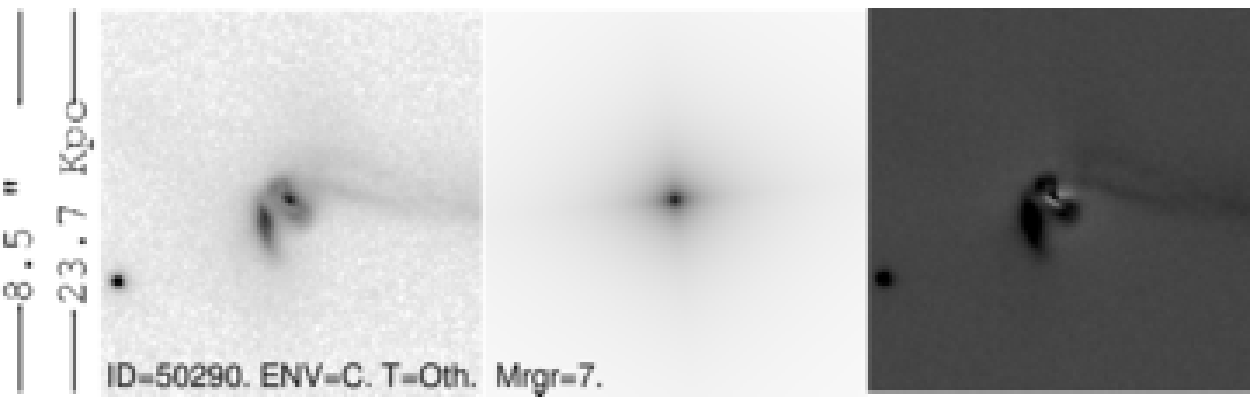}\\
\includegraphics[scale=0.64]{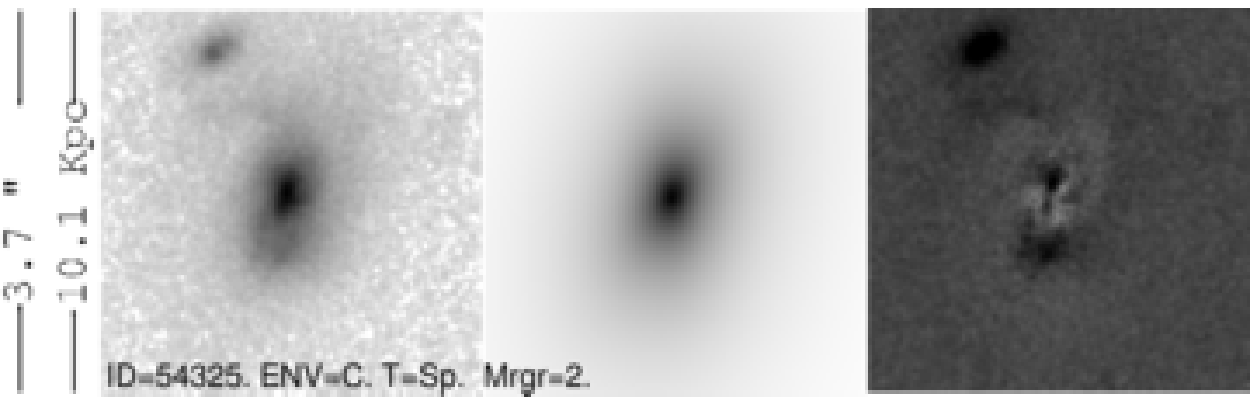}&\includegraphics[scale=0.64]{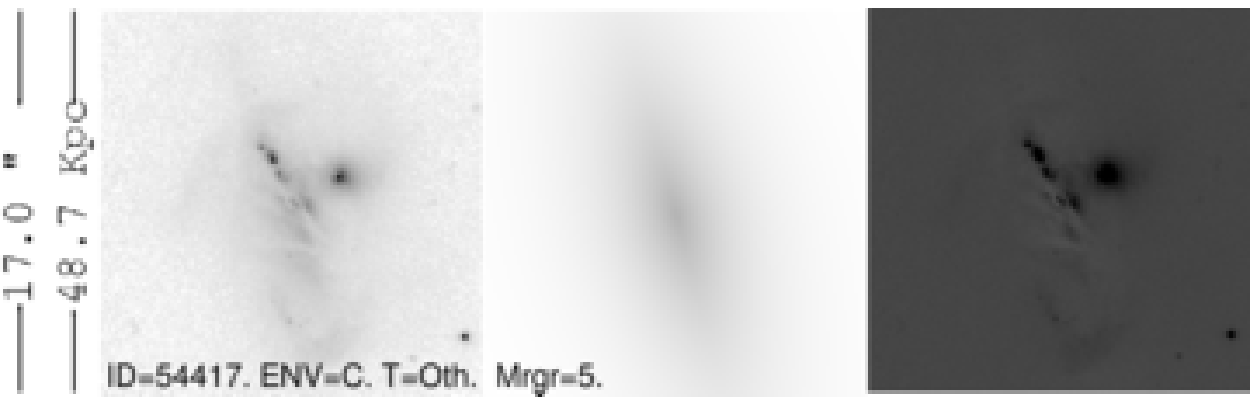}\\
\includegraphics[scale=0.64]{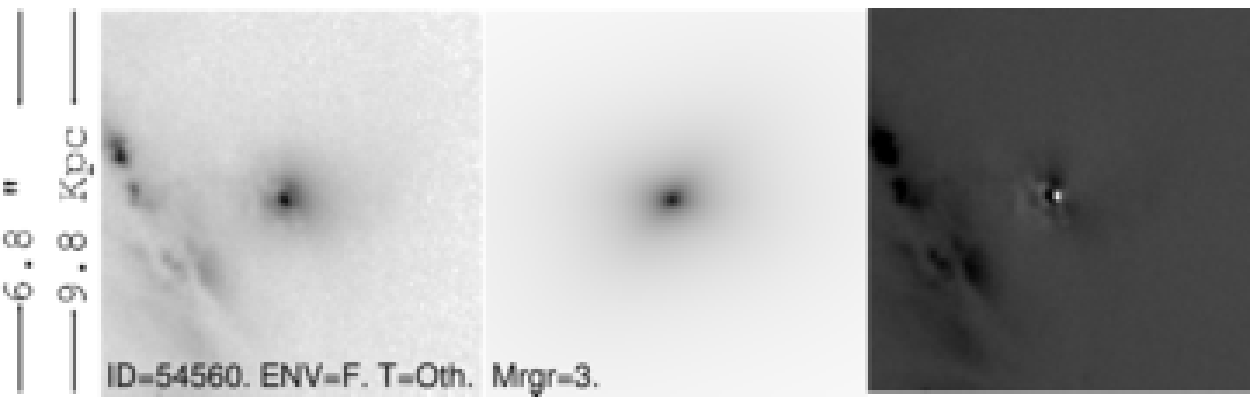}&\includegraphics[scale=0.64]{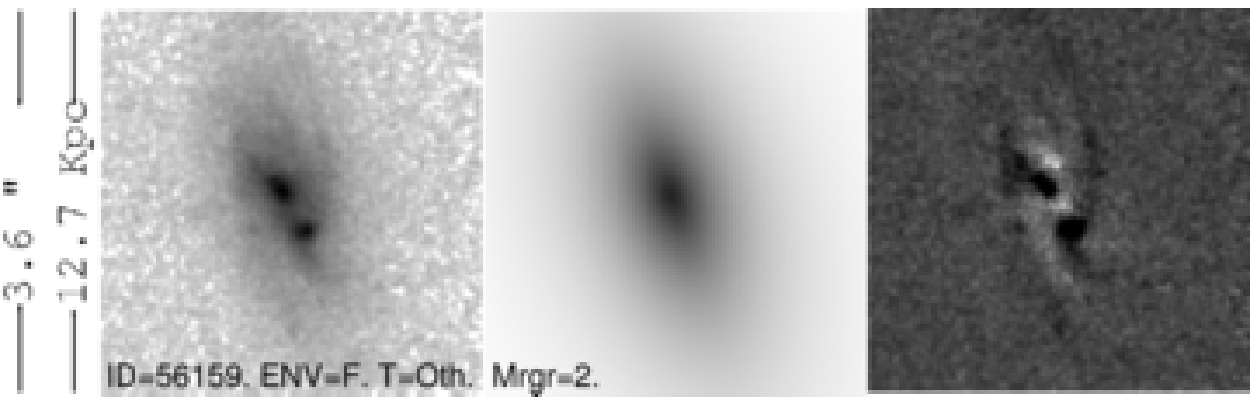}\\
\includegraphics[scale=0.64]{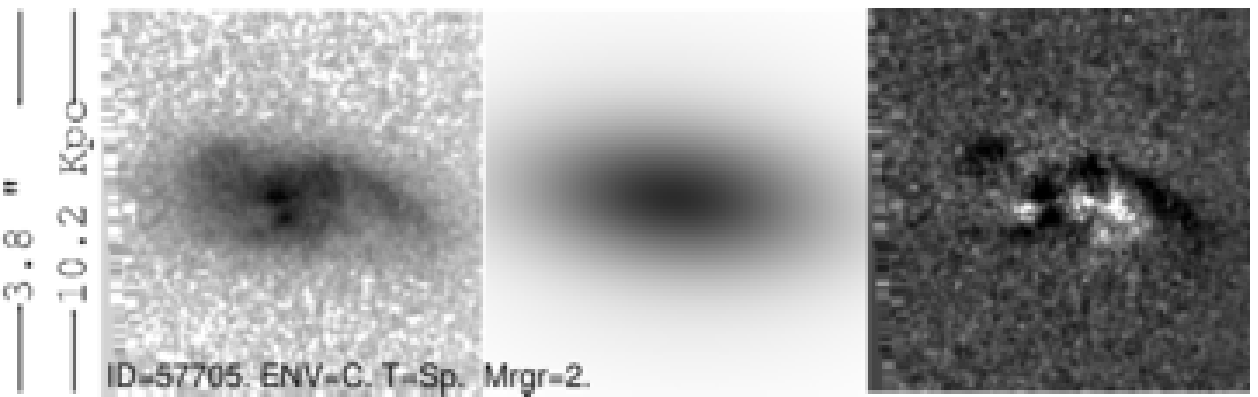}&\phantom{Nothing to see here.}\\
\end{tabular}
\end{figure*}

Figure~\ref{fig:showroom_trainingsample} shows that most of the 39 galaxies used 
as the training set are indeed bona-fide mergers. It is also true that a small fraction might
be more marginal. These latter galaxies need to be included in the merger set
because one of the main goals of the current work is to explore whether the
analysis of the structural parameters of the residuals can be used to detect minor mergers reliably.

\subsection{Comparison Between Merger Diagnostics.}
\label{subsec:compa_diagnostics}

Given the wealth of structural parameters that have been calculated for the galaxies in the parent population, it
is better to start presenting how the $F-\mathrm{score}$ method works for the 
$G(\mathrm{Obj})-M_{20}(\mathrm{Obj})$ plane. This is a very well known merger 
diagnostic that has already been used in the literature.
In this subsection, the image for which the structural parameters are calculated
is denoted inside parentheses after the name of the structural parameter itself. ``Obj'' means that 
the morphological parameter was obtained in the original image, ``MDL'' refers to the S\'ersic 
model, and ``Res'' implies that the parameter was obtained for the residual image.

\begin{figure}
%\begin{center}
\includegraphics[scale=0.35,angle=90]{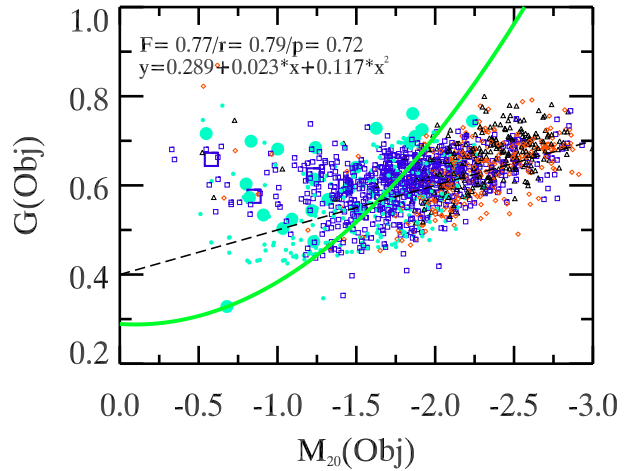}
\caption{The well-known $G(\mathrm{Obj})-M_{20}(\mathrm{Obj})$ merger criteria applied 
to the \textsc{STAGES} sample. Large symbols are sources that were marked as mergers 
by the \textsc{STAGES} team observers. Blue squares show spiral galaxies, black triangles denote lenticular
systems and red diamonds present elliptical galaxies. Irregular and disturbed systems are represented as
beige filled circles. The green line is the \textit{best} border found by the Amoeba 
algorithm, and the black, dashed line is the initial guess this algorithm is given.
The optimal value of $F-\mathrm{score}$, together with the sample completeness and specificity
are given within the figure.}
\label{fig:metodo_GObjwMObj}
%\end{center}
\end{figure}

Figure~\ref{fig:metodo_GObjwMObj} shows the $G(\mathrm{Obj})-M_{20}(\mathrm{Obj})$ plane
used as a merger diagnostic.
Large symbols are the galaxies that were marked as mergers by the \textsc{STAGES} 
observers\footnote{This is thus 
the training set the Amoeba algorithm will use to find the \textit{best} border
to separate mergers from non-mergers.}, and the
smaller symbols show the galaxies that were not regarded as mergers. Cyan filled circles
represent irregular objects, blue squares give the location of
spiral galaxies, black triangles represent lenticular systems and red diamonds denote elliptical galaxies.
This panel shows the ``best'' border as a thick, green solid line.
The dashed black line is just the initial border that the Amoeba algorithm
is given to start its iterations. This is just a rough guess given
the location of the larger symbols in the diagram.
The final \textit{best} border the Amoeba algorithm obtains
does not depend on the initial guess as long as this initial guess
is reasonable. The polynomial that defines the \textit{best} border is also given within the figure, together with
the resulting completeness, contamination, and $F-\mathrm{score}$ values.
Here, the completeness is defined as the number of clear visual mergers above the border, divided
by the total number of visual mergers (i.e., the sensitivity).

It is seen that the $F-\mathrm{score}$ maximization algorithm has been able to
find most of the merging systems that constitute the training set.
The statistical quality of this sample is $F=0.77$.
The completeness is $r=0.79$, and the contamination by
objects not classified as mergers is $1-p=0.28$. 
It is interesting to note that this method rejects most of the lenticular
objects from the training sample. 

\begin{figure}
%\begin{center}
\includegraphics[scale=0.35,angle=90]{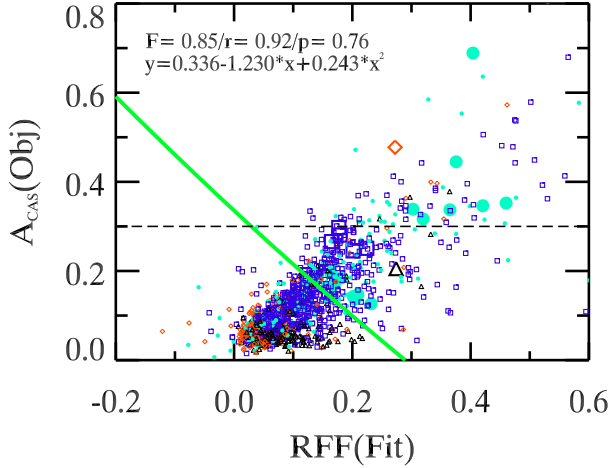}
\caption{The $A(\mathrm{Obj})-RFF$ merger diagnostic.
Symbols and information as in figure \ref{fig:metodo_GObjwMObj}.}
\label{fig:metodo_AObjwRFF}
%\end{center}
\end{figure}

The next merger diagnostic presented is the $A(\mathrm{Obj})-RFF$
plane. This merger indicator is motivated in the more common
$A(\mathrm{Obj})-S(\mathrm{Obj})$, using the \textit{RFF}
instead of the clumpiness since these two quantites are very similar.
Figure~\ref{fig:metodo_AObjwRFF} presents this test. This is the first test 
that makes use of the structural parameters of the residual images.
In this test, the initial border is $A(\mathrm{Obj})=0.30$, simply because
the asymmetry is expected to bear the highest predictive power in this
test.

Figure~\ref{fig:metodo_AObjwRFF} shows that the statistical quality
of the sample obtained using this criterion has improved significantly 
with respect to the results achieved by the $G(\mathrm{Obj})-M_{20}(\mathrm{Obj})$
diagnostic. The sample purity is $F=0.85$, the sensitivity is $r=0.92$, and the specificity is
$p=0.76$. Also, Figure~\ref{fig:metodo_AObjwRFF} shows a clear correlation between
$A(\mathrm{Obj})$ and the \textit{RFF}.  It might be objected that the
usual limit in the asymmetry introduced in \cite{2003ApJS..147....1C} is 
$A(\mathrm{Obj})>0.35$, while
the limit suggested by this test is $A(\mathrm{Obj})>0.20$.
This is not only caused by the training sample used, but by the current
choice of the $\beta$ parameter, which is designed to weight completeness more
than specificity. Had these condition been different, the
resulting \textit{best} borders would have changed.
This method therefore imperatively requires an objective calibration in order to produce
meaningful, physically motivated borders and hence reliable merger fractions.
This objective calibration will be produced in a forthcoming paper, using
full fledged N-body simulations of galaxy mergers.
Also, it has to be borne in mind that the $A(\mathrm{Obj})>0.35$ criterion
is tuned to detect major mergers, and one of the aims of this work is
to improve the morphological detection of minor merger episodes.
Figure~\ref{fig:metodo_AObjwRFF} also shows that the $F-\mathrm{score}$ maximization
algorithm has indeed found that the ``best'' border is very different from the
initial, flat guess. In particular, the Amoeba algorithm has discovered the correlation
between $A(\mathrm{Obj})$ and \textit{RFF}, and takes advantage of it by converging towards a line that cuts
the correlation in a perpendicular way. It is also tempting to think that the use
of B+D decompositions could produce better overall fits for the S0 and Sp galaxies, leading
to lower RFF numbers. This in turn could reduce the number of false positives for these sources, increasing
the potential of this structural parameter as a merger diagnostic. This will be explored in the future.

\begin{figure}
\includegraphics[scale=0.35,angle=90]{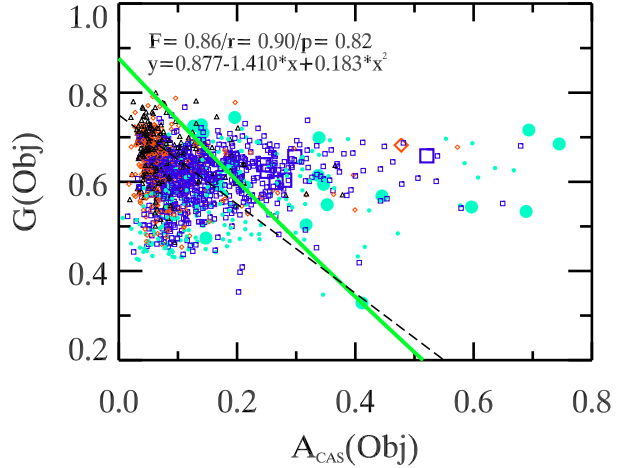}
\caption{This figure presents the $G(\mathrm{Obj})-A(\mathrm{Obj})$ plane used as a merger
diagnostic. Symbols and information as in figure \ref{fig:metodo_GObjwMObj}.}
\label{fig:metodo_GobjAobj}
\end{figure}

The results shown in Figures~\ref{fig:metodo_GObjwMObj} and~\ref{fig:metodo_AObjwRFF}
make it possible to think that a good merger diagnostic could be put to the test
by combining $G(\mathrm{Obj})$ with $A(\mathrm{Obj})$. Figure \ref{fig:metodo_GobjAobj} presents this 
investigation, which confirms the previous ideas with a sample purity of $F=0.86$, a
recall $r=0.90$ and a fairly high specificity $p=0.82$. The algorithm has correctly identified that 
the \textit{best} border in this plane is a diagonal, as expected. The final border is found
to be very close to initial border tried by the minimization algorithm.

\begin{figure}
\includegraphics[scale=0.35,angle=90]{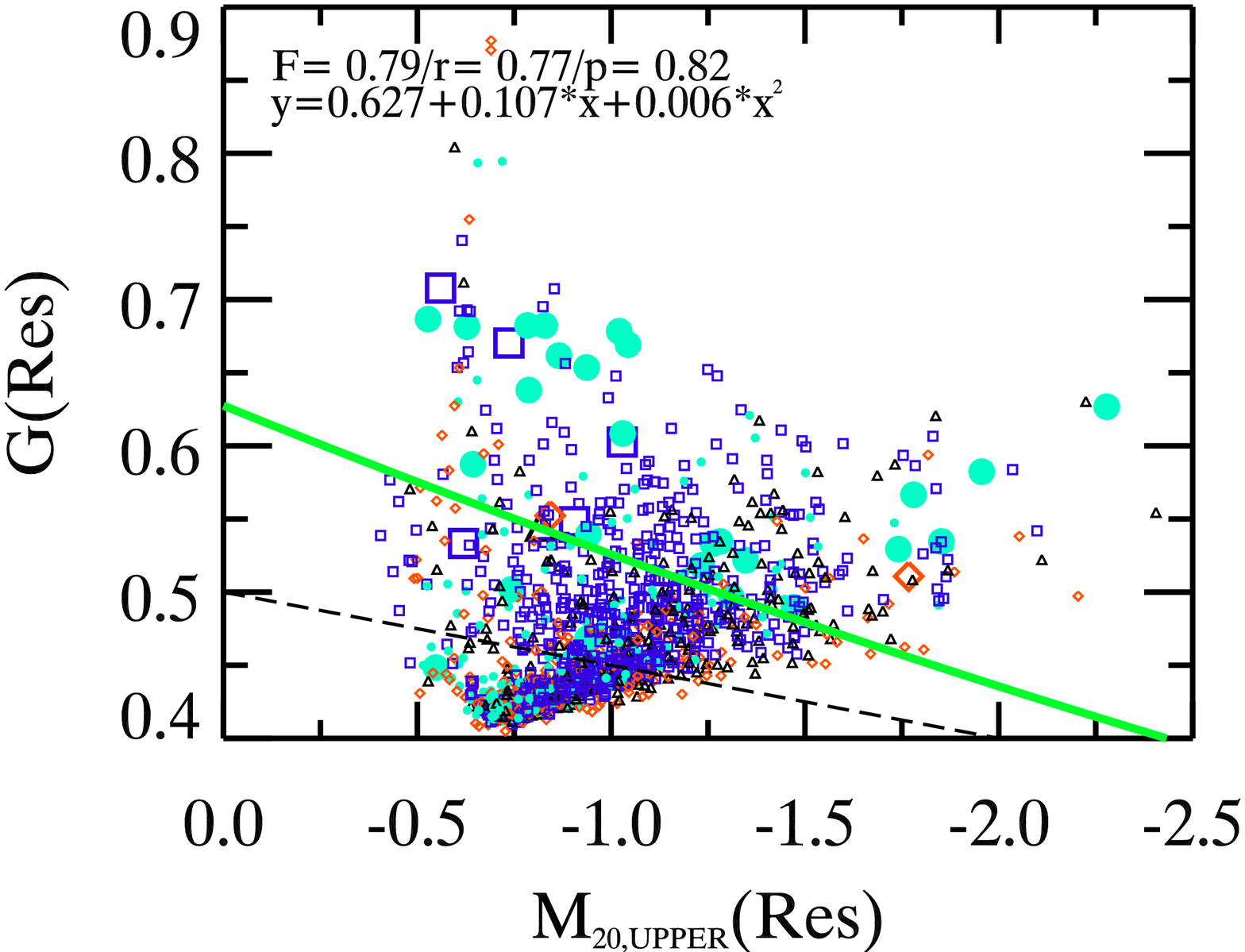}
\caption{The $G(\mathrm{Res})-M_{20}(\mathrm{Res})$ plane as a merger diagnostic. Symbols as in Fig. \ref{fig:metodo_GObjwMObj}.}
\label{fig:metodo_GRes-M20Res}
\end{figure}

The following merger criterion considered continues with the exploration of the morphological parameters of the residual images.
Figure~\ref{fig:metodo_GRes-M20Res} shows the $G(\mathrm{Res})-M_{20}(\mathrm{Res})$ merger test 
plane. This diagnostic was motivated by two main ideas.

\begin{itemize}
\item{If the galaxy that was fitted and removed by the smooth S\'ersic model was indeed involved in 
a merger episode in a very late stage, the residual image should expose the effect of the fainter component 
as a bump. This bump will then be easily detected in $G(\mathrm{Res})$ because it will 
form a reduced number of high-intensity pixels, surrounded by a large number of pixels with very low 
intensity values with an average value of 0.0. This will boost the value of $G(\mathrm{Res})$. If, on the 
other hand, the galaxy is well described by the smooth model, the pixel intensity of the residuals will 
all cluster around 0.0, and $G(\mathrm{Res})$ will be very close to 0.5.}

\item{In the case of the horizontal axis, which is $M_{20}(\mathrm{Res})$, the situation is very similar.
If there is an off-centre bump in the residual image, it will probably be among the 20\% brightest
pixels of the residual image. This will enhance the value of $M_{20}$ with respect to the situation
in which no substructure is seen in the residuals. In this latter situation, the brightest 20\% pixels of the residuals
will be randomly distributed in the image, albeit with a preference for the central values.}
\end{itemize}

\begin{figure}
%\begin{center}
\includegraphics[scale=0.35,angle=90]{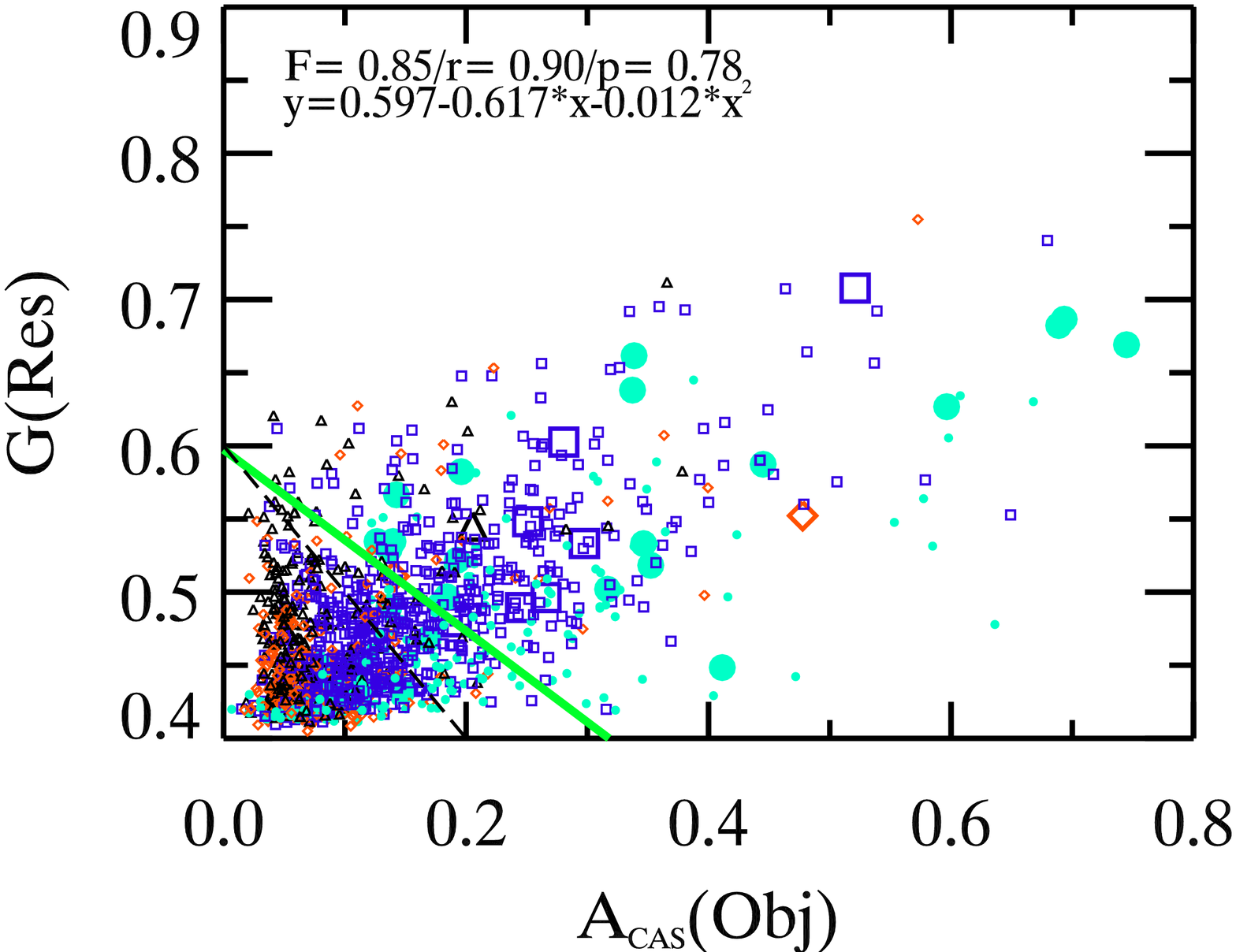}
\caption{The $G(\mathrm{Res})-A(\mathrm{Obj})$ plane as a merger diagnostic. Symbols as in Fig. \ref{fig:metodo_GObjwMObj}.}
\label{fig:metodo_GRes-AObj}
%\end{center}
\end{figure}

Figure \ref{fig:metodo_GRes-M20Res} confirms all these expectations. It is seen that the sample purity 
is $F=0.79$, with $r=0.77$. The specificity is $p=0.82$.
It is also clearly seen that the predicting power of this diagnostic is mostly associated with
$G(\mathrm{Res})$. The results of this test suggest
the use of the $G(\mathrm{Res})-A(\mathrm{Obj})$ plane as a merger diagnostic. This idea is presented
in Figure \ref{fig:metodo_GRes-AObj}. The correlation between the two quantities presented is evident in this figure.
This alone indicates that $G(\mathrm{Res})$ is bound to be a good
merger tracer, even when used by itself. For this test, the sample purity is $F=0.85$, with
a good completeness $r=0.90$ and a better specificity $p=0.78$. This merger indicator
is then comparable to the $G(\mathrm{Obj})$ with $A(\mathrm{Obj})$ indicator.

\begin{figure}
\includegraphics[scale=0.35,angle=90]{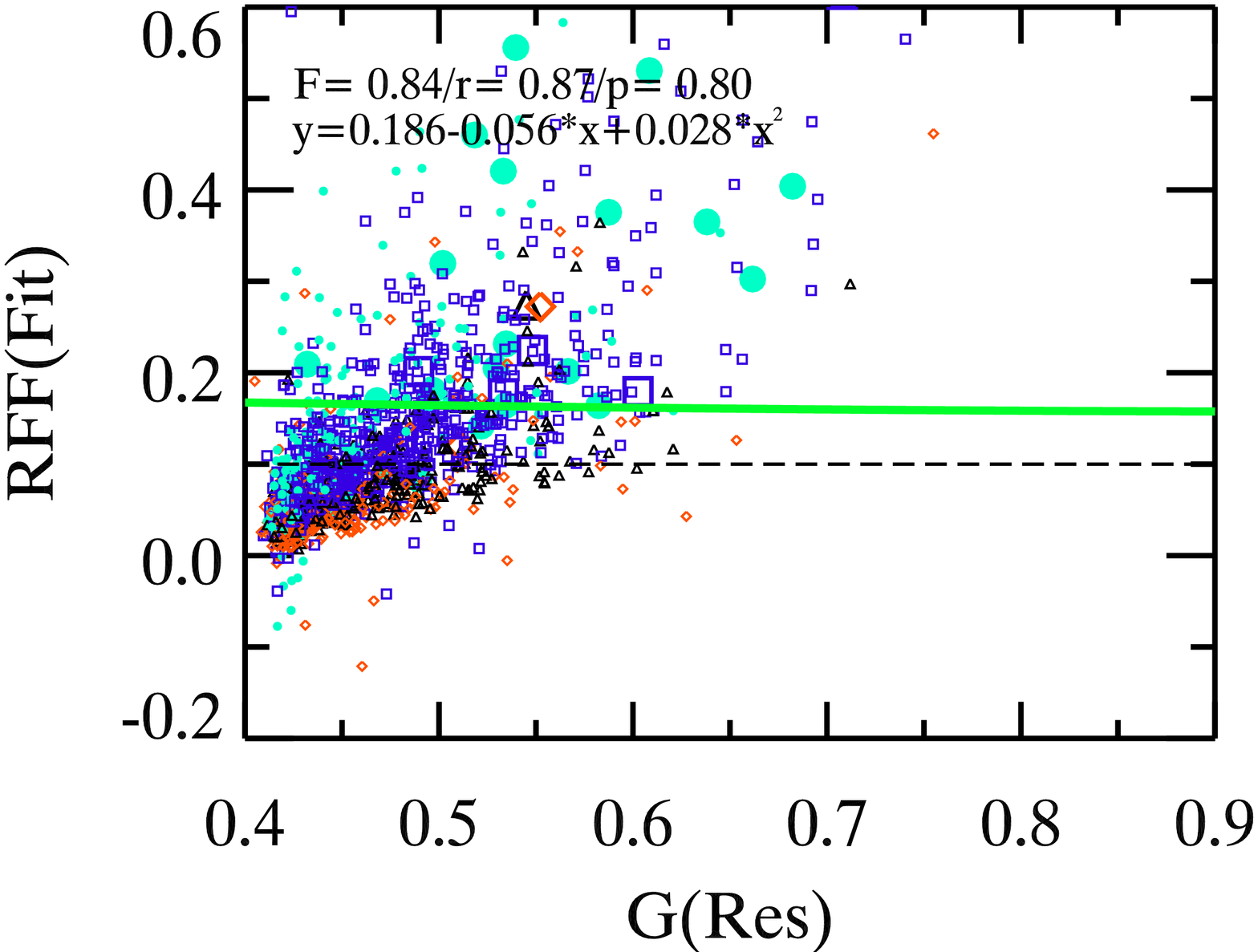}
\caption{The $\mathrm{RFF}-G(\mathrm{Res})$ plane, used as a merger criterion. Symbols as in Fig. \ref{fig:metodo_GObjwMObj}.}
\label{fig:metodo_RFF-Gres}
\end{figure}

The next natural merger criterion that the conclusions drawn from Figures~\ref{fig:metodo_GRes-M20Res}
and~\ref{fig:metodo_AObjwRFF} lead to explore is the $G(\mathrm{Res})-RFF$ plane, shown in 
Figure~\ref{fig:metodo_RFF-Gres}.
These two parameters have been selected because the \textit{RFF} correlates 
with $A(\mathrm{Obj})$. It is therefore expected that this indicator will also 
be able to separate mergers from other galaxies.
The $G(\mathrm{Res})$ was used because Figure~\ref{fig:metodo_GRes-M20Res} shows
it could single out mergers almost as a standalone indicator.
Figure~\ref{fig:metodo_RFF-Gres} shows that the \textit{RFF} is indeed correlated
with $G(\mathrm{Res})$, and that this correlation could be exploited to identify
mergers. The sample quality is $F=0.84$, with a specificity
$p=0.80$. The completeness is $r=0.87$. It is however the case that
the \textit{best} border in Figure~\ref{fig:metodo_RFF-Gres} is approximately horizontal, indicating
that the Amoeba algorithm does not take advantage of the correlation between the \textit{RFF} and the
Gini index of the residuals.

\begin{figure}
\begin{tabular}{l}
\includegraphics[scale=0.35,angle=90]{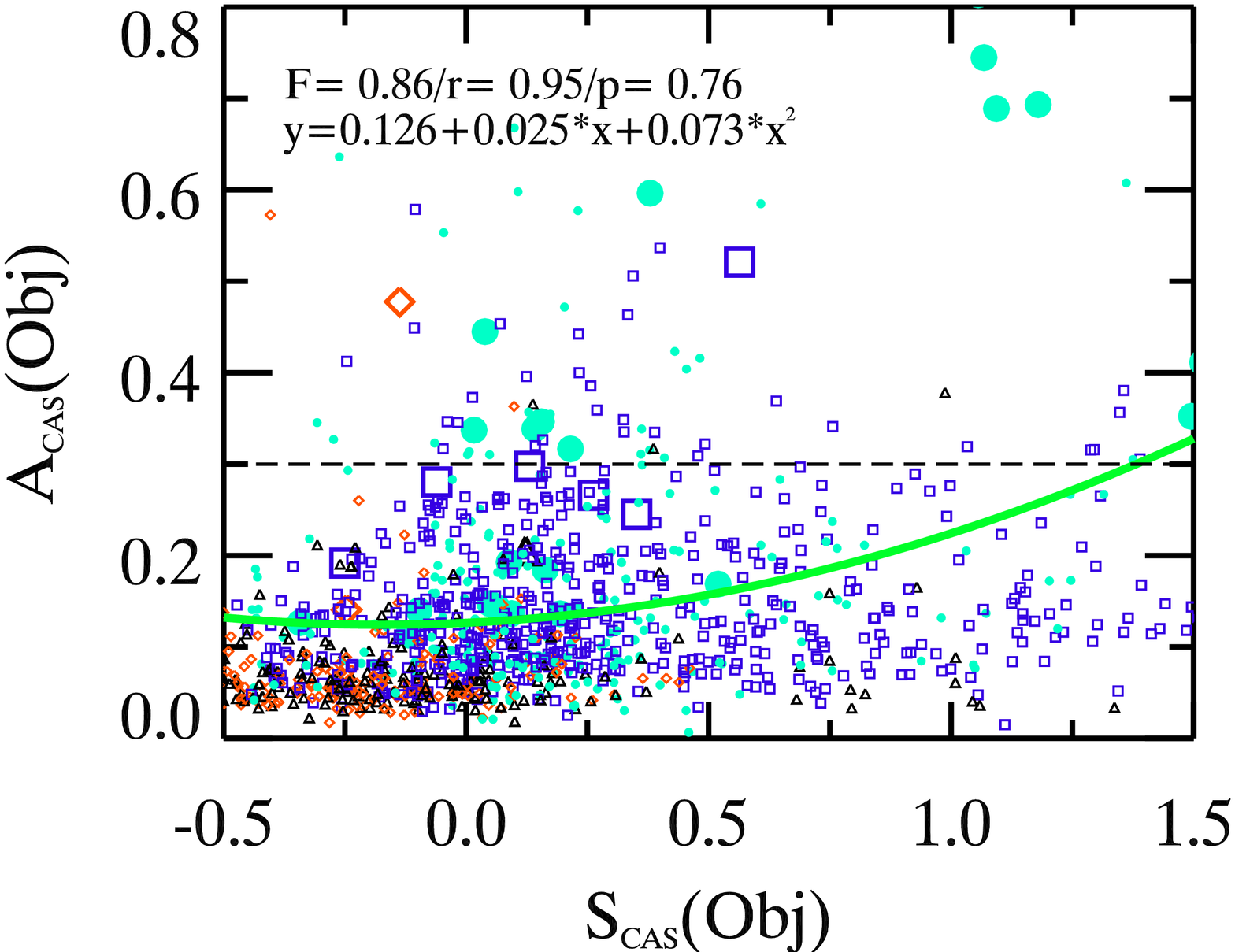}\\ \hline
\includegraphics[scale=0.35,angle=90]{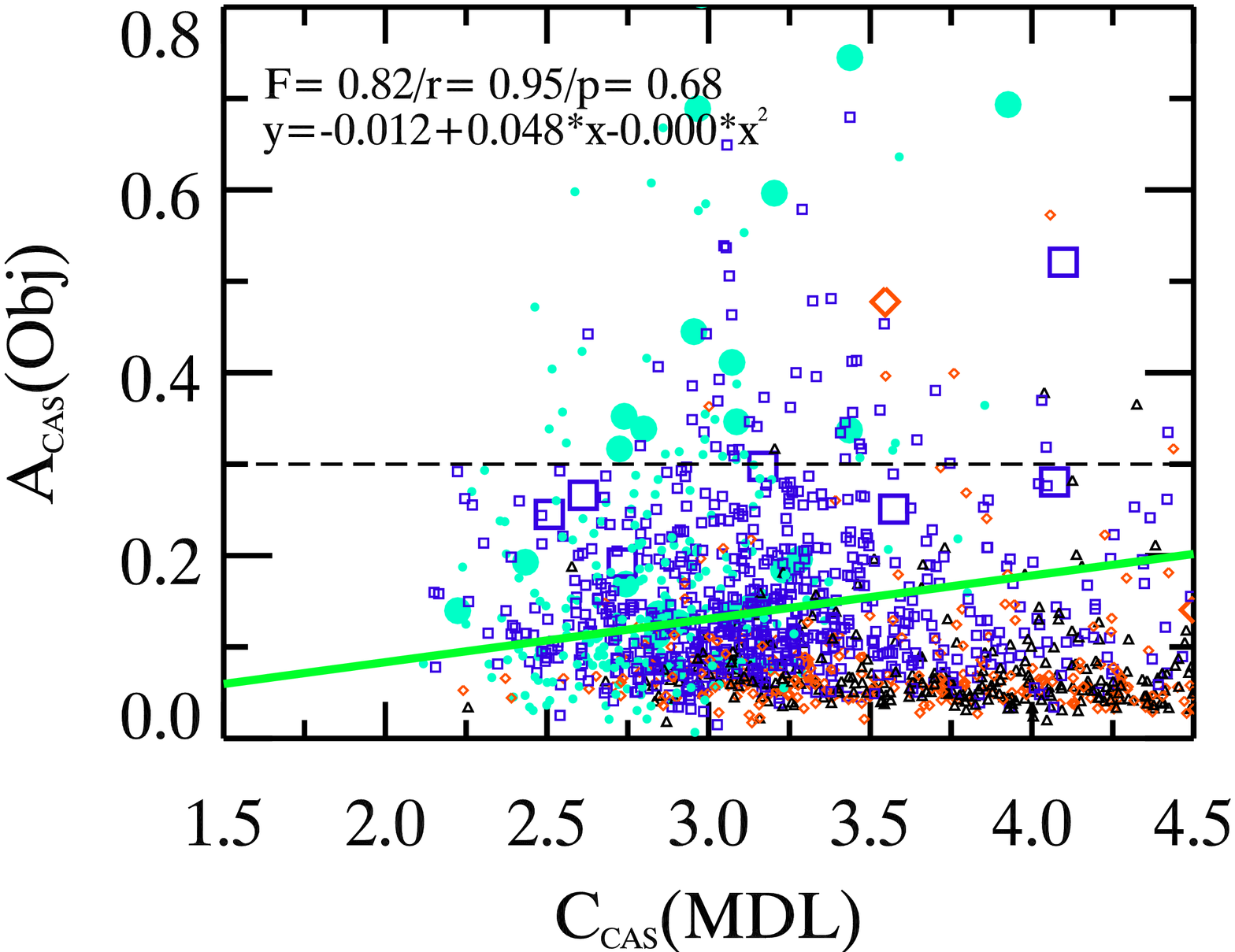}\\ \hline
\includegraphics[scale=0.35,angle=90]{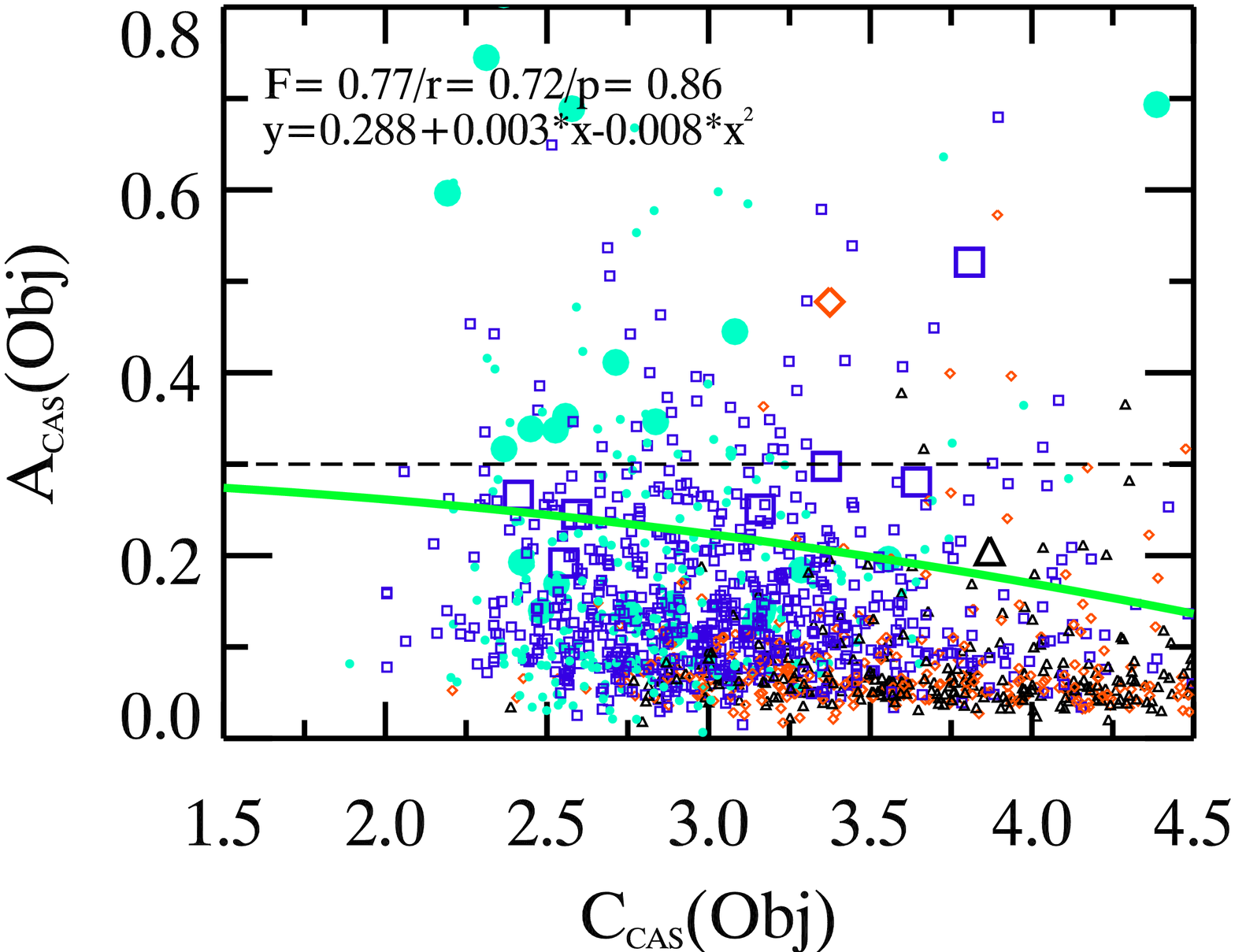}\\
\end{tabular}
\caption{Upper panel: The $A(\mathrm{Obj})-S(\mathrm{Obj})$ plane as a merger diagnostic.
Middle panel: The $A(\mathrm{Obj})-C(\mathrm{Mdl})$ indicator.
Lower panel: The $A(\mathrm{Obj})-C(\mathrm{Obj})$ merger test.
Symbols as in Fig. \ref{fig:metodo_GObjwMObj}.}
\label{fig:metodo_AObj-Cm}
\end{figure}

This work would not be complete without studying the CAS parameter space.
Figure~\ref{fig:metodo_AObj-Cm} explores the $A(\mathrm{Obj})-S(\mathrm{MDL})$,
$A(\mathrm{Obj})-C(\mathrm{MDL})$, and the $A(\mathrm{Obj})-C(\mathrm{Obj})$ planes.
In all the panels in Figure~\ref{fig:metodo_AObj-Cm}, the vertical axis is $A(\mathrm{Obj})$. However, the 
horizontal axis of the upper panel is the clumpiness $S(\mathrm{Obj})$, the horizontal axis of the middle 
panel is $C(\mathrm{MDL})$ and the horizontal axis of the lower panel is $C(\mathrm{Obj})$.
In all these cases, the initial guess line
is obviously motivated  by the classical $A(\mathrm{Obj})>0.35$, $A(\mathrm{Obj})>S(\mathrm{Obj})$.
It is seen that the classical CAS criterion works very well, achieving
a sample purity $F=0.86$, with a good recall $r=0.95$ and a fairly high
specificity $p=0.76$. It is also seen that the Amoeba algorithm has found that the
clumpier mergers need also be more asymmetric in order to be classified as such.
Comparison between the $A(\mathrm{Obj})-C(\mathrm{MDL})$ and $A(\mathrm{Obj})-C(\mathrm{Obj})$ 
highlights that the use of a smooth model to calculate $C(\mathrm{MDL})$ is slightly beneficial
for merger detection. This comes from the higher sample purity
obtained with the $A(\mathrm{Obj})-C(\mathrm{MDL})$ plane.
For this latter test, $F=0.82$, with a specificity $p=0.68$. The
completeness is $r=0.95$. This is clearly a success of 
the $A(\mathrm{Obj})$ parameter.
The contamination rate obtained by the use of the $A(\mathrm{Obj})-S(\mathrm{Obj})$, as
presented here, will be shown in \S \ref{subsec:contaminacion}.

It is interesting to complete the analysis of the CAS parameter space
by deriving the values of the $r$, $p$, and \textit{F}, parameters
for the $A(\mathrm{Obj}) \geq 0.35$, $A(\mathrm{Obj})>S(\mathrm{Obj})$ criterion, which
is a frequently used criterion for CAS based major merger studies. Using these 
limits, $r=0.21$, $p=0.98$, and $F_{\beta}=0.30$.
It is clear that with this criterion the merger sample obtained
discards the majority of the non-mergers, but is very incomplete.
The contamination of this latter sample will be presented in \S \ref{subsec:contaminacion}, but
it has to be borne in mind that the optimization used in this work aims towards completeness, while
the goal of the $A(\mathrm{Obj}) \geq 0.35$, $A(\mathrm{Obj})>S(\mathrm{Obj})$ limits
is a clean sample of major mergers. The two diagnostics cannot be compared directly, then.

\begin{figure}
\begin{tabular}{l} 
\includegraphics[scale=0.35,angle=90]{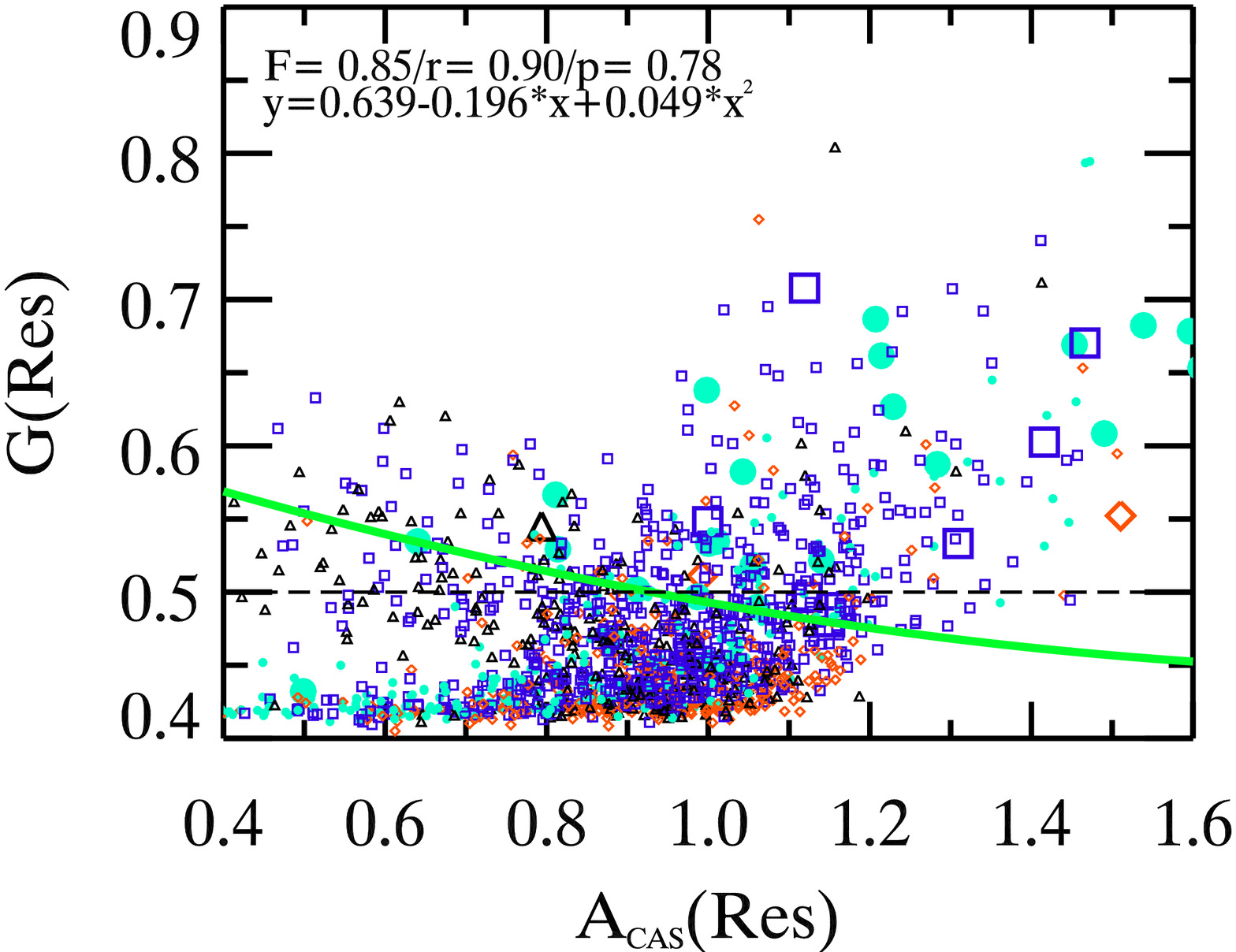}\\ \hline
\includegraphics[scale=0.35,angle=90]{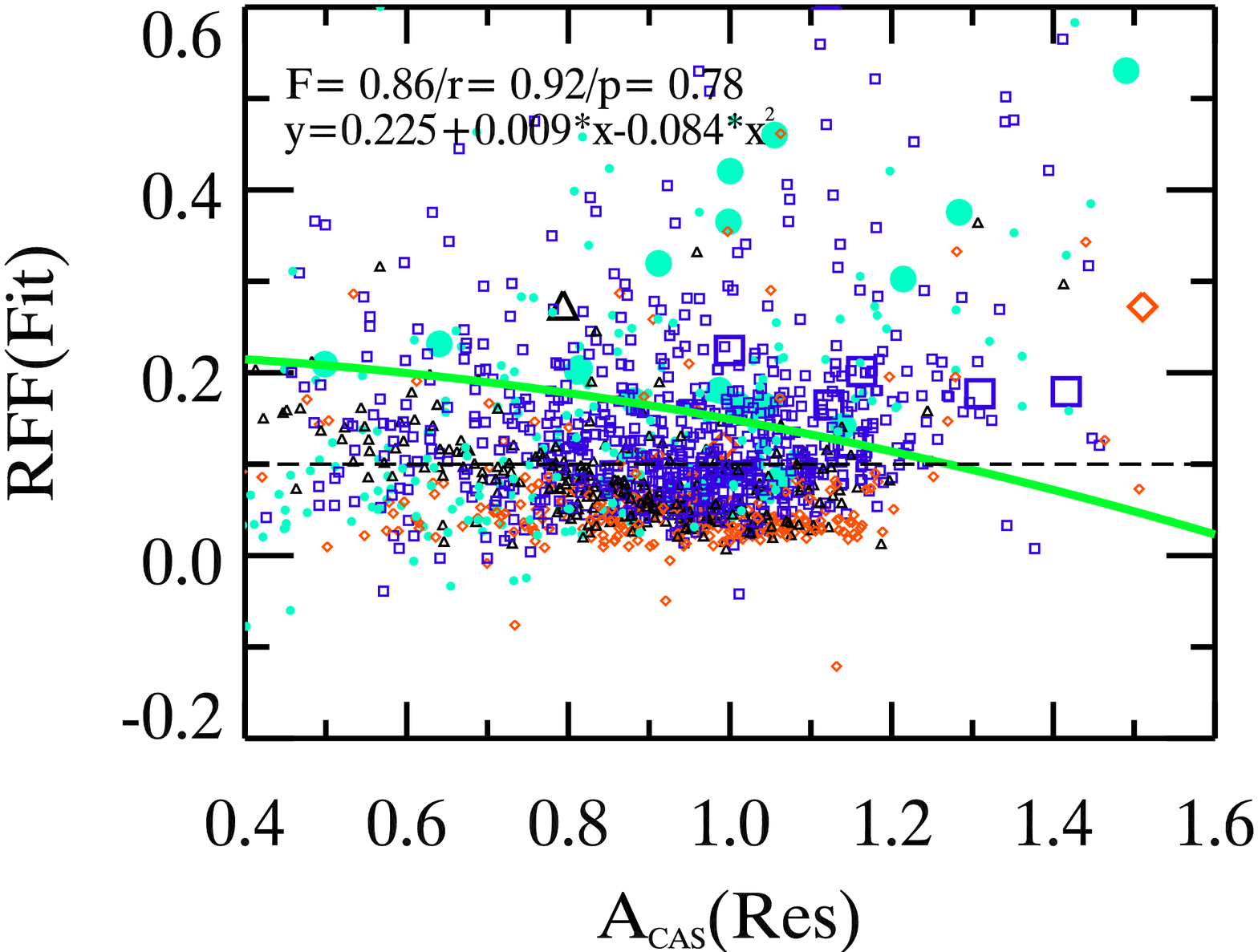}\\
\end{tabular}
\caption{Upper panel: The $A(\mathrm{Res})-G(\mathrm{Res})$ merger test.
Lower panel: The $A(\mathrm{Res})-RFF$ plane as a merger diagnostic.
Symbols as in Fig. \ref{fig:metodo_GObjwMObj}.}
\label{fig:metodo_ACreswFinal}
\end{figure}

The results presented so far suggest that the asymmetry is a very good merger indicator.
It is also the case that the $G(\mathrm{Res})$ and the \textit{RFF} can also
be used as merger indicators by themselves. In the spirit
of this work, the following planes studied will include the asymmetry 
\emph{of the residuals} $A(\mathrm{Res})$.
This morphological parameter is explored because the asymmetry of a system is expected
to be boosted after the subtraction of an intrinsically symmetric profile such as the
S\'ersic profile used here. Figure~\ref{fig:metodo_ACreswFinal} shows
the success of this approach, in particular on its lower panel.

Both diagnostics shown in Figure~\ref{fig:metodo_ACreswFinal} are found to work very well.
In particular, the $RFF-A(\mathrm{Res})$ produces the best sample purity
of all the diagnostics tested here that exploit the structural parameters of the residuals, with 
a very high completeness. As expected, the 
asymmetry of the residuals is boosted with respect to the asymmetry of the original images.
The \textit{best} border for the $RFF-A(\mathrm{Res})$ diagnostic is no longer 
horizontal, implying that $A(\mathrm{Res})$
has enough predicting power so as to curve the \textit{best} line.
The \textit{best} border line of the $RFF-A(\mathrm{Res})$ test appears to be more curved
than that of the $G(\mathrm{Res})-A(\mathrm{Res})$ merger test. This is consistent
with the previous conclusion that $G(\mathrm{Res})$ could also be used as a standalone
merger diagnostic.
The purity of the sample obtained using the $G(\mathrm{Res})-A(\mathrm{Res})$ indicator
has $F=0.84$, with a recall $r=0.90$ and a specificity $p=0.78$.
The purity of the sample obtained using the $RFF-A(\mathrm{Res})$ indicator
has $F=0.86$, with a $r=0.92$ and $p=0.78$. This is then the best test, and this
confirms that the structural parameters of the residuals can indeed be competitive if used
as merger diagnostics.
The corresponding contamination stemming from the $RFF-A(\mathrm{Res})$ diagnostic will be shown in \S \ref{subsec:contaminacion}, where
a sample of visually classified mergers will be used to establish the merger prevalence.
Whether the use of these parameters can be used to probe deeper in 
the luminosity function in order to detect mergers with mass ratios \emph{larger} than \texttt{10:1}
or if these parameters allow to trace the merger event up to later stages in which the
less luminous galaxy is almost engulfed by the host will be presented in a forthcoming paper.

The quality of the merger samples obtained using the $RFF-A(\mathrm{Res})$ test
is thus the best one of all the residual-based merger diagnostics that have been explored in this work.
The discussion presented in \S \ref{sec:further} will be based on the merger sample obtained using this merger test.
Also, \S \ref{subsec:minors} shows some objects that are thought to be involved in minor merger events
that have been detected in the $RFF-A(\mathrm{Res})$ plane as mergers that would have been missed
by the usual CAS criterion $A(\mathrm{Obj}) \geq 0.35$, $A(\mathrm{Obj})>S(\mathrm{Obj})$.

Table~\ref{tab:resumen_tests} gathers a summary of the different merger diagnostics
used and their respective statistical performances, for convenience.

\begin{table}
\begin{tabular}{|ll|ccc|} \hline
Diagnostic. & Figure. & $F-\mathrm{score}$ & $r$ & $p$ \\ \hline
$G(\mathrm{Obj})-M_{20}(\mathrm{Obj})$ & \ref{fig:metodo_GObjwMObj} & 0.77 & 0.79 & 0.82 \\
$A(\mathrm{Obj})-RFF$ & \ref{fig:metodo_AObjwRFF} & 0.85 & 0.92 & 0.76 \\
$G(\mathrm{Obj})-A(\mathrm{Obj})$ & \ref{fig:metodo_GobjAobj} & 0.86 & 0.90 & 0.82 \\ 
$G(\mathrm{Res})-M_{20}(\mathrm{Res})$ & \ref{fig:metodo_GRes-M20Res} & 0.79 & 0.77 & 0.82 \\
$G(\mathrm{Res})-A(\mathrm{Obj})$ & \ref{fig:metodo_GRes-AObj} & 0.85 & 0.90 & 0.78 \\
$\mathrm{RFF}-G(\mathrm{Res})$ & \ref{fig:metodo_RFF-Gres} & 0.84 & 0.87 & 0.80 \\
$A(\mathrm{Obj})-S(\mathrm{Obj})$ & \ref{fig:metodo_AObj-Cm} & 0.86 & 0.95 & 0.76 \\
$A(\mathrm{Obj})-C(\mathrm{Mdl})$ & \ref{fig:metodo_AObj-Cm} & 0.82 & 0.95 & 0.68 \\
$A(\mathrm{Obj})-C(\mathrm{Obj})$ & \ref{fig:metodo_AObj-Cm} & 0.77 & 0.72 & 0.86 \\
$A(\mathrm{Res})-G(\mathrm{Res})$ & \ref{fig:metodo_ACreswFinal} & 0.85 & 0.90 & 0.78 \\
$A(\mathrm{Res})-RFF$ & \ref{fig:metodo_ACreswFinal} & 0.86 & 0.92 & 0.78 \\ \hline
\end{tabular}
\caption{Summary of the different merger diagnostics tried and their
statistical performance.}
\label{tab:resumen_tests}
\end{table}

\subsection{The Effect of the $\beta$ Parameter.}
\label{subsec:otrobeta}

\begin{figure}
\begin{tabular}{l}
\includegraphics[scale=0.35,angle=90]{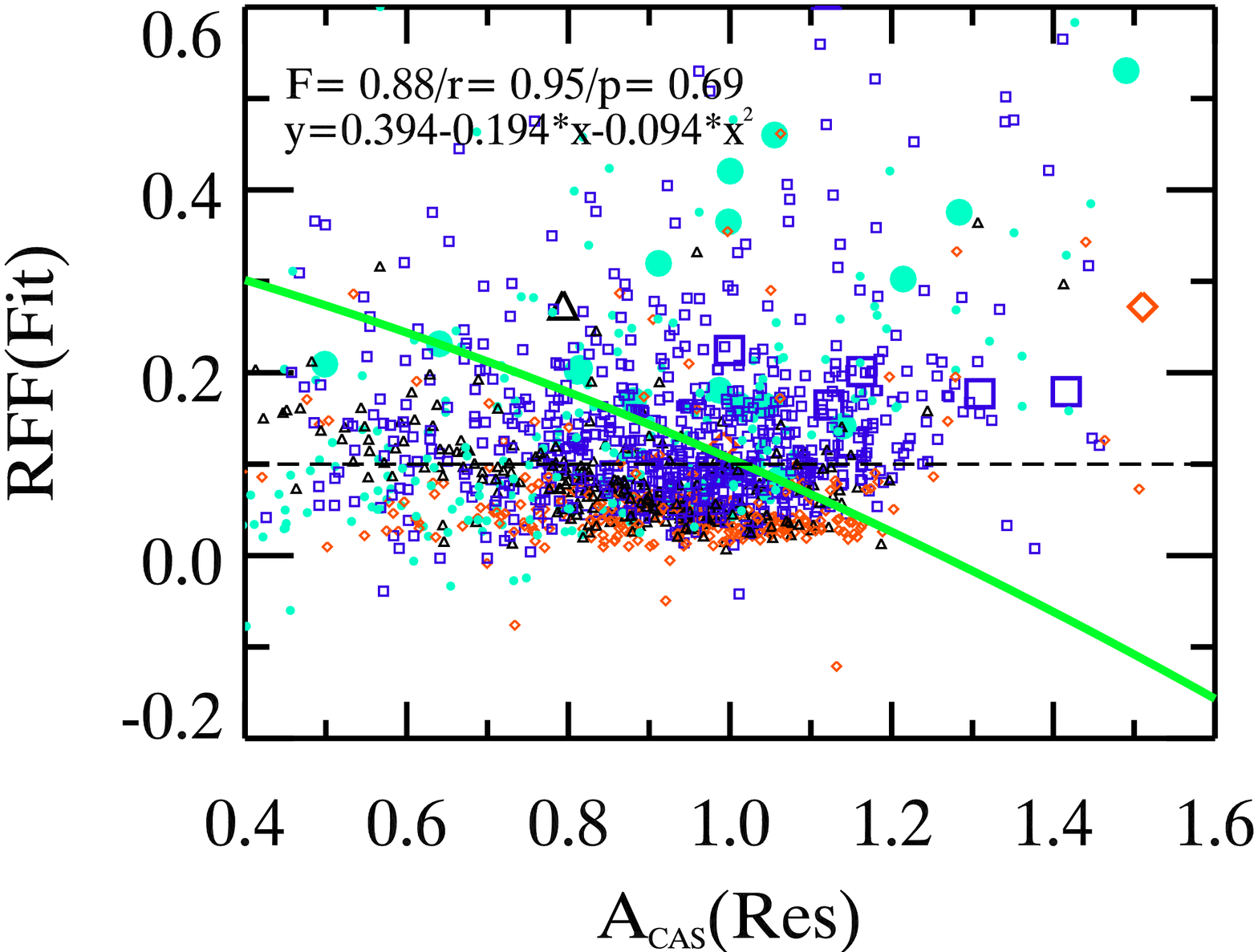} \\ \hline
\includegraphics[scale=0.35,angle=90]{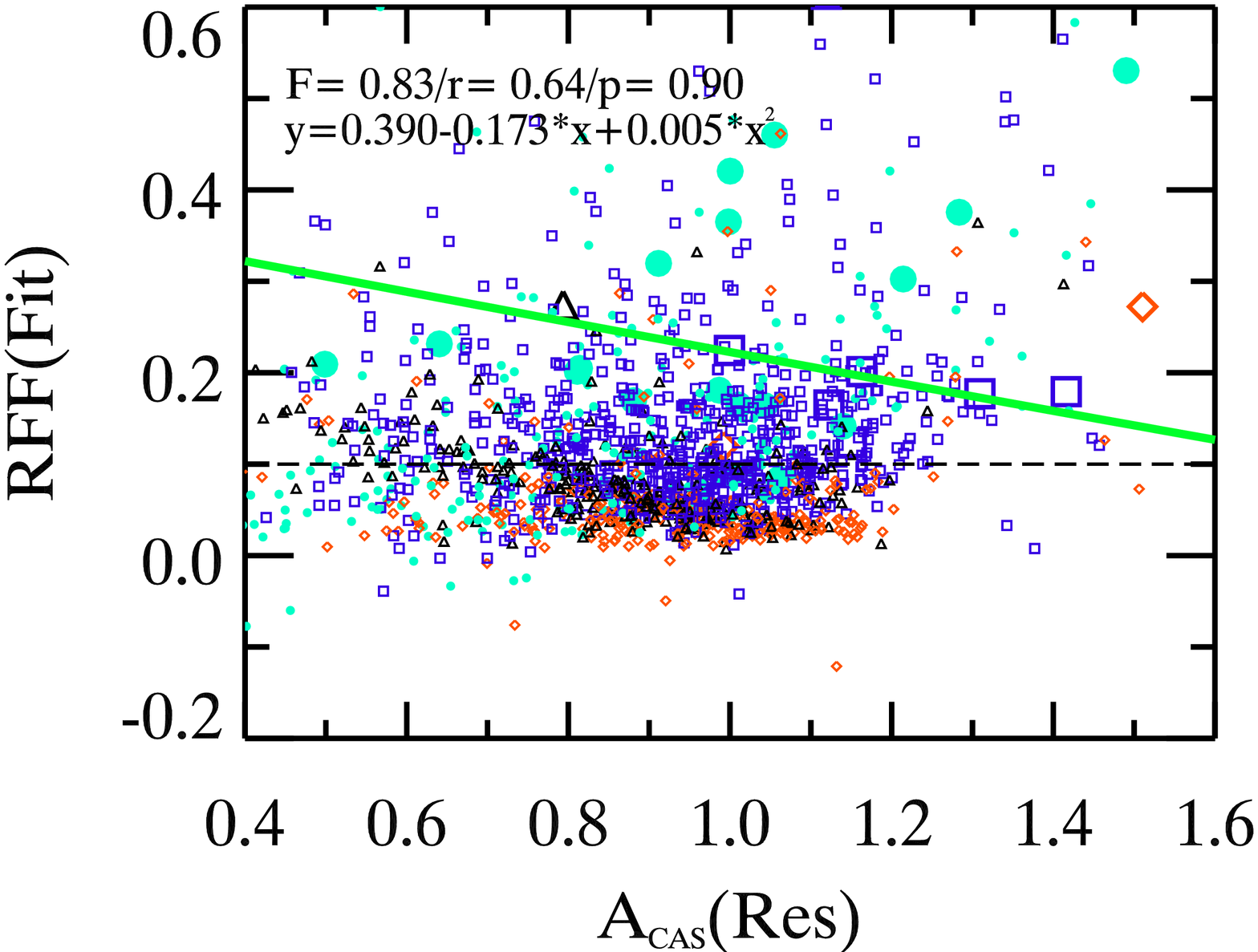} \\
\end{tabular}
\caption{Recalculation of the $A(\mathrm{Res})-RFF$ merger diagnostic using
$\beta=2.0$ (upper panel) and $\beta=0.5$ (lower panel).
Symbols as in Fig. \ref{fig:metodo_GObjwMObj}.}
\label{fig:otrosbetas}
\end{figure}

The $F-\mathrm{score}$ maximization technique used in the current work is simply
a way to select a number of objects from a parent population whose structural properties are similar 
to those of the training set. The method is thus only as good as the training set.
The $\beta$ parameter in the definition of $F_{\beta}$ is the ingredient
used by this statistic to decide whether or not the structural properties of an
object are close enough to the structural properties of the objects in the training set to be
considered as a merger candidate. The $\beta$ parameter is therefore 
determined by the scientific needs of the sampling process. Figure~\ref{fig:otrosbetas}
presents a recalculation of the ``best'' border of the $A(\mathrm{Res})-RFF$
diagnostic using two different values of $\beta$, $\beta=2.0$ (upper panel) and $\beta=0.5$ (lower panel).

Comparison of Figure~\ref{fig:otrosbetas} with Figure~\ref{fig:metodo_ACreswFinal}
clearly indicates that the $\beta$ parameter has a decisive impact on the sample of
potential mergers obtained by the Amoeba algorithm. If completeness is considered to be
much more important than specificity, as is done in the upper panel of Figure~\ref{fig:otrosbetas}, the
Amoeba algorithm increases the recall ratio by 3\% with respect to the value achieved
in Figure~\ref{fig:metodo_ACreswFinal}. This amounts to the inclusion of just an additional
object from the training sample, at the price of littering the sample with
150 non-mergers. This decreases the precision by 10\%.
On the other hand, the lower panel of Figure~\ref{fig:otrosbetas} shows what happens
if the specificity is weighted more than the completeness.
It is first seen that the new merger diagnostic excludes a higher number of non-mergers, raising the 
precision by 12\%. This comes at the expense of missing 11 objects from the training set.
The choice of $\beta=1.25$, adopted in \S \ref{subsec:Fscore} is thus a good compromise between
these two options.

It is highlighted here that no single value of $\beta$ can be considered to be ``correct''. The
value of this parameter is set by the goals of the test. If, for instance, the
objects selected by the method are to be the targets of a spectroscopic follow-up programme where
targeting only true mergers is deemed essential, it would
be advisable to give a higher weight to the specificity.

\section{Visual Assessment of the Merger Samples Obtained.}
\label{sec:further}

As it is clear from the previous discussion, the method presented in the current work
is optimized to be very complete, hopefully detecting minor mergers thanks to
the use of the structural parameters of the residual images. It is therefore
needed to establish both the contamination by non-mergers and the success in recovering minor mergers.
The study of the contamination by non-mergers is gathered in \S \ref{subsec:contaminacion}.
On the other hand, \S \ref{subsec:minors} presents a number of examples of minor merger candidates that have been detected through the
use of the structural parameters of the residuals that would otherwise have been missed by the traditional
CAS diagnostics.

\subsection{Contamination by Non-Mergers.}
\label{subsec:contaminacion}

The different galaxy samples obtained by the \emph{sheer} use of the various
indicators that have been put to the test in \S \ref{subsec:compa_diagnostics} need to be
further evaluated. It is clear that not all the objects that test positive (i.e., sources
that fell in the merger side of the merger tests used) to these criteria can be mergers.
In particular, the contamination is the key statistic that needs to be included in this appraisal.
This section presents the non-merger contamination of the galaxy sets that are obtained
from the use of the merger diagnostics that attained the highest $F-\mathrm{score}$ values in \S \ref{subsec:compa_diagnostics}.

The contamination is here defined as:

\begin{equation}
C=\frac{\mathrm{\#Non-mergers\ that\ however\ test\ positive.}}{\mathrm{\#All\ positives.}}
\end{equation}

\noindent
where the objects that test positive are those in the merger side
of the ``best'' borders calculated above. The denominator of this fraction includes both
mergers and non-mergers, which implies that the contamination depends on the
merger fraction in a non-linear way.
\indent

The next step is thus oriented towards obtaining a complete and accurate estimate
of the fraction of mergers found in the parent population studied. The total number of good 
mergers is here calculated by adding together the objects included in the training set of 
galaxies used above, and a number of merger galaxies that were recovered during an additional
visual classification which will be described below. This further observational classification
is justified because the \textsc{STAGES} morphological catalogue used to define
the training set of objects for the $F-\mathrm{score}$ maximization technique is a general
morphological catalogue that is not designed to split the parent population
into mergers and non-mergers. In particular, the \textsc{STAGES} observers were not
specifically looking for the \emph{minor} mergers whose detection is the goal of this study. This additional
study thus serves to check whether minor mergers do share the structural properties of major mergers, which
is the main assumption behind the use of the $F-\mathrm{score}$ number as a diagnostic discriminator.

\begin{figure}
\vspace{0.5cm}
\includegraphics[scale=0.32,angle=0]{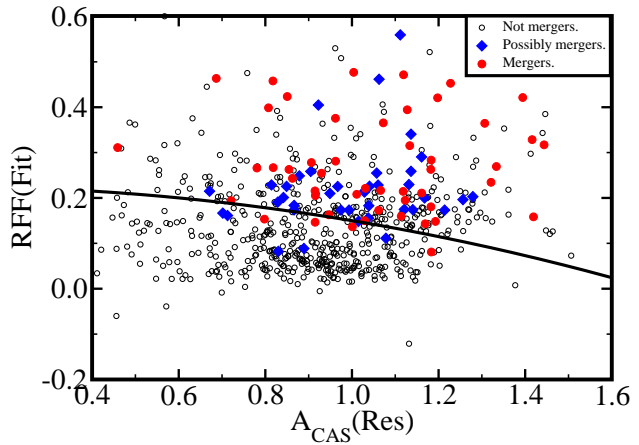}
\caption{Visual assessment of the galaxy sample obtained using the $RFF-A(\mathrm{Res})$ 
merger test. Red, solid dots represent the location of the new best merger cases (55 points), blue diamonds
give the location of the dubious cases (41 points) and the open, black points show the location of the non
mergers (568 points). The black line shown is the same line shown in Figure~\ref{fig:metodo_ACreswFinal}.}
\label{fig:establecer}
\end{figure}

The new visual assessment examines the set of galaxies obtained by 
the blind application of the $A(\mathrm{Res})-RFF$ diagnostic as shown in 
Figure~\ref{fig:metodo_ACreswFinal}. This merger test was shown in \S \ref{subsec:compa_diagnostics}
to yield the highest $F-\mathrm{score}$ and \texttt{specificity} numbers of all the diagnostics that make 
use of the structural parameters of the \emph{residual} images and is therefore
more likely to produce a clean list of mergers. This set of galaxies is 
made of $36=0.92\times 39$ objects from the original training set and $332=(1-0.78)\times 1498$ 
objects that were not included in the training set. These latter
systems are the False Positives involved in the calculation of the $p$ statistic. The majority of 
these objects (282) have zero merger marks in the \textsc{STAGES} morphological catalogue and 50 of them
only have one merger mark.
These simple statistics motivated us to examine more closely the 332 objects not included in the original training 
set that however fell in the merger side of the ``best'' border in the $A(\mathrm{Res})-RFF$ plane.
The purpose of this further investigation is to establish whether or not those
objects could be mergers that escaped the original assessment of the \textsc{STAGES}
team observers. This will allow an accurate and non-parametric determination
of the merger prevalence in the parent population of galaxies studied.
To this end, four of the authors of this paper (AAS, CH, EFB, MEG) reinspected
the 332 False Positives together with 332 randomly selected True Negatives
as a control sample. These four independent assessments
were then combined into a single trinary outcome, splitting these
664 sources into three different sets, (i) clear mergers, (ii) clear non-mergers, and (iii) the dubious cases.
These 664 objects were then placed in the $A(\mathrm{Res})-RFF$ plane. This is presented in Figure~\ref{fig:establecer}.

%%%%%%%%%%%%%%%%%%%%%%%%%%%%%%%%%%%%%%%%%%%%%%%%%%%%%%%%%%%%%%%%%%%%%%%%%%%
%%%%%%%%%%%%%%%%%%%%%%%%%%%%%%%%%%%%%%%%%%%%%%%%%%%%%%%%%%%%%%%%%%%%%%%%%%%%%%

Figure~\ref{fig:establecer} immediately shows a correlation between the location of the mergers that
have been recovered by the new visual merger re-assessment and the  $RFF-A(\mathrm{Res})$  ``best'' border.
It is seen that the majority of the recovered mergers and dubious mergers lies above the diagnostic line.
It is also highlighted that \emph{none} of the objects shown in Figure~\ref{fig:establecer} belongs to the original training set.

\begin{figure}
\includegraphics[scale=0.80,angle=0]{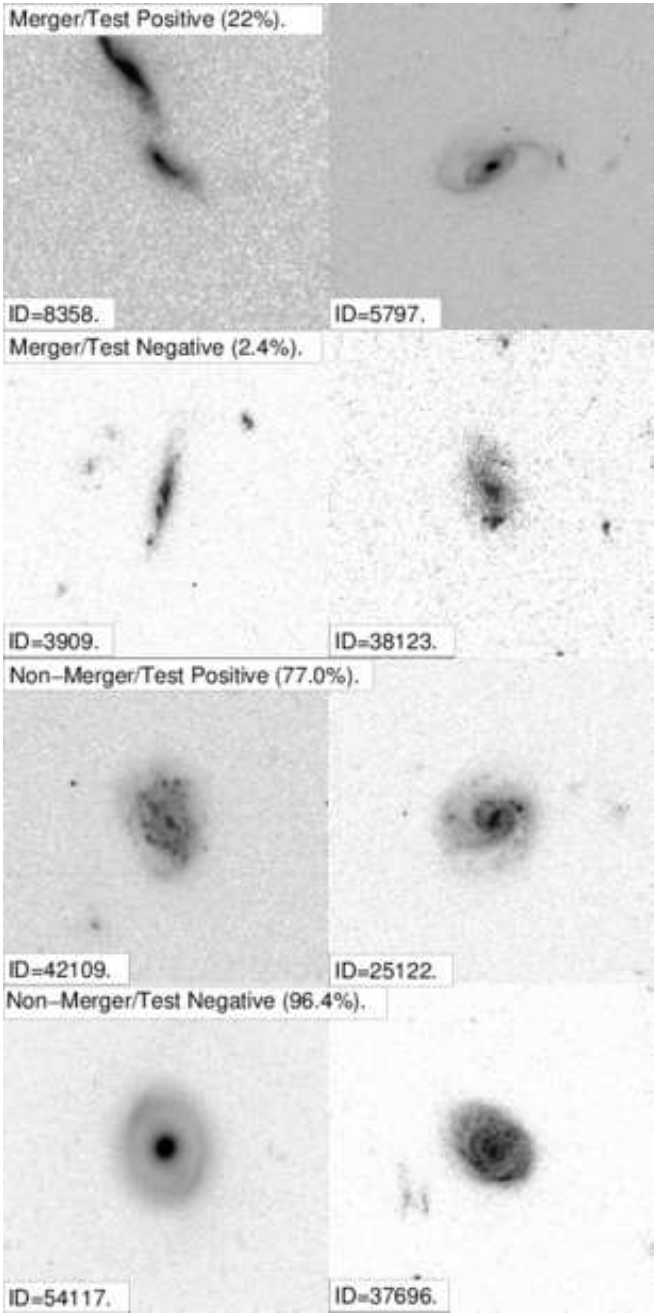}
\caption{Examples of galaxies examined in the new visual assessment. COMBO-17 IDs are shown for each galaxy.
The percentages given reflect the frequencies of the four different possibilities shown.}
\label{fig:muestruario_reala}
\end{figure}

Figure~\ref{fig:muestruario_reala} presents eight of the galaxies that were reinspected
during this further classification. This figure presents four rows of two objects each, marked
with their respective COMBO-17 IDs. The upper row shows the objects that tested as positives
to the $RFF-A(\mathrm{Res})$ diagnostic and were subsequently classified as mergers
in the new visual assessment. The second row presents mergers that were regarded as such 
that however test negative to the diagnostic. These latter objects are very rare.
The third row is made of non-mergers that nevertheless tested positive in the automated test, and the bottom
row presents the non mergers that fell below the ``best'' border in Figure~\ref{fig:metodo_ACreswFinal}.
The percentages given in each row are the frequencies with which each of the different
possibilities appears. These percentages do not add up to 100\% because the
galaxies that were classified as dubious mergers are not shown.

The new visual inspection has recovered a total of 55 clear merger systems together with 41 dubious cases.
The remaining 568 objects were classified as non mergers. It is interesting to note
that 36\% of the 55 clear merger cases had just one merger mark in the original \textsc{STAGES} 
morphological catalogue, 18\% of the new possible mergers were considered
as mergers by only one of the \textsc{STAGES} classifiers, and only 6\% of the non-mergers
had received a single merger mark. This adds up to a total of
60 objects with only one merger mark in the whole pool of the 664 objects 
inspected. A total of 50 sources out of these 60 objects are found above the ``best'' border 
defined in Figure~\ref{fig:metodo_ACreswFinal} for the $RFF-A(\mathrm{Res})$ merger test.
In addition, the $RFF-A(\mathrm{Res})$ merger test as presented in 
Figure~\ref{fig:metodo_ACreswFinal} detects 93\% of the recovered clear merger cases, 88\% of the new 
dubious mergers and 43\% of the non-mergers.

Thus, the total number of objects selected by the $RFF-A(\mathrm{Res})$  test can be 
broken down in the following way:

\begin{enumerate}

\item{The $36=0.92 \times 39$ objects included from the training set.}
\item{The $51=0.93 \times 55$ new mergers recovered by the new visual inspection.}
\item{An unknown number of merger systems between 0 and $36=0.88 \times 41$ that were classified as 
dubious mergers by the visual re-assessment. Under the assumption of a flat probability
distribution this number could be represented as $18 \pm 10$.}
\item{A total of approximately $260=332-57-19$ non mergers.}

\end{enumerate}

Therefore, the total number of bona-fide mergers above the $RFF-A(\mathrm{Res})$ ``best'' border line is 105, and the
total number of objects of all classes above this line is $368=332+36$.
The final contamination ratio is $71\%=(368-105)/368$, which is fairly high since
it means that 70\% of all the objects set aside by the blind use of the diagnostic
are non-merger contaminants. This number is close, but conceptually different
from the fraction of non-mergers that are found to test positive in during the additional 
visual classification, which is $74\%$.

In the same way, the non-merger contamination associated with the
$A(\mathrm{Obj})-S(\mathrm{Obj})$ test as shown in Figure~\ref{fig:metodo_AObj-Cm} is
$71 \pm 3 \%=(397-(37+(360/286) \times (49 + (0.5 \pm 0.3  )\times 28)))/397$. This calculation takes into account that there
are $397=39\times 0.95+1498\times (1-0.76)$ objects above the ``best'' border defined in Figure~\ref{fig:metodo_AObj-Cm}
and that not all the dubious mergers will indeed be mergers. This latter consideration is again made assuming
a flat probability distribution.
This contamination that affects the sample selected by this method is seen to be fully compatible with the
contamination that is calculated for the $RFF-A(\mathrm{Res})$ test. If a similar analysis is carried out for the
$A(\mathrm{Obj})-S(\mathrm{Obj})$ criterion using the traditional 
CAS limits $A(\mathrm{Obj})\geq 0.35$, $A(\mathrm{Obj}) \geq S(\mathrm{Obj})$, the contamination is
$50\%=19/(19+17)$. Thus, the contamination by non-mergers found in this set of galaxies is
lower than the one found for the other two merger diagnostics explored, but its completeness is obviously
much lower.

The next step in this analysis is then the study of the negative detections, which
focuses on the 1169 sources below the ``best'' border line presented in Figure~\ref{fig:metodo_ACreswFinal}.
These 1169 galaxies include 3 galaxies from the original training set (these are the ``False Negatives''
in the $F-\mathrm{score}$ analysis). These 1169 galaxies will also include an indeterminate number
of mergers that can be estimated from the visual re-assessment results by multiplying
the fraction of mergers found in the 332 objects from the control sample by the total number
of objects below the ``best'' line (1169). The new visual classification discovered
4 clear merger and 5 dubious cases in the 332 objects in the control set. If these latter
sources are given a weight of $0.5$, the fraction of mergers below the ``best'' line
of the $RFF-A(\mathrm{Res})$ test is $2.2 \pm 0.4\%=(3+(1169/332)\times(4+(0.5 \pm 0.3 )\times 5))/1169$, where
the error interval again assumes a flat probability distribution for the dubious mergers.
If the galaxies from the original training set and the newly identified dubious mergers are not 
included in this calculation, the fraction drops to $1\%=((1169/332)\times 4)/1169$, which is the number given 
in Figure~\ref{fig:muestruario_reala} for the negative contamination ratio of the visual re-assessment.
The negative contamination ratio is then very low, indicating that this
technique is \emph{very} powerful as a negative merger test.
Furthermore, if this negative contamination ratio is derived for the $A(\mathrm{Obj})-S(\mathrm{Obj})$ test as
presented in Figure~\ref{fig:metodo_AObj-Cm}, the percentage 
is $4.0 \pm 1\%=(2+(1140/378)\times(6+(0.5 \pm 0.3 )\times 13))/1140$, which is
compatible but slightly worse than the result for the $RFF-A(\mathrm{Res})$ 
test. This latter calculation uses that there are 378
objects in the total pool of objects inspected during the second visual classification 
that fell below the ``best'' border shown in the upper panel of Figure~\ref{fig:metodo_AObj-Cm}.
Finally, the corresponding negative contamination for the traditional 
CAS criterion is $5\%=77/(1424+77)$. This is compatible with the negative 
contamination for this diagnostic as presented in Figure~\ref{fig:metodo_AObj-Cm}.

In summary, despite the fact that the contamination ratio is fairly high for the positive detections, it 
is very low for the negative detections. The above considerations therefore lead us to conclude that 
the $RFF-A(\mathrm{Res})$ \emph{minor} merger diagnostic presented in 
Figure~\ref{fig:metodo_ACreswFinal} works best as a \emph{negative} test. In particular, it could be possible
to use this automated technique with large area surveys such as 
the APM \citep{1990MNRAS.242P..43M}, 2dFGRS \citep{1999MNRAS.308..459F}, SDSS \citep{2009ApJS..182..543A},
UKIDSS \citep{2006MNRAS.372.1227D}, KIDS\footnote{\texttt{http://www.astro-wise.org/projects/KIDS/}},
VIKING\footnote{http://www.astro-wise.org/Public/viking10.pdf}, GAMA \citep{2009A&G....50e..12D},
GEMS \citep{2004ApJS..152..163R}, COSMOS \citep{2005NewAR..49..461K},%
DES\footnote{\texttt{http://www.darkenergysurvey.org}}, or the 
LSST \citep{2008arXiv0805.2366I}
by calibrating the ``best'' borders to the new observational conditions and wavebands 
using a reduced and manageable number of objects within a predefined redshift window up to a certain
magnitude. These new ``best'' borders should then be used to split the target galaxy population.
The technique illustrated here would then produce two
different sets of galaxies. One set would be almost completely free of mergers and the
other set would include the overwhelming majority of mergers. This latter sample
would need further purification in order to produce a clean sample of mergers.
This recalibration step is particularly needed in the case of surveys including
U band observations. In this case, even very minor wet mergers will leave a larger
impact on the general appearance of their host galaxies because the youngest stars
will be much clearly seen. This step could also help understand the effect of the
photometric band chosen on the optimization process used here.
It is clear that the additional pruning can not be done using the structural
properties of the galaxies only, since these have been fully exploited here.
In addition, one clear conclusion from this visual assessment which
will be strengthened in \S \ref{subsec:minors} is that even a visual
inspection cannot unambiguously tell whether a particular object is involved
in a merger episode or is just the product of a by-chance alignment or an \textsc{HII} region.
On the contrary, additional information such as colours or kinematical information obtained using 
Integral Field Units spectrographs needs to be included. Note that this problem affects all automated
methods based on structural parameters that we have explored in this paper and are commonly used
in the literature. This issue is merely a consequence of the empirical existence
of visually classified mergers with undisturbed morphologies, which was also
observed in \cite{2009ApJ...705.1433H}.
The above considerations make it advisable to seek an inclusive criterion, with a high recall ratio, rather
than a specific one, which is reflected in the choice of the $\beta$ parameter.

\subsection{The Search for Minor Mergers.}
\label{subsec:minors}

As it has been mentioned, the main driver behind the very high completeness that
the method presented here is tuned to achieve is the detection of \emph{minor} mergers.
It is therefore interesting to test the ability of the methodology introduced here
to detect minor mergers. This is done by selecting a set of visually identified mergers
that would not have been detected by the usual cuts that are applied to the 
$A(\mathrm{Obj})-S(\mathrm{Obj})$ plane that were however recovered by the
$RFF-A(\mathrm{Res})$ diagnostic. This set of galaxies
is made of 69 galaxies. As a comparison, there are 17 visually classified mergers
that would have been detected by the usual CAS diagnostic.

Figure~\ref{fig:minormergercandidates1} presents a number of examples
taken from the 69 visually classified galaxies that would have not been retrieved
by the $A(\mathrm{Obj})-S(\mathrm{Obj})$ diagnostic using its usually adopted limits. For each of 
the selected galaxies, this figure
presents the COMBO-17 ID, its environment, its morphological type, and its B-band
absolute magnitude. Each inset also presents an estimate of the contribution
from the less luminous component to the total \texttt{FLUX\_AUTO} \textsc{SEXTRACTOR}
measurement, expressed as a fraction. This is calculated simply by dividing
the flux enclosed in the aperture shown by the automatic flux measurement
that is performed by \textsc{SEXTRACTOR}.

%\begin{figure*}
%\includegraphics[scale=0.30,angle=0]{minormergers.eps}
%\begin{tabular}{lll} \hline
%\includegraphics[scale=0.32]{MinorMuestra/minormuestra_8_20998.eps}&\includegraphics[scale=0.32]{MinorMuestra/minormuestra_51_36819.eps}&\includegraphics[scale=0.32]{MinorMuestra/minormuestra_25_28623.eps} \\
%\hline
%\end{tabular}
%\caption{Visually classified minor mergers found by the $RFF-A(\mathrm{Res})$ diagnostic but missed
%by the CAS method adopting its usual limits. COMBO-17 IDs, environments, morphological types and B-band absolute
%magnitudes from the \textsc{STAGES} public  catalogue are shown for each galaxy. The panels also give an indication
%of the angular extent of the image insets and the approximate fractional contribution of the
%light included in the circle to the total \textsc{SEXTRACTOR} automatic flux measurement.}
%\label{fig:minormergercandidates1}
%\end{figure*}

%\input{atlas_menor.tex}
\begin{figure}
\caption{Visually classified minor merger candidates found by the $RFF-A(\mathrm{Res})$ diagnostic but missed
by the CAS method adopting its usual limits. COMBO-17 IDs, environments, morphological types and B-band absolute
magnitudes from the \textsc{STAGES} public  catalogue are shown for each galaxy. The panels also give an indication
of the angular extent of the image insets and the approximate fractional contribution of the
light included in the circle to the total \textsc{SEXTRACTOR} automatic flux measurement. Objects 20213 and 7479
might be localized star formation episodes in irregular galaxies. Also, it is not straightforward to identify
the putative satellite in object 40654, which presents an alternative satellite marked in red.}
\label{fig:minormergercandidates1}
\begin{tabular}{cc} \hline
\includegraphics[scale=0.48]{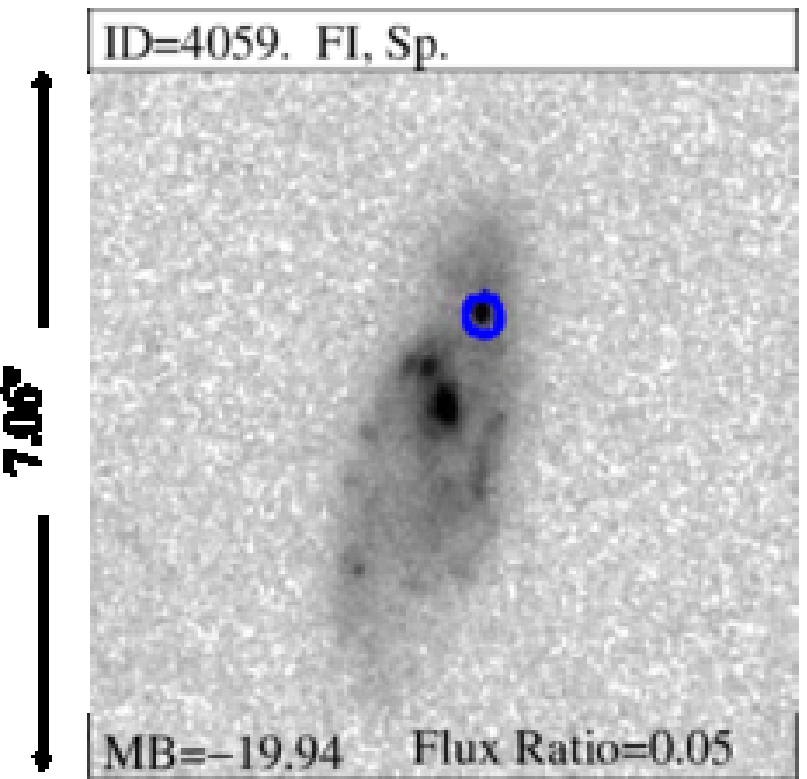}&\includegraphics[scale=0.48]{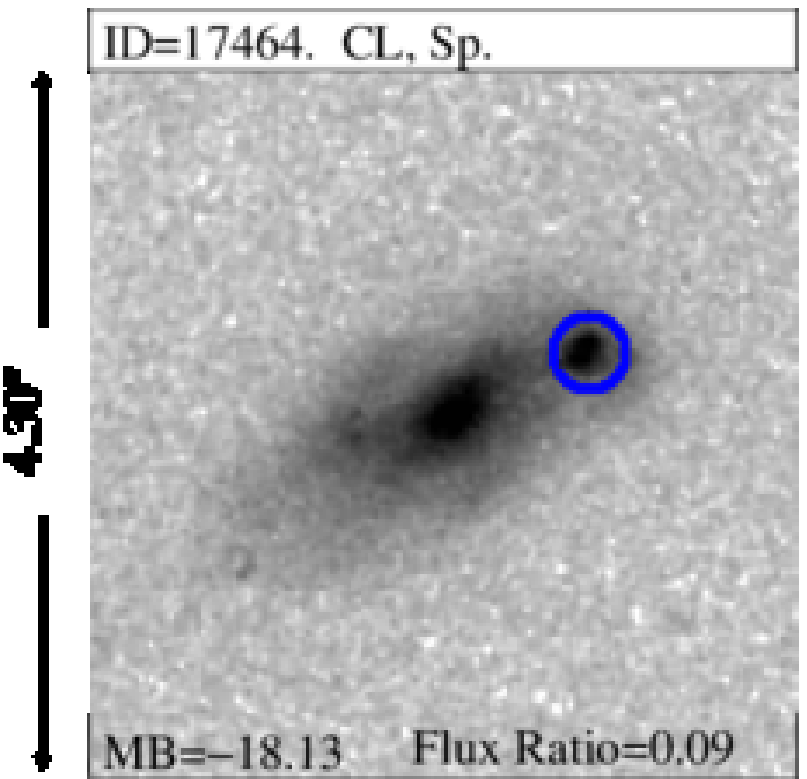} \\   
\includegraphics[scale=0.48]{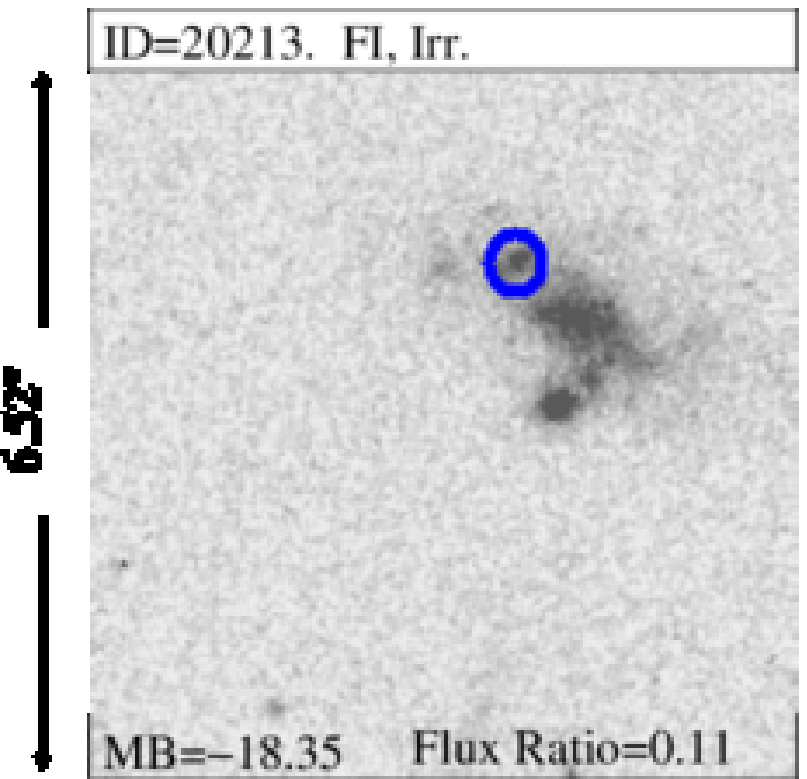}&\includegraphics[scale=0.48]{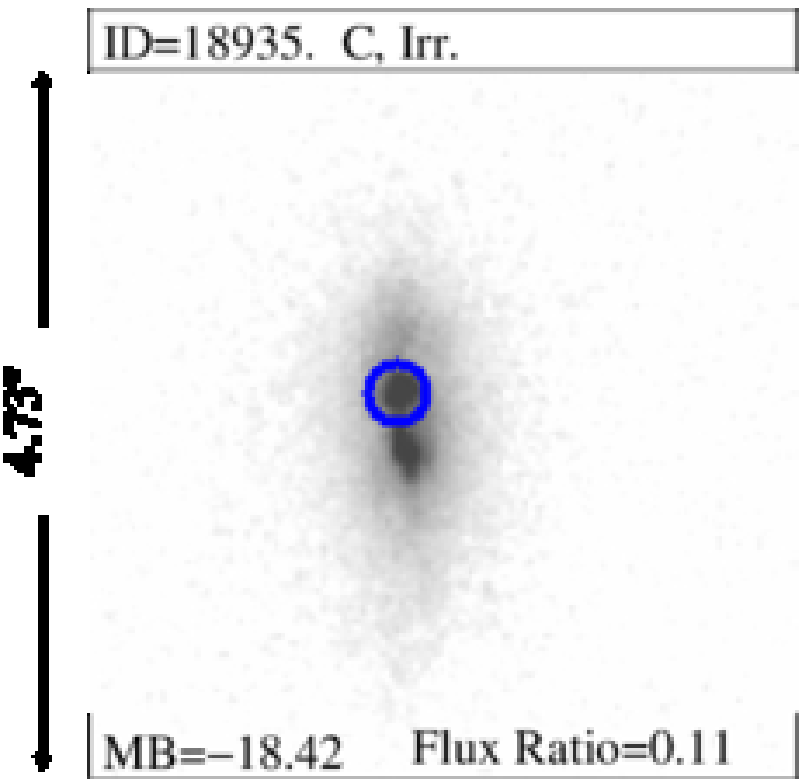} \\   
\includegraphics[scale=0.48]{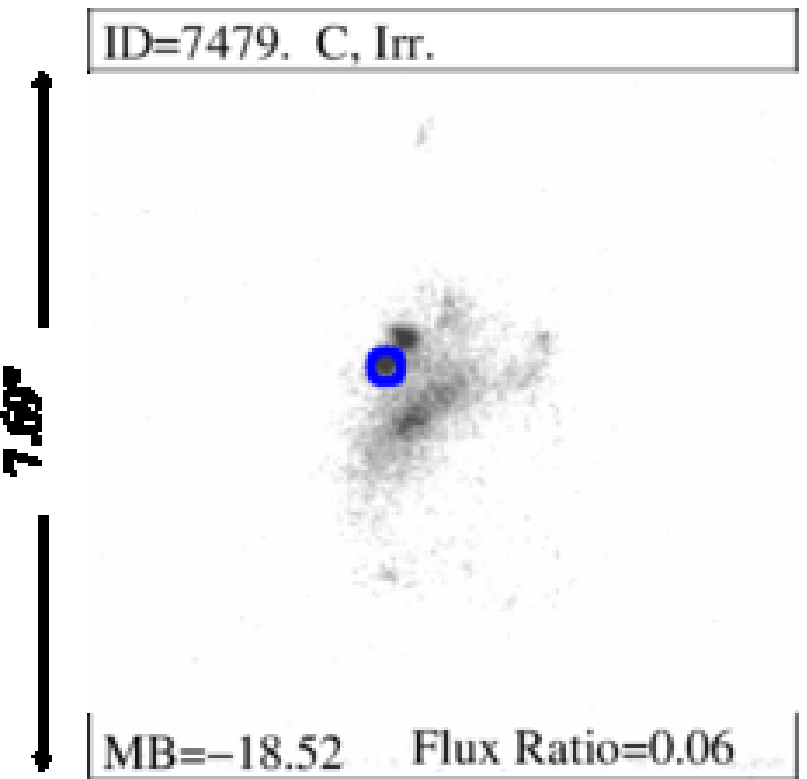}&\includegraphics[scale=0.48]{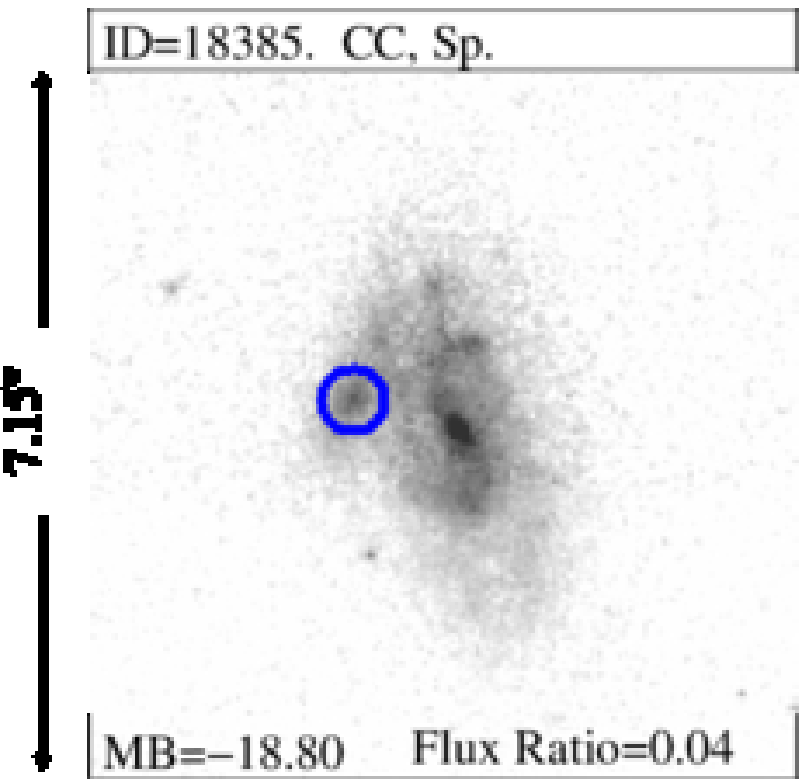} \\   
\includegraphics[scale=0.48]{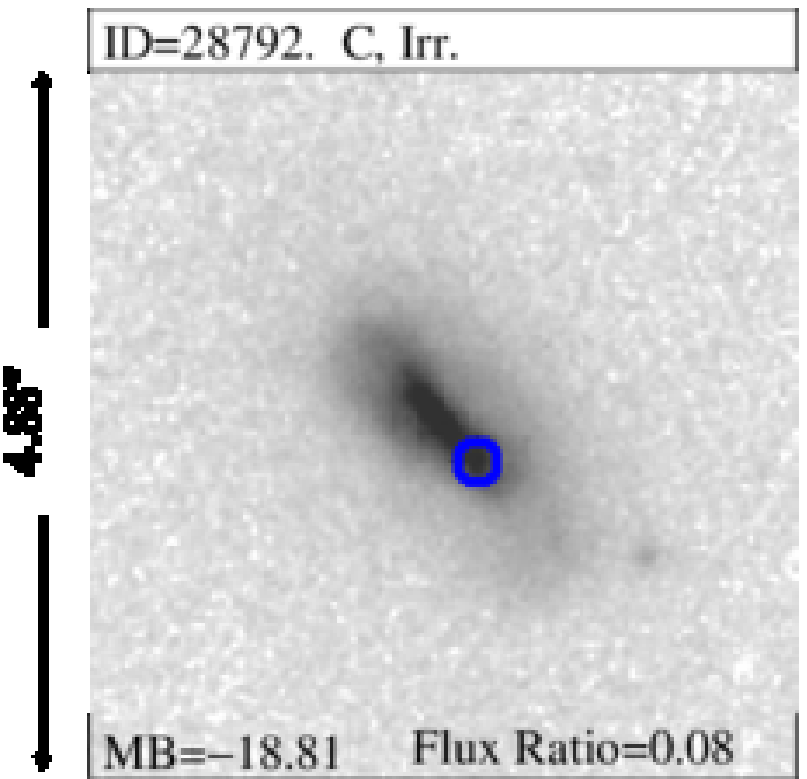}&\includegraphics[scale=0.48]{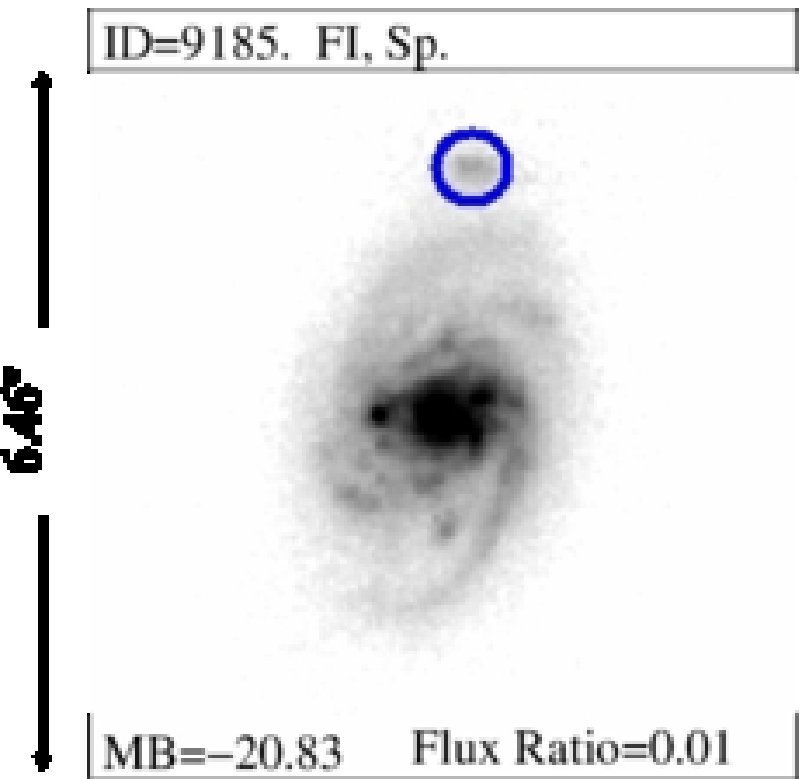} \\   
\end{tabular}
%\vspace{0.2cm}
\end{figure}

\begin{figure}
\contcaption{Visually classified minor mergers found by the $RFF-A(\mathrm{Res})$ diagnostic but missed
by the CAS method adopting its usual limits. COMBO-17 IDs, environments, morphological types and B-band absolute
magnitudes from the \textsc{STAGES} public  catalogue are shown for each galaxy. The panels also give an indication
of the angular extent of the image insets and the approximate fractional contribution of the
light included in the circle to the total \textsc{SEXTRACTOR} automatic flux measurement.}
\label{fig:minormergercandidates1_2}
\begin{tabular}{cc}
\includegraphics[scale=0.48]{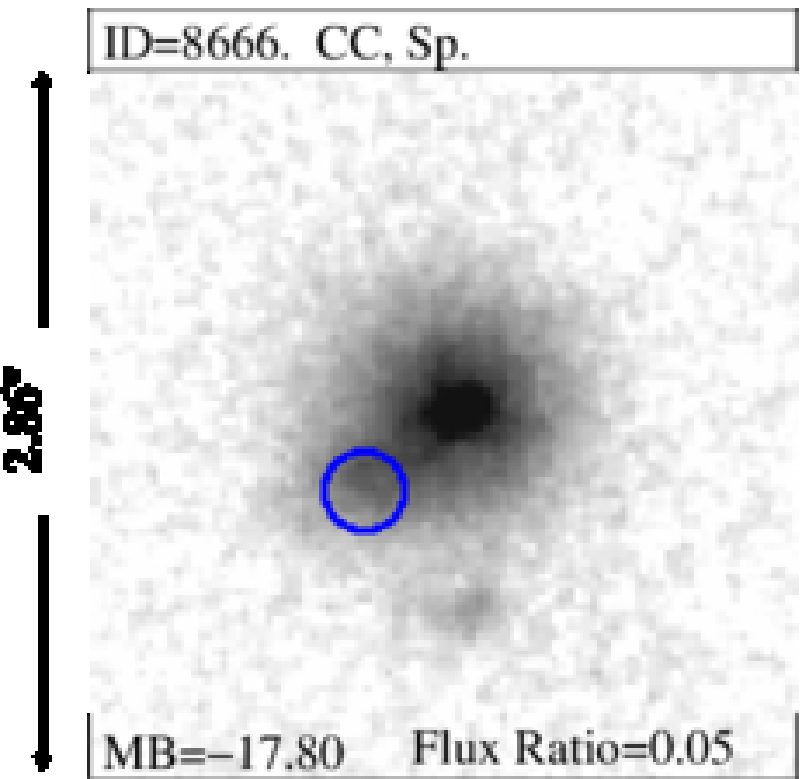}&\includegraphics[scale=0.48]{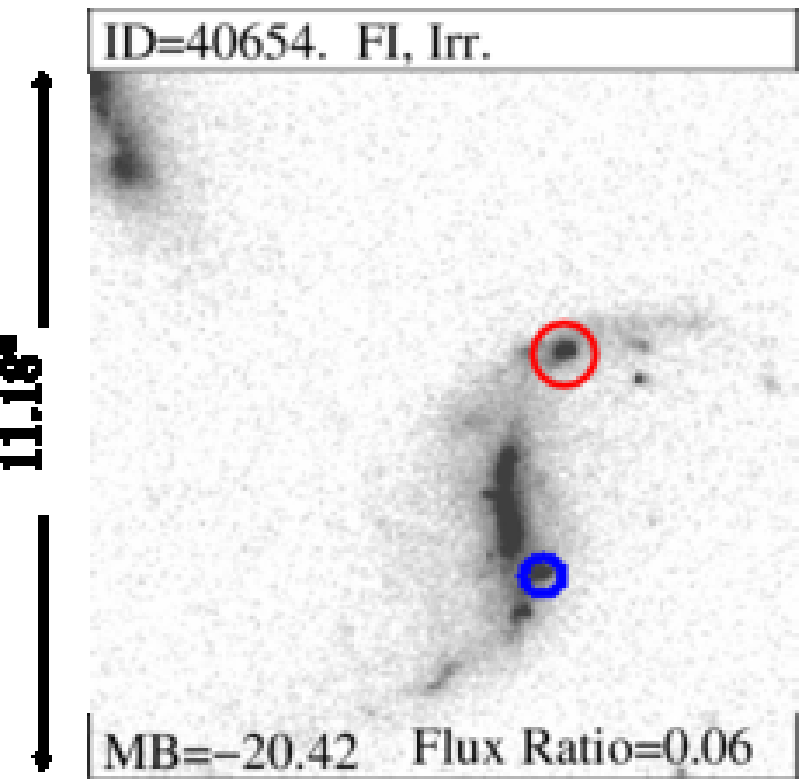} \\  
\includegraphics[scale=0.48]{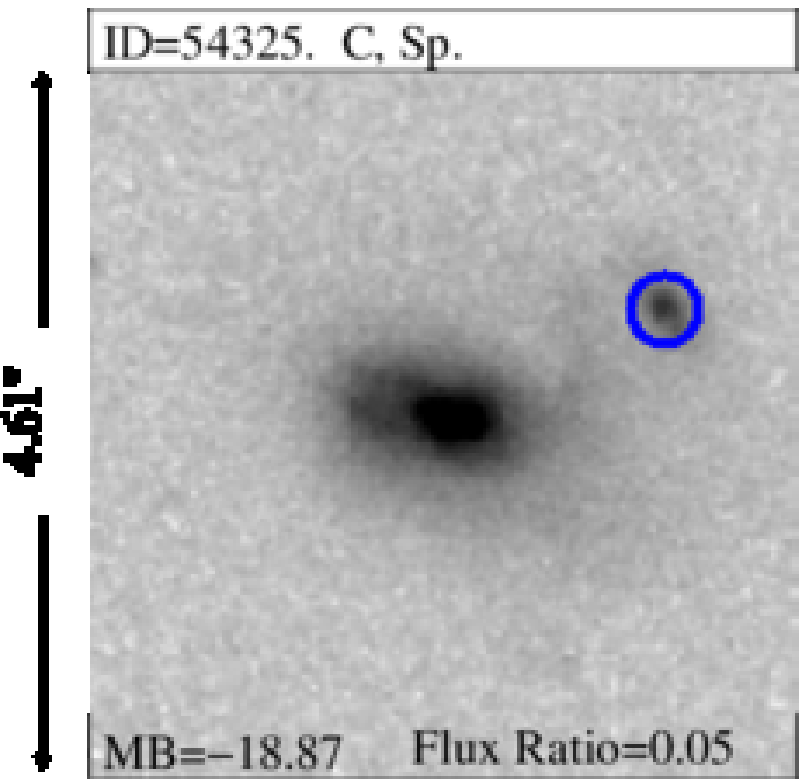}&\includegraphics[scale=0.48]{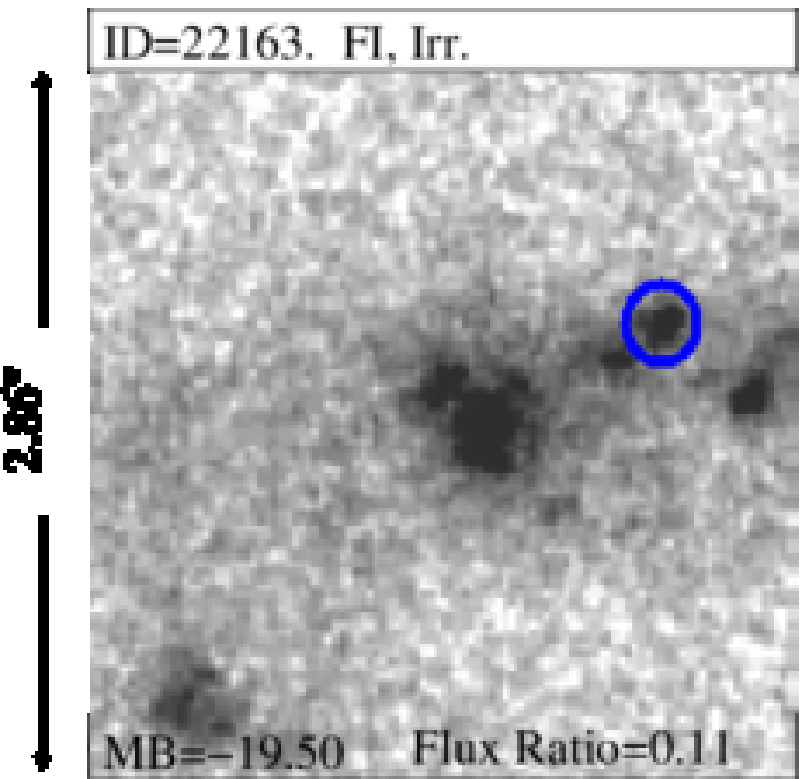} \\   
\includegraphics[scale=0.48]{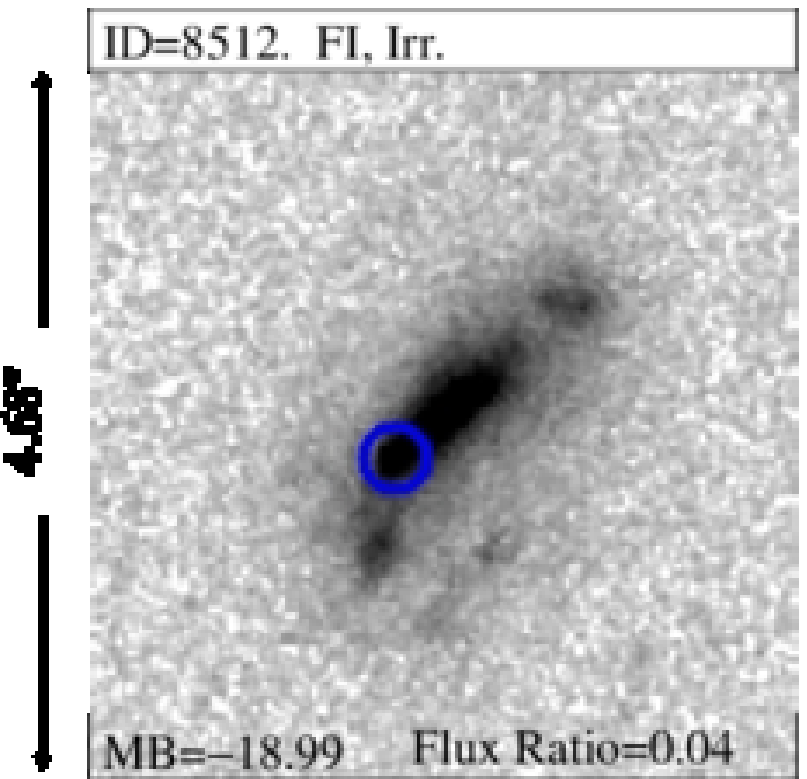}&\includegraphics[scale=0.48]{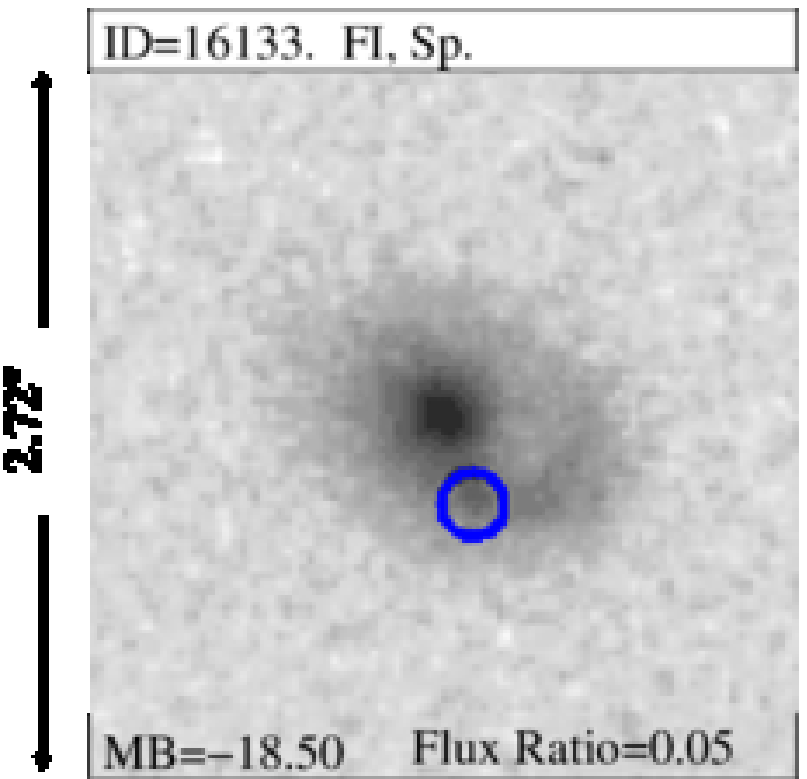} \\    
\hline
\end{tabular}
\end{figure}

Figure~\ref{fig:minormergercandidates1} shows that the $RFF-A(\mathrm{Res})$ method presented here
has detected galaxies that do present inhomogeneities in their light distribution. These inhomogeneities
typically amount to 10\% of the total flux received from the parent galaxy, as estimated
by the crude but robust flux estimate given by \textsc{SEXTRACTOR}. This confirms
the potential validity of the approach presented in the current paper to separate minor mergers
from the whole parent population of galaxies.
However, Figure~\ref{fig:minormergercandidates1}
also makes manifest the reason of the fairly high contamination by non-mergers
that has been found in \S \ref{subsec:contaminacion}. It is very difficult, even with the eye, to tell
apart galaxies undergoing a minor merger episode from galaxies that are experiencing other local
phenomena such as star formation in H\textsc{II} regions. 
This is particularly true in the case of very late 
minor mergers, in which the less massive galaxies have been almost entirely dissolved amidst the larger galaxies.
However, for less evolved merger systems for which the less luminous object has not entirely lost its
individuality, it is easier to separate mergers and non-mergers.
Specifically, objects 20213 and 7479 represent two cases of objects selected by the $RFF-A(\mathrm{Res})$
that could indeed be star formation enhancements.
Finally, Figure~\ref{fig:minormergercandidates1} also shows that, although the structural merger
diagnostic used here can indeed select galaxies with inhomogeneous light distributions, this method alone
cannot identify which light clump is to be identified as the potential satellite. This is best seen in the
case of ID=40654.

\section{Conclusions}
\label{sec:conclusiones}

We present a new structural merger diagnostic geared towards the structural detection of
minor mergers which is entirely based in the morphological
properties of the \emph{residual} images of galaxies after the subtraction of a 
smooth S\'ersic model. The new indicator makes use of the asymmetry of the residuals and of
the Residual Flux Fraction of the fit, both calculated over the Kron aperture
of the galaxies. This diagnostic has been objectively proven to be able of producing
merger samples of equal or better statistical quality than samples obtained using other
well established methods based on the morphological properties of the original images.
In particular, objects with symmetric residuals for which the Residual Flux Fraction is larger 
than 0.2 or objects with more asymmetric residuals for which the RFF is larger than 0.1 are 
very good candidates to be mergers.
We have also found that the Gini index of the residuals could also produce merger 
samples of high statistical purity. In this case, objects for which the Gini index of the 
residual image calculated within the Kron aperture is higher than 0.5 are also good merger
candidates.

Using the structural parameters of the residuals and the limits provided
by the $F-\mathrm{score}$ optimization process shown here, we have split the whole population of galaxies into
two different set of galaxies. The first set, sharing the structural trends and properties of
the mergers included in the training set is shown to contain the majority of \emph{major and minor} mergers. The second set
has been shown to be almost completely free of mergers, as exposed by the very low \emph{negative} contamination rates.
However, given the relative dearth of mergers among the general galaxy
population and the self-imposed goal of detecting the more elusive \emph{minor}
mergers, it turns out that the $RFF-A(\mathrm{Res})$ diagnostic introduced in this paper
works best as a negative merger test. In other words, it is very effective at
selecting non-merging galaxies. In common with all the currently-available
automatic methods, the sample of both \emph{major and minor} merger candidates selected by our test is
heavily contaminated by non-mergers, and further steps are needed to produce
a clean merger sample from the first set of galaxies. Nevertheless, the methodology introduced in this paper
can be very useful when applied to the huge datasets provided by modern large
area surveys such as SDSS or UKIDSS. By visually classifying a relatively
small and manageable number of galaxies, one can derive the resulting cuts and
“best” borders which can then be applied to the whole sample. This would
produce two statistically well-defined sets of galaxies, a smaller one containing
the vast majority of \emph{both major and minor} mergers, and a much larger one almost completely
devoid of them. To identify bona-fide mergers, only the first set would need to
be processed further to remove the non mergers. This could be done either by
visual inspection or by using additional information such as colours or 3-D spectra.

This work also suggests that the use of the structural parameters of the residual images
could indeed be used as a tool to study the properties of minor mergers. We argue that this is 
due to the fact that by removing the bulk of the host galaxy light we might able to detect much 
fainter merging galaxies and over much longer timescales. This will be further studied by using N-body merger 
simulations in a forthcoming paper, which will eventually provide the way to link
the structural parameters measured over the HST/ACS images with the underlying properties
of the observed mergers.

\section*{Acknowledgments}

CH acknowledges financial support from the proyect ``Estallidos de Formaci\'on Estelar. Fase III'', under 
the Spanish Ministerio de Ciencia e Innovaci\'on grant AYA2007-67965-C03-03.
CH also thanks a Spanish MICINN postdoctoral grant.
MEG acknowledges an STFC advanced Fellowship.
SJ acknowledges support from the National Aeronautics and Space
Administration (NASA) LTSA grant NAG5-13063, NSF grant AST-0607748,
and $HST$ grants GO-11082  from STScI, which is operated by
AURA, Inc., for NASA, under NAS5-26555.

%\appendix
%\section{Sextractor Parameters Used in the Second Run.}
%\label{app:SextractorParameters}

\label{lastpage}

\end{document}